\documentclass[letterpaper,twocolumn,10pt]{article}
\usepackage{usenix}

\usepackage{tikz}
\usetikzlibrary{arrows.meta, positioning, calc, shapes.geometric, decorations.pathreplacing}

\usepackage{amsmath}
\numberwithin{equation}{section}
\usepackage{amsfonts}
\usepackage{tabularx,xcolor,booktabs}
\usepackage{threeparttable}
\usepackage{multirow}
\usepackage{makecell}
\usepackage{xurl}
\usepackage{array}
\usepackage{graphicx}
\usepackage{subcaption}
\usepackage{textcomp}
\usepackage{stfloats}
\usepackage{url}
\usepackage{verbatim}
\usepackage{graphicx}

\usepackage{booktabs}
\usepackage{hyperref}
\usepackage{subcaption}
\usepackage{xurl}

\newcommand{\shorturl}[1]{\href{https://#1}{\nolinkurl{#1}}}

\def\eip{EIP-1559}
\usepackage{hyperref}

\begin{document}

\title{On the (In)stability of Adaptive Fee Mechanisms in Multi-Chain Systems}

\author{
{\rm Yifan Zhou\textsuperscript{1}\qquad
Jiang Xiao\textsuperscript{1}\qquad
Xuechao Wang\textsuperscript{2}}\\
{\rm Ru Cheng\textsuperscript{3}\qquad
Shijie Zhang\textsuperscript{4}}\\[0.5em]
\textsuperscript{1}Huazhong University of Science and Technology\\
\textsuperscript{2}HKUST(GZ)\\
\textsuperscript{3}City University of Hong Kong\\
\textsuperscript{4}Nanjing University of Posts and Telecommunications
}

\maketitle

\begin{abstract}

EIP-1559 has become the dominant transaction fee mechanism in permissionless blockchains, regulating scarce block space through an adaptive base fee. The design balances a trade-off between responsiveness to demand fluctuations and stability of long-term resource usage. On a single chain, this trade-off is well understood and empirically stable. In multi-chain systems, however, cross-chain flows introduce delayed economic interactions across otherwise independent fee markets. These interactions alter the control dynamics and create stability risks that existing models do not capture. Deployed systems have already polarized. Polkadot enforces conservative update rates, effectively disabling fee adaptivity and reverting the market to inefficient first-price auctions. Cosmos adopts aggressive update rates without formal guarantees.

To fill this gap, we propose a control-theoretic framework that formalizes multi-chain fee markets as a nonlinear dynamical system with delayed feedback. We identify \emph{delay-amplified over-correction} as the structural root of instability. Extensive experimental evaluations yield quantitative and qualitative findings: cross-chain coupling erodes stability relative to monolithic blockchains, reducing the safe parameter space for base fee update rates by up to 84.8\%. 
Moreover, instability propagates across chains, contaminating the otherwise stable peer chains.
At Ethereum's update rate, which leaves a monolithic chain stable, the multi-chain system exhibits extreme load oscillation between full and near-empty blocks, with over half of all blocks at full capacity. Mainstream Cosmos chains already exceed their safe bounds under their default configurations, and the resulting load oscillation raises their block failure rate by an order of magnitude. Guided by this mechanism, we design a lightweight protocol-level mitigation that removes the over-correction and stabilizes dynamics in multi-chain environments.

\end{abstract}

\section{Introduction}\label{sec: intro}
The transaction fee mechanism~\cite{roughgarden2020transaction, chung2023foundations} is a fundamental component of permissionless blockchains~\cite{nakamoto2008bitcoin, wood2014ethereum, ren2019analysis}. It allocates scarce block space and deters spam, serving both security and usability. 
On security~\cite{buterin2018blockchain, liu2022empirical}, the mechanism uses \textit{gas} as a universal unit of computation and enforces a strict per-block resource cap, namely the \textit{gas limit}, which prevents system overload and preserves decentralization, consensus security, and protocol resilience~\cite{garay2015bitcoin, croman2016scaling, sompolinsky2015secure}. On usability, each transaction consumes gas determined by its execution, and its initiator pays a fee. The fee per unit of gas acts as a market-based price signal that incentivizes validators to include transactions and regulates user demand under congestion. Ethereum's original first-price auction produced highly volatile fee markets, because users competing for inclusion had no pricing guidance and relied only on discretionary tips to the block proposer.

This limitation motivated Ethereum Improvement Proposal 1559 (\eip)~\cite{eip-1559-proposal}. 
EIP-1559 replaces the first-price auction with a market-driven adaptive \textit{base fee}, and introduces elastic block capacity that absorbs short-term demand spikes by allowing blocks to temporarily exceed the target gas usage up to twice the pre-\eip~limit. 
At its core, \eip~is a negative feedback control loop. The base fee increases when block usage exceeds the target, scaled by a critical update parameter $\delta$, and decreases symmetrically otherwise, driving the long-run average load toward a fixed gas-usage target. The mechanism improves usability, because fees become predictable and less volatile, and preserves security, because the effective operating load converges to the target~\cite{liu2022empirical}.

The effectiveness of \eip~hinges on $\delta$, which presents a fundamental trade-off between usability and security. If $\delta$ is set too conservatively, the mechanism responds sluggishly and collapses into an unpredictable first-price auction~\cite{roughgarden2020transaction}. If $\delta$ is set too aggressively, the negative feedback effects can destabilize the system and induce load oscillations~\cite{leonardos2021dynamical}. 
Moreover, stability is a security requirement. Stable operation keeps the realized long-run normalized load at the target and thus at the pre-\eip~baseline. 
Instability drives the load into a persistent oscillation~\cite{leonardos2021dynamical, leonardos2023optimality, leonardos2026dynamics}, which deepens from a pattern that never fills a block to one
in which blocks alternate between the gas limit and near-emptiness, with more than half
of all blocks full in the extreme case.
A monolithic chain never enters this load oscillation at any deployed update rate. We discuss the security 
implications in Section~\ref{sec: security-eip-1559}.

Monolithic blockchains such as Ethereum currently operate within a stable regime, and empirical evidence confirms that it remains robust despite high responsiveness~\cite{eip-1559-experiment, liu2022empirical}. 
This stability has led to widespread adoption of \eip controllers beyond Ethereum~\cite{leonardos2021dynamical}, and also in multi-chain systems such as Polkadot~\cite{Polkadot} and Cosmos~\cite{cosmoswhitepaper}. These systems scale through functional specialization. They run multiple independent blockchains in parallel, referred to as \textit{parachains} and \textit{zones}, respectively, and each chain typically implements an independent instance of \eip. Unlike monolithic chains, however, the chains are coupled through interoperability protocols~\cite{herlihy2019cross, xie2022zkbridge} that enable cross-chain composition, introducing new interactions between otherwise isolated fee markets.

\noindent\textbf{Motivation.} 
Transplanting the responsiveness-stability trade-off of \eip~into multi-chain architectures creates a critical dilemma, and deployed systems have already polarized. At one extreme, Polkadot enforces update rates as small as $10^{-5}$ (Table~\ref{table: data}) and ensures stability, which turns the base fee into a quasi-\textit{static tax}. Users fall back on tips under congestion and the market regresses to an inefficient first-price auction~\cite{roughgarden2020transaction}. 
At the other extreme, Cosmos zones such as Osmosis adopt aggressive Ethereum-like update rates without formal guarantees. Cosmos has not yet exhibited instability in production, because weak demand keeps these chains out of the high-load regime where the feedback dynamics become visible, so their traces characterize only the low-load regime and cannot establish stability when user demand is active.

This dilemma reflects a structural shift introduced by multi-chain systems. Unlike atomic transactions in monolithic blockchains, a single user action unfolds as a sequence of sub-transactions executed across blockchains. 
Such workflows introduce delayed economic coupling, two effects absent from monolithic chains. A load disturbance on one chain is regulated by its own \eip~feedback, but it also propagates through cross-chain flows into the feedback loops of connected chains. Second, cross-chain delay makes the fee mechanism respond to outdated states. The two effects transform stability from a local property into a systemic concern.

\noindent\textbf{Research problem.} Existing models~\cite{leonardos2021dynamical, leonardos2023optimality} fail to capture the dynamics introduced by cross-chain flows. It is therefore unclear whether the monolithic stability still applies, how the safe regime of the update rule changes, and what security consequences arise when stability is lost. 
This motivates a dedicated multi-chain analysis. However, the multiplicative update rule makes the dynamics nonlinear, and cross-chain latencies convert them into a delayed discrete-time dynamical system. The stability depends on historical states and is difficult to analyze~\cite{gu2003stability}. We therefore linearize the dynamics around the load target to derive a local asymptotic analysis, and simulate the full nonlinear system to quantify the security consequences for realistic multi-chain architectures.

\noindent\textbf{Our contributions.} We systematically analyze the stability of \eip~in 
multi-chain systems.
The instability we identify is \emph{structural}. Within the unstable regime defined by the system 
parameters, it arises without any adversary, and once a single chain crosses its own safe bound, it 
propagates to conservatively configured chains. Building on the validated mechanisms that incur 
instability, we further propose a protocol-level mitigation.
Our contributions are threefold.
\begin{itemize}
    \item \textbf{Trade-off characterization.} We characterize the responsiveness-stability trade-off, which is manageable in monolithic settings but destabilized in multi-chain systems. Suppressing responsiveness, as in Polkadot, degrades into an inefficient first-price auction, while retaining high responsiveness, as in Cosmos, can incur severe security outcomes once the chain becomes active.

    \item \textbf{Stability analysis (Section~\ref{sec: stability-analysis}).} 
    We formalize the dynamics induced by cross-chain interactions through a multi-chain model. We conduct an asymptotic stability analysis. The analysis identifies \textit{delay-amplified over-correction} as the instability mechanism, where cross-chain latencies cause negative corrective feedback to act on stale, larger past deviations. 
    This over-correction is stronger with higher cross-chain inbound loads.

    \item \textbf{Experimental validation (Section~\ref{sec: oscillation-analysis}).} 
    The safe regime of the \eip~update rate, $\Delta_{\text{safe}}$, shrinks by 84.8\% relative to 
    the monolithic baseline.
    The standard rate of $0.125$, which keeps a monolithic chain stable, causes instability that reaches the deepest stage of the oscillation. About half of all blocks saturate the gas limit, creating consistently large blocks that the 
    system does not provision.
    The deployed rates of Cosmos Hub and Osmosis already exceed their safe bounds, and the 
    saturation raises their block failure rate by an order of magnitude.

\end{itemize}

\section{Background}\label{sec: background}

\subsection{Security Implications of Instability}\label{sec: security-eip-1559}

Stability of \eip~is a security requirement explicitly specified in the \eip~official document~\cite{eip-1559-proposal}.
Ethereum adopted \eip~and doubled its gas limit, setting the target normalized load to $0.5$, so the target load equals the pre-\eip~capacity.
Under stable operation, the load asymptotically converges to the target.
Under instability, the load instead alternates between two levels,
one below the target and one above it, and the two levels spread apart as the
configuration moves further past the stability boundary.
Finally, blocks alternate between the gas limit and near-emptiness~\cite{leonardos2021dynamical}, and the share of blocks at the limit grows as the instability deepens. In the extreme case, more than half of the blocks reach the gas limit.
\eip~sets the gas limit at twice the target, so occasional full blocks are expected by design, but it does not anticipate that more than half of the blocks are full.

This saturation has two consequences. First, it intensifies \textit{centralization}. Hardware-constrained validators are priced out, shrinking the active validator set~\cite{croman2016scaling}. Second, it degrades \textit{consensus resilience}. Larger blocks take longer to propagate and execute, and a validator that has not finished the current block cannot react to the next one. The result is an honest power split, and each consensus family pays for it in its own timing budget. 
The honest-majority guarantee of Nakamoto consensus weakens as blocks grow~\cite{garay2015bitcoin, sompolinsky2015secure, gervais2016security, dembo2020everything}. Under LMD-GHOST, an attester must vote before a deadline, but only after execution, so a late attester casts no vote, and a block with too few timely votes is easier to override. Synchronous BFT~\cite{abraham2019synchronous} assumes message delays below a threshold, and added latency can break this security threshold.
Partial-synchronous BFT~\cite{yin2019hotstuff, gelashvili2022jolteon, lewis2024lumiere, malkhi2023hotstuff} triggers costly view synchronization~\cite{lewis2024lumiere} once delays exceed the timeout.

Two past incidents show that unprovisioned load is severe. The 2016 Shanghai DoS attacks exploited Ethereum's underpricing of the I/O cost of \texttt{EXTCODESIZE} and continuously created slow-to-execute blocks, reducing the block-production rate\footnote{\url{https://eips.ethereum.org/EIPS/eip-150}}.
Ethereum's gas refund mechanism allowed a load of up to twice the gas limit\footnote{\url{https://eips.ethereum.org/EIPS/eip-3529}}, so EIP-3529 removed most of the refund due to security concerns, and EIP-7778\footnote{\url{https://eips.ethereum.org/EIPS/eip-7778}} was proposed to eliminate the residual refund that still introduces unprovisioned load. Both cases required either an adversary or a mispriced opcode. Multi-chain instability produces the unprovisioned load with more than half of the blocks being full just from an update rate where monolithic chains are stable.

\subsection{Multi-Chain Architectures}\label{sec: multi-chain-blockchain}
Polkadot and Cosmos are two representative systems. At a high level, such systems execute a single user action as a \textit{cross-chain workflow}, namely a sequence of dependent sub-transactions distributed across distinct chains. This introduces structural constraints across chains.
Consider a two-chain workflow. A sub-transaction is initiated on source chain $i$ at time $t$ (measured in block steps). The system generates a cryptographic inclusion proof and relays it to destination chain $j$, incurring a rigid physical latency $d$ determined by consensus finality and relay overhead~\cite{herlihy2019cross, chervinski2023analyzing}, and the destination-side sub-transaction commits on $j$ at time $t+d$.

\section{Warmup: Single-Chain Baseline}\label{sec: warmup}

\subsection{The \eip~Mechanism}
\eip~\cite{eip-1559-proposal} introduces an algorithmic \textit{base fee} $P(t)$ as a mandatory toll that tracks user demand, where $t$ is the discrete block number. Users still attach a self-specified priority tip for miner incentives, and the gas price is the sum of the two. The protocol expands the gas limit and steers the \textit{normalized load} $L(t) \in [0,1]$, the fraction of the gas limit consumed, toward a target $T$ through the update rule:
\begin{equation}\label{eq: eip1559-update}
    P(t+1) = P(t)\left(1 + \delta \frac{L(t) - T}{T}\right).
\end{equation}
This update rule implements a negative feedback loop. The base fee rises when $L(t) > T$ and falls otherwise, and is expected to become the dominant price signal that regulates user demand and shapes the subsequent block load.
The update rule has two configurable parameters: the update rate $\delta$ and the target load $T$. $\delta$ controls the feedback force and embodies a responsiveness-stability trade-off.
A small $\delta$ lets a full block change the fee by only $\delta(1-T)/T$, forcing users onto priority tips.
A large $\delta$ triggers fee spikes that over-suppress demand and may induce instability. Ethereum sets $\delta = 0.125$, which is empirically stable~\cite{eip-1559-experiment, liu2022empirical}.
The target $T$ is coupled to the expanded block gas limit. Given an original gas limit $S$, a protocol may use $T=0.5$ with $g_{\text{lim}}=2S$ (Ethereum), or $T=0.25$ with $g_{\text{lim}}=4S$. Both configurations yield $T \cdot g_{\text{lim}}=S$ and thus maintain security if stability is ensured. 

\subsection{Definitions}\label{sec: monolithic-definition}
We introduce key definitions in the monolithic setting, which the multi-chain model later extends. 
The \eip~update rule steers load toward $T$, implying an \textit{equilibrium price} $P^*$, also called the market-clearing price or stationary point~\cite{roughgarden2020transaction, leonardos2021dynamical}, at which user demand matches the target block load:
$$
    \mathcal{G}(P^*) = T \cdot g_{\text{lim}},
$$
where $\mathcal{G}(P)$ maps the gas price to block demand. We model the gas price as the algorithmic base fee, excluding priority tips. This abstraction isolates the \eip~control and attributes the instability only to its feedback loop.
We define the normalized price as $x(t)=P(t)/P^*$, so the equilibrium is $x^*=1$. This normalization removes chain-specific price units and lets the analysis depend only on deviations from equilibrium. The update rule~\eqref{eq: eip1559-update} becomes:
\begin{equation}\label{eq: normalized-update}
    x(t+1)=x(t)\left(1+\delta\frac{L(t)-T}{T}\right).
\end{equation}

\noindent\textbf{Demand elasticity.} At $P^*$, the load equals the target and the base fee remains unchanged. However, the fixed point does not guarantee convergence after an arbitrarily small load perturbation. Convergence depends on how sensitively the next-step price responds to load deviations (governed by $\delta$), and on how demand responds to price changes \emph{around $P^*$}.
We characterize the latter via demand elasticity:
\begin{equation}\label{eq: elasticity-def}
    \lambda=-\frac{\partial \mathcal{G}/\mathcal{G}}{\partial x/x}\bigg|_{x=1}.
\end{equation}
Intuitively, this measures the local sensitivity of demand at equilibrium, and the global shape of $\mathcal{G}$ does not enter. A $1\%$ price increase reduces demand by approximately $\lambda\%$.

\subsection{Monolithic Stability Analysis}\label{sec: monolithic-analysis}
We study \emph{local asymptotic stability}, namely whether the trajectory returns asymptotically after a small perturbation from equilibrium. The fee mechanism is intended to operate around the market-clearing point, and an equilibrium from which arbitrarily small perturbations do not decay cannot serve as a stable operating point.
Since the analysis is local, the generated load stays near the target, so $\mathcal{G}$ does not reach the limit and directly maps to the next-step load.
Let $\Delta x(t) = x(t) - 1$ denote the price deviation.

\noindent\textbf{Linearization around $P^*$.} With demand elasticity $\lambda$, we expand the demand function around $x^*=1$ to first order:
$$
    \mathcal{G}(x(t))\approx \mathcal{G}(1)+\frac{\partial \mathcal{G}}{\partial x}\bigg|_{x=1}\Delta x(t)=T\,g_{\text{lim}}-\lambda\,T\,g_{\text{lim}}\,\Delta x(t).
$$
A price above equilibrium ($\Delta x > 0$) suppresses load below the target, with the magnitude governed by $\lambda$. Moreover, around $P^*$, load convergence and price convergence are equivalent: $\Delta x(t)\to 0$ if and only if $\Delta L(t)=L(t)-T\to 0$. We therefore study stability through the price dynamics.
Substituting the linearized demand into the normalized update rule~\eqref{eq: normalized-update} with $L(t)=\mathcal{G}(x(t))/g_{\text{lim}}$ yields the linear recurrence:
\begin{equation}\label{eq: monolithic-recurrence}
    \Delta x(t+1) \approx (1 - \delta\lambda)\,\Delta x(t).
\end{equation}

\noindent\textbf{Stability condition.} The price perturbation decays to zero if and only if $|1 - \delta\lambda| < 1$:
\begin{equation}\label{eq: monolithic-boundary}
    \delta\lambda < 2.
\end{equation}
The set of stable update rates is $(0,\; 2/\lambda)$. With Ethereum's $\delta = 0.125$, the system remains stable for all $\lambda < 16$.
On-chain estimates place Ethereum's elasticity between $9.09$ and $16.29$, with a preferred estimate of $12.45$~\cite{ndiaye2024blockchain}, so Ethereum operates inside this bound, consistent with empirical stability~\cite{liu2022empirical}.

\noindent\textbf{Load pattern.} Note that instability does not make the price 
diverge. A price far above $P^*$ drives the load toward zero and is pulled back down, and a price far below  $P^*$ fills blocks and is pulled back up, so the price stays in a bounded band even when it never settles. 
The load keeps oscillating around the target.
Prior work turns this into a bound on the load inflation level, the fraction by which the long-run average load exceeds the target. The bound is $6.27\%$ at $\delta = 0.125$~\cite{leonardos2023optimality}.
The bound comes from an asymmetry. A full block multiplies the price by $1+\delta$ and an empty block by $1-\delta$, and $(1+\delta)(1-\delta) < 1$, so full blocks must outnumber empty ones.
In the extreme case where every block is either full or empty, keeping the price bounded requires a 
fraction $f = [1-\ln(1+\delta)/\ln(1-\delta)]^{-1} = 53.13\%$ of full blocks, which is exactly the 
$6.27\%$ bound. Any other load pattern inflates the load by less, because $\ln(1+\delta u)$ is concave in the normalized deviation $u = (L-T)/T$.

Prior work also discusses the load pattern and traces the oscillation progression in its bifurcation diagrams~\cite{leonardos2021dynamical, leonardos2023optimality}. Past the stability
boundary, the load splits into two branches around the target, and the branches spread
apart as $\delta$ grows, until they pin at the gas limit and near zero. The progression contains two stages. In the \textit{unsaturated stage}, right past the boundary, the peak load stays below the gas limit, and no block is ever full. In the \textit{saturated stage}, the peak load reaches the gas limit, and a share of blocks are full. 
The unsaturated stage is faint in the prior diagrams, and the reason is that the prior analyses assume users have similar valuations for block inclusion. 
Therefore, a fee below the common level lets every user in, and a fee above it lets none in, so the load pattern enters the saturated stage shortly after instability.
However, valuations on deployed chains are instead dispersed. The fee response spreads over a wide price range, and on-chain estimates measure it as an elasticity~\cite{ndiaye2024blockchain}. We therefore model the fee response through the demand elasticity and a non-strategic demand function. Monolithic chains sit inside the stable region and enter neither stage. We find that cross-chain coupling pushes configurations that are stable as a single chain into the saturated stage, and
Section~\ref{sec: boundary} shows that some reach the far end of that stage at Ethereum's $\delta = 0.125$.

\section{Motivation}\label{sec: multi-chain-problem}

\subsection{The Polarization Problem}\label{sec: motivation-polarization}
Despite the robust stability of \eip~in monolithic systems, its deployment across multi-chain architectures has led to polarization. We survey the two largest multi-chain systems, Polkadot and Cosmos. We select chains that explicitly adopt the \eip~feedback mechanism and carry substantial cross-chain traffic, and examine the dynamics on their peak-activity day in 2025. Since \eip~updates the base fee per block and block intervals are below tens of seconds on those chains, a single day yields thousands of block-level observations, sufficient to reveal the pattern.

\begin{figure}[t]
    \centering
    \includegraphics[width=0.95\columnwidth]{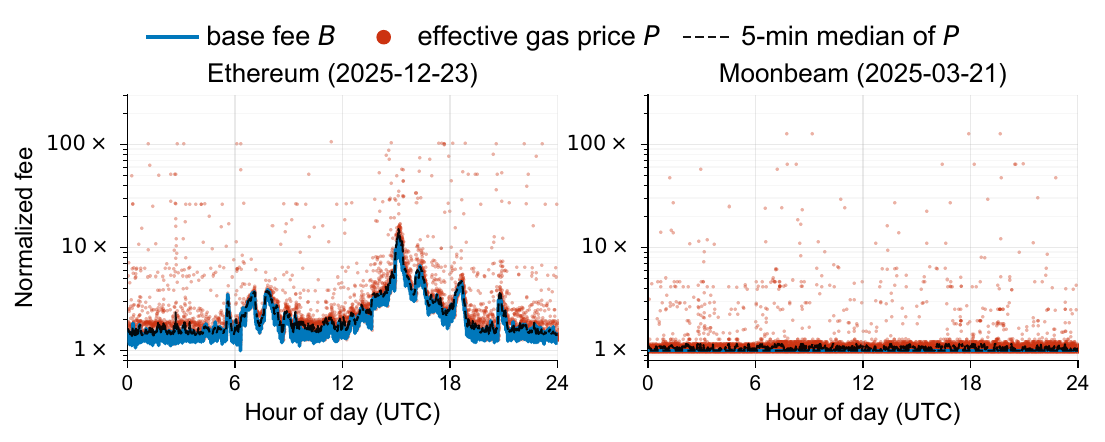}
    \includegraphics[width=0.95\columnwidth]{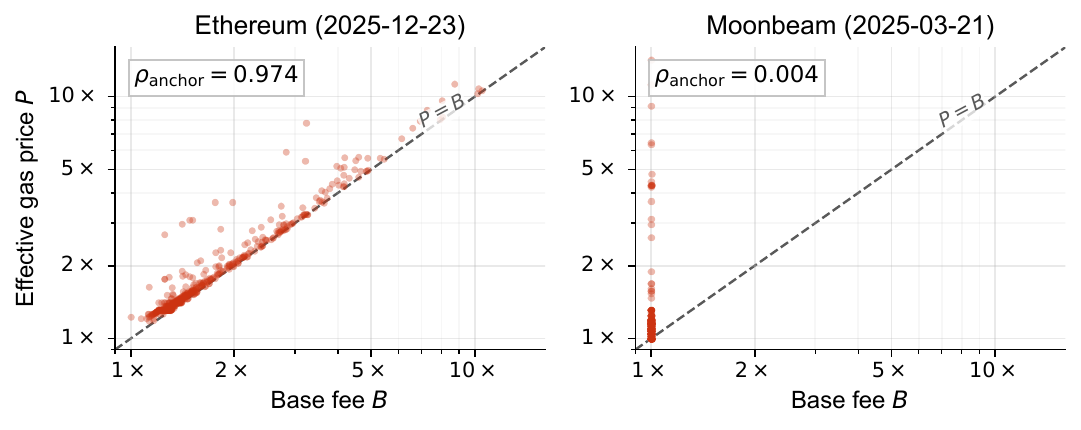}
    \caption{Fee dynamics on Ethereum and Moonbeam. The upper panels show the full-day fee trajectories. The lower panels show the relationship between the base fee and effective gas price for the top \(10\%\) of blocks by observed load. All fee values are normalized by the minimum base fee observed on the corresponding chain during the sampled day.}
    \label{fig: fee-comparison}
\end{figure}

\noindent\textbf{Polkadot.} Polkadot enforces extremely small update rates, $\delta \approx 7.5 \times 10^{-5}$ on the standard Substrate chains and $\delta=0.004$ on the EVM-compatible Moonbeam (sources in Table~\ref{table: data}). The base fee may no longer function as an effective price anchor.
Figure~\ref{fig: fee-comparison} compares fee dynamics on Ethereum and Moonbeam at the block level. Since tip fees vary across transactions, we report for each block the base fee alongside an effective gas price, defined as the base fee plus the median tip in that block. In the upper panels, the blue curve connects the per-block base fees in timestamp order, and the black curve shows the five-minute median trajectory of the effective gas price. On Ethereum, the base fee varies continuously, and the effective gas price closely follows it. On Moonbeam, the base fee remains almost unchanged, and the effective gas price still stays close to it, possibly due to low demand.

The two chains separate under high load (top $10\%$). The lower panels measure the coupling between the base fee \(B_i\) and the effective gas price \(P_i\) as \(\rho_{\mathrm{anchor}}=\operatorname{Corr}(\log B_i,\log P_i)\) in that load regime. Ethereum yields \(0.974\) and forms a diagonal band, because the effective gas price moves with the base fee. Moonbeam yields \(0.004\), because the base fee stays nearly fixed while the effective gas price varies. The price variation on Moonbeam is thus carried almost entirely by the priority fee \(T_i\), a pattern consistent with first-price-like competition for block inclusion above a nearly fixed base-fee floor. Note that this does not contradict our exclusion of priority tips from the analytical model. The exclusion isolates the algorithmic base-fee loop, and the observation demonstrates exactly the \emph{consequence} when this loop fails, namely that tips absorb all price discovery and volatility returns.

\noindent\textbf{Cosmos.} Several Cosmos zones, such as Cosmos Hub and Osmosis, replicate Ethereum's EIP-1559 controller parameters. However, the measured chains lack sufficient demand and rarely sustain load above the controller target, so their on-chain traces provide little evidence about the controller dynamics after the chain becomes active. 
As Polkadot enforces conservative settings and Cosmos is inactive, our analysis aims to address this unobserved regime ex ante.

\subsection{Delayed Economic Coupling}\label{sec: delayed-economic-coupling}
The workflow structure in Section~\ref{sec: multi-chain-blockchain} introduces two systemic effects that the single-chain model does not capture.
First, the workflow from chain $j$ to $i$ consumes gas twice, on the source chain $j$ at time $t$ and on the destination chain $i$ at time $t+d$ as inbound load. The cost of a user action therefore depends on the fees of all chains involved, and a fee change on one chain alters cross-chain demand and the loads on connected chains. We term this effect \textit{economic coupling}.
Second, the two consumptions are separated by the rigid latency $d$, so inbound load on chain $i$ at time $t$ reflects price conditions on $j$ at time $t-d$ rather than the current state.
We refer to this temporal misalignment as \textit{feedback lag}.

These two effects compose into a closed feedback loop across chains. When chain $i$'s fee change shifts the load on a connected chain $j$, chain $j$'s \eip~controller adjusts its own base fee, which in turn modulates the cross-chain traffic flowing from $j$. Each chain's local controller remains a negative-feedback loop, but the loops are wired together through delayed cross-chain traffic. Stability becomes a \emph{systemic} property of the coupled network, governed by coupling strength and delay magnitude, two factors absent from $\delta\lambda < 2$.

Capturing these dynamics requires three extensions, including load decomposition into local, outbound, and inbound components, a demand specification responding to multiple prices, and an explicit latency model.

\subsection{Research Questions}\label{sec: problem}
The above discussion leads to two central questions.

\begin{itemize}
     \item \textbf{RQ1:} How does delayed economic coupling alter the stability condition, and which mechanism drives it or propagates it across chains?

    \item \textbf{RQ2:} Does this coupling erode the safe update rate regime, and if instability occurs, how severe is the resulting load pattern and what harm does it do to consensus?
\end{itemize}

We answer RQ1 through formal modeling (Section~\ref{sec: multi-chain-model}) and stability analysis (Section~\ref{sec: stability-analysis}), and validate the driving and propagating channels experimentally (Sections~\ref{sec: validation} and~\ref{sec: non-uniform-delta}). We answer RQ2 through evaluation of the mechanism that causes instability in Section~\ref{sec: oscillation-analysis}.

\section{Multi-Chain Model}\label{sec: multi-chain-model}

\subsection{General Assumptions}\label{sec: core-assumptions}
We consider a system of $N$ chains, each implementing \eip~and satisfying standard SMR properties~\cite{blum2023abraxas, dai2023parbft, dai2024ipotane}. Each chain is connected to other chains with cross-chain flow, which involves one source chain and one destination chain. Multi-hop workflows add further delayed dependencies that only amplify the instability we investigate. 
For local asymptotic stability analysis, we abstract away the mempool of pending transactions, because the demand remains near the target and creates no backlog. The simulation includes a mempool to capture backlog effects.
The threat we identify is an \textit{emergent systemic instability}, determined by the system structure and protocol parameters rather than by an attacker. As Section~\ref{sec: oscillation-analysis} shows, a one-block $1\%$ perturbation suffices to expose it.

Each chain $i$ has gas limit $g_{\text{lim}}^i$, target $T_i \in (0,1)$, equilibrium base fee $P_i^*$, and update rate $\delta_i > 0$.
Let $\mathbf{P^*}=(P_1^*, \ldots, P_N^*)$ denote the equilibrium prices at which aggregate demand matches each chain's target load. Following Section~\ref{sec: warmup}, we use the normalized price $x_i(t) = P_i(t)/P_i^*$, so the equilibrium is $\mathbf{x}^*=\mathbf{1}$. The update rule preserves its form:
\begin{equation}\label{eq:multi-chain-eip-1559}
    x_i(t+1) = x_i(t)\left(1 + \delta_i\frac{L_i(t) - T_i}{T_i}\right).
\end{equation}

\subsection{Model Specification}\label{sec: multi-chain-specification}

\noindent\textbf{Load composition.} Without loss of generality, we specify the model with respect to chain $i$. Chain $i$'s load has three components: local flows ($\mathcal{G}_{i\to i}$), outbound flows ($\mathcal{G}_{i\to j}$), and delayed inbound flows ($\mathcal{G}_{j\to i}$). Our analysis targets the local area around equilibrium, so we define the equilibrium load shares $\alpha_{i,\text{local}}$, $\alpha_{i\to j,\text{out}}$, and $\alpha_{j\to i,\text{in}}$ as the fractions of chain $i$'s target load at $\mathbf{x}^* = \mathbf{1}$, with
$$
\alpha_{i,\text{local}} + \sum_{j\neq i}\alpha_{i\to j,\text{out}} + \sum_{j\neq i}\alpha_{j\to i,\text{in}} = 1.
$$
These shares are structural constants and are not tunable, determined by load topology.  Cross-chain ratios of representative chains range from $18\%$ to $43\%$ (Appendix~\ref{sec: statistics}).
Realistic systems exhibit heterogeneous block intervals across chains. This can be absorbed into $\alpha$. If a source chain produces blocks twice as fast as the destination, the cumulative inbound load per destination block is captured by a larger $\alpha_{j\to i,\text{in}}$. This absorption avoids sub-step simulation and enables a common block-step abstraction without loss of generality.

\noindent\textbf{Multi-chain demand elasticity.} An intuitive calibration of cross-chain demand is to sum all fees into an aggregate cost and apply a single elasticity. 
However, as chains denominate fees in distinct native tokens with floating exchange rates, no canonical aggregate cost exists. We therefore treat the source-side and destination-side prices as separate variables, each carrying its own elasticity. 
The treatment loses no generality. Linearizing any aggregate-cost demand around $\mathbf{x}^* = \mathbf{1}$, even one with a single uniform elasticity, still yields one sensitivity per price, so any aggregate specification reduces to a special case of the per-price elasticities.
We specify the demand functions around equilibrium as:
\begin{align*}
     \mathcal{G}_{i\to i}(x_i) &\approx \alpha_{i,\text{local}}\,T_i\,g_{\text{lim}}^i\,x_i^{-\lambda_{ii}}, \\
     \mathcal{G}_{i\to j}(x_i, x_j) &\approx \alpha_{i\to j,\text{out}}\,T_i\,g_{\text{lim}}^i\,x_i^{-\lambda_{ij}}\,x_j^{-\epsilon_{ij}}, \\
     \mathcal{G}_{j\to i}(x_j, x_i) &\approx \alpha_{j\to i,\text{in}}\,T_i\,g_{\text{lim}}^i\,x_j^{-\lambda_{ji}}\,x_i^{-\epsilon_{ji}}.
 \end{align*}
Here $\lambda_{ii}$ is the local demand elasticity, $\lambda_{ij}$ captures the elasticity to the source price, and $\epsilon_{ij}$ captures the elasticity to the destination price. Demand is calibrated so that at $\mathbf{x}^* = \mathbf{1}$, the generated load matches the structural baseline defined by $\alpha$.

\noindent\textbf{Feedback lag.}
We model cross-chain latency with delay weights $w_d$ for $d=1,\ldots,d_{\max}$, where $w_d\geq0$, $\sum_d w_d=1$, and $w_d$ is the fraction of cross-chain traffic in step $t$ arriving at $t+d$. We assume all chains share the same delay distribution. Like the cross-chain ratio, latency is a physical constraint, and typical Cosmos chains have a delay of tens of block steps (Appendix~\ref{sec: statistics}), placing production systems well beyond the instantaneous-feedback regime.
The delay introduces a crucial asymmetry in the load equation. Local and outbound components respond to current prices $x_i(t)$ and $x_j(t)$, while inbound components depend on historical prices $x_i(t-d)$ and $x_j(t-d)$. The \eip~controller thus mixes fresh local feedback with delayed cross-chain feedback in the same block step, and cannot distinguish between the two.
The delay shape also matters. A spike concentrates a past deviation into one step, while a dispersed distribution spreads it over time.

\noindent\textbf{Overall load equation.} The load on chain $i$ at time $t$ is:
\begin{equation}\label{eq:multi-chain-load}
\begin{aligned}
L_i(t)=\frac{1}{g_{\mathrm{lim}}^i}\Big(&
\mathcal{G}_{i\to i}(x_i(t))
+\sum_{j\neq i}\mathcal{G}_{i\to j}(x_i(t),x_j(t)) \\
&+\sum_{j\neq i}\sum_{d=1}^{d_{\max}}
w_d\,\mathcal{G}_{j\to i}(x_j(t-d),x_i(t-d))
\Big).
\end{aligned}
\end{equation}
The first two terms are the local transactions and the outbound workflows initiated from $i$, and the third term is the delayed inbound load from workflows initiated on other chains. 
A workflow $i\to j$ contributes to $L_i(t)$ through $\mathcal{G}_{i\to j}$ at time $t$, and later to $L_j(t+d)$ as inbound load under the weight $w_d$.
This equation defines the coupled dynamics studied in the subsequent stability analysis.

\section{Stability Analysis of Multi-Chain \eip}\label{sec: stability-analysis}
In this section, we explain how the multi-chain architecture makes \eip~stability harder through local asymptotic stability analysis near the target load. In Section~\ref{sec: oscillation-analysis}, we validate the load pattern and stability boundary experimentally.

\subsection{Linearization of Multi-chain Dynamics}\label{sec: decomposition}

As in the single-chain analysis, we study local asymptotic stability around $\mathbf{x}^* = \mathbf{1}$. Let $\Delta x_i(t) = x_i(t) - 1$ and $\Delta L_i(t) = L_i(t) - T_i$ denote the perturbations from equilibrium.

\noindent\textbf{Load decomposition.} We perform a first-order expansion of the load equation (Eq.~\ref{eq:multi-chain-load}) around $\mathbf{x}^* = \mathbf{1}$. The expansion is structured by the four groups of prices:

\begin{align}\label{eq:decomposition}
&\Delta L_i(t) \approx
\underbrace{\tfrac{\partial L_i}{\partial x_i(t)}\Delta x_i(t)}_{\text{(I)}}
+ \underbrace{\sum_{j \neq i}\tfrac{\partial L_i}{\partial x_j(t)}\Delta x_j(t)}_{\text{(II)}} \\
&\quad+ \underbrace{\sum_{d=1}^{d_{\max}}\tfrac{\partial L_i}{\partial x_i(t\!-\!d)}\Delta x_i(t\!-\!d)}_{\text{(III)}}
+ \underbrace{\sum_{j \neq i}\sum_{d=1}^{d_{\max}}\tfrac{\partial L_i}{\partial x_j(t\!-\!d)}\Delta x_j(t\!-\!d)}_{\text{(IV)}} \nonumber
\end{align}

These four channels arise from two dimensions of the load structure: whether the price belongs to chain $i$ or to a peer chain $j$, and whether it enters at the current time with delay.

\noindent\textbf{Channel (I): instantaneous self-feedback.} Chain $i$'s price $x_i(t)$ affects its own load through local demand and through the source-side response of outbound cross-chain demand. Both components act as negative feedback.

\noindent\textbf{Channel (II): instantaneous feedback from peer chains.} A peer chain's current price $x_j(t)$ enters $L_i(t)$ through the destination-side component of outbound demand from $i$ to $j$. Unlike Channel~(I), this coupling reacts to the price of the neighbor chain rather than $i$'s own price.

\noindent\textbf{Channel (III): delayed self-feedback.} Chain $i$'s past price $x_i(t-d)$ re-enters through delayed inbound traffic. Users on chain $j$ initiating transactions toward $i$ at time $t-d$ responded to $x_i(t-d)$ via the destination-side elasticity, and this traffic arrives at $i$ at time $t$.

\noindent\textbf{Channel (IV): delayed feedback from peer chains.} A peer chain's past price $x_j(t-d)$ enters through the source-side elasticity of inbound demand. Users on chain $j$ responded to $x_j(t-d)$ when initiating transactions toward $i$.

Note that economic coupling is present in all four channels, while feedback lag appears only in Channels~(III) and~(IV). Without cross-chain traffic, only the local component of Channel~(I) survives, and stability reduces to the single-chain bound $\delta\lambda < 2$. The remaining terms of Eq.~\ref{eq:decomposition} are the structural additions of multi-chain systems.

\noindent\textbf{Channel coefficients and governing recurrence.} 
The demand function is a power law in $x$, and differentiating at $\mathbf{x}^*=\mathbf{1}$ yields constant coefficients comprising elasticities ($\lambda, \epsilon$) and equilibrium load shares ($\alpha$). For example, differentiating the delayed inbound terms with respect to $x_i(t\!-\!d)$ yields the Channel~(III) coefficient $\Lambda_{i}^{\text{delay}} = \sum_{j \neq i} \epsilon_{ji}\,\alpha_{j \to i, \text{in}}$, which captures the delayed effect of $x_i(t\!-\!d)$. We defer the derivation to Appendix~\ref{sec: derivation-of-forces}, and Table~\ref{table: forces} summarizes them. 

\begin{table}
\renewcommand\arraystretch{1.15}
\caption{Channel coefficients for chain $i$}
\label{table: forces}
\centering
\small
\setlength{\tabcolsep}{4pt}
\begin{tabular}{@{}lll@{}}
    \toprule
    \textbf{Coefficient} & \textbf{Parameters} & \textbf{Formula} \\
    \midrule
    I. $\Lambda_i^{\text{inst}}$ 
    & $\lambda_{ii}, \lambda_{ij}, \alpha_{i,\text{local}}, \alpha_{i\to j,\text{out}}$ 
    & $\lambda_{ii}\alpha_{i,\text{local}} + \sum_{j\neq i}\lambda_{ij}\alpha_{i\to j,\text{out}}$ \\
    \addlinespace[0.3em]
    II. $\Lambda_{ij}^{\text{inst}}$ 
    & $\epsilon_{ij}, \alpha_{i\to j,\text{out}}$ 
    & $\epsilon_{ij}\alpha_{i\to j,\text{out}}$ \\
    \addlinespace[0.3em]
     III. $\Lambda_i^{\text{delay}}$ 
    & $\epsilon_{ji}, \alpha_{j\to i,\text{in}}$ 
    & $\sum_{j\neq i}\epsilon_{ji}\alpha_{j\to i,\text{in}}$ \\
    \addlinespace[0.3em]
    IV. $\Lambda_{ji}^{\text{delay}}$ 
    & $\lambda_{ji}, \alpha_{j\to i,\text{in}}$ 
    & $\lambda_{ji}\alpha_{j\to i,\text{in}}$ \\
    \bottomrule
\end{tabular}
\end{table}

Note that $g_{\text{lim}}^i$, $P_i^*$, and $T_i$ cancel from the coefficients. $P_i^*$ cancels because derivatives are taken with respect to the normalized $x_i = P_i/P_i^*$. The factor $g_{\text{lim}}^i$ in every demand function is divided out by the $1/g_{\text{lim}}^i$ in the load definition, and the common prefactor $T_i$ is canceled by the $1/T_i$ in the update rule. The recurrence below therefore contains only the update rates, the channel coefficients, and the delay weights.
Substituting Eq.~\ref{eq:decomposition} into the normalized update rule, expanding to first order ($\Delta x_i(t\!+\!1) \approx \Delta x_i(t) + \delta_i\,\Delta L_i(t)/T_i$), and discarding second-order terms yields the governing recurrence:
\begin{align}\label{eq:governing-recurrence}
\Delta x_i(t\!+\!1) \approx{}\;
& (1 - \delta_i\,\Lambda_i^{\text{inst}})\,\Delta x_i(t)
 - \delta_i\!\sum_{j\neq i}\Lambda_{ij}^{\text{inst}}\,\Delta x_j(t) \nonumber\\
& - \delta_i\!\sum_{d=1}^{d_{\max}} w_d\,\Lambda_i^{\text{delay}}\,\Delta x_i(t\!-\!d) \nonumber\\
& - \delta_i\!\sum_{j\neq i}\sum_{d=1}^{d_{\max}} w_d\,\Lambda_{ji}^{\text{delay}}\,\Delta x_j(t\!-\!d).
\end{align}

\subsection{System-wide Stability Representation} 
Equation~\ref{eq:governing-recurrence} illustrates the load for any specific chain $i$, and couples $N$ chains across a delay window of $d_{\max}$ steps. To express this as a standard linear system, we define the augmented state vector:
$$
\mathbf{X}(t)=\begin{bmatrix}
\Delta\mathbf{x}(t)^T & 
\Delta\mathbf{x}(t\!-\!1)^T & 
\cdots & 
\Delta\mathbf{x}(t\!-\!d_{\max})^T
\end{bmatrix}^T.
$$
where $\Delta\mathbf{x}(t)=[\Delta x_1(t),\dots,\Delta x_N(t)]^T$.

The system evolves as $\mathbf{X}(t+1)=\mathbf{M}_{\mathrm{ext}}\,\mathbf{X}(t)$, where
$$
\mathbf{M}_{\mathrm{ext}}=\begin{bmatrix}
\mathbf{A}_0 & \mathbf{A}_1 & \mathbf{A}_2 & \cdots & \mathbf{A}_{d_{\max}}\\
\mathbf{I} & \mathbf{0} & \mathbf{0} & \cdots & \mathbf{0}\\
\mathbf{0} & \mathbf{I} & \mathbf{0} & \cdots & \mathbf{0}\\
\vdots & \vdots & \ddots & \ddots & \vdots\\
\mathbf{0} & \mathbf{0} & \cdots & \mathbf{I} & \mathbf{0}
\end{bmatrix}
$$
\begin{itemize}
\item $(\mathbf{A}_0)_{ii} = 1 - \delta_i\,\Lambda_i^{\text{inst}}$, \;
      $(\mathbf{A}_0)_{ij} = -\delta_i\,\Lambda_{ij}^{\text{inst}}$ \;($i\neq j$)
\item $(\mathbf{A}_d)_{ii} = -\delta_i\,w_d\,\Lambda_i^{\text{delay}}$, \;
      $(\mathbf{A}_d)_{ij} = -\delta_i\,w_d\,\Lambda_{ji}^{\text{delay}}$ \;($i\neq j$)
\item $\sum_{d=1}^{d_{\max}} w_d = 1$, \; $w_d \geq 0$.
\end{itemize}
Since $\mathbf{M}_{\mathrm{ext}}$ contains the update rates of all chains, stability is a joint property of the vector $(\delta_1,\ldots,\delta_N)$. 
The matrix structure further suggests that each chain's feedback is governed by its own $\delta_i$ and structural configuration, as every term in row $i$ is scaled by $\delta_i$. Section~\ref{sec: non-uniform-delta} validates this observation and shows that a single aggressive chain suffices to destabilize the coupled system and inflate other chains' load.

The augmented matrix $\mathbf{M}_{\mathrm{ext}}$ has dimension $N(d_{\max}\!+\!1) \times N(d_{\max}\!+\!1)$, and the system is stable if and only if $\rho(\mathbf{M}_{\mathrm{ext}}) < 1$. We cross-check this criterion against our experimental stability outcomes in Section~\ref{sec: oscillation-analysis}. 
The criterion yields no closed-form boundary, because even at $d_{\max}=1$ the characteristic polynomial has degree $2N$ and admits no tractable roots unless chains are fully symmetric, which real deployments do not satisfy. A numerical check of $\rho(\mathbf{M}_{\mathrm{ext}})$ then returns only a binary verdict for a concrete configuration, without revealing which structural parameter or mechanism drives or propagates instability.
We next identify the potential instability pathways embedded in the governing recurrence.

\subsection{Potential Instability Pathways}\label{sec: pathways}
Beyond the local Channel~(I), the recurrence (Eq.~\ref{eq:governing-recurrence}) includes three additional channels that the monolithic case does not capture. Each opens a potential route to instability. We group them into two candidate pathways. Whether either is practically dominant is outlined in Section~\ref{sec: validation}.

\begin{figure}[t]
  \centering
  \tikzset{
    fnt/.style  = {font=\footnotesize},
    eq/.style   = {draw=gray!70,densely dashed,line width=0.4pt},
    mirr/.style = {draw=black!55,densely dashed,line width=0.4pt},
    dimc/.style = {<->,draw=blue!65!black,line width=0.4pt},
    dims/.style = {<->,draw=red!72!black,line width=0.4pt},
    fa/.style   = {->,draw=blue!65!black,line width=1.1pt},
    fb/.style   = {->,draw=red!72!black,line width=1.3pt},
    fall/.style = {draw=gray,dotted,line width=0.6pt},
    tagc/.style = {fnt,text=blue!65!black,inner sep=1pt,align=left},
    tags/.style = {fnt,text=red!72!black,inner sep=1pt,align=left},
    tagg/.style = {fnt,text=black!55,inner sep=1pt,align=center},
    ttl/.style  = {fnt,text=black,inner sep=1pt,align=center},
    bud/.style  = {draw=green!55!black,line width=0.5pt},
    tagb/.style = {fnt,text=green!45!black,inner sep=1pt,align=right},
  }
  \begin{tikzpicture}
    \draw[eq] (0.15,0) -- (3.95,0);
    \node[tagg,anchor=west] at (0.2,-0.3) {equilibrium};
    \draw[dimc] (1.15,0.02) -- (1.15,0.98);
    \node[tagc,anchor=east] at (1.02,0.5) {$\Delta x_i(t)$};
    \node[tagc] at (1.7,1.3) {current price};
    \fill[blue!65!black] (1.7,1.0) circle (1.6pt);
    \draw[fa] (1.7,1.0) -- (1.7,-0.55);
    \node[tagc,anchor=west] at (1.88,0.45)
      {correction force\\ $\delta\lambda\,\Delta x_i(t)$};
    \node[tagc,anchor=west] at (1.88,-0.75) {stays above};
    \draw[mirr] (0.3,-1.0) -- (3.95,-1.0);
    \node[tagg] at (2.05,-1.32) {mirror point $-\Delta x_i(t)$};
    \node[tagg] at (2.05,-2.45)
      {stability condition:\\
       $\delta\lambda\,\Delta x_i(t)<2\,\Delta x_i(t)\Leftrightarrow\delta\lambda<2$};
    \draw[bud] (3.85,1.0) -- (3.85,-1.0);
    \draw[bud] (3.77,1.0) -- (3.93,1.0);
    \draw[bud] (3.77,-1.0) -- (3.93,-1.0);
    \node[tagb,rotate=90,anchor=center] at (4.08,0) {budget $2\,\Delta x_i(t)$};
    \node[ttl] at (2.05,-3.1) {(a) Monolithic};
    \draw[eq] (4.75,0) -- (8.55,0);
    \node[tagg,anchor=north] at (5.25,-0.08) {$t{-}d$};
    \node[tagg,anchor=north east] at (6.17,-0.08) {$t$};
    \draw[dims] (4.9,0.02) -- (4.9,1.98);
    \node[tags,anchor=east] at (4.77,1.35) {$\Delta x_i(t{-}d)$};
    \draw[dimc] (6.05,0.02) -- (6.05,0.98);
    \node[tagc,anchor=east] at (5.92,0.45) {$\Delta x_i(t)$};
    \fill[red!30] (5.25,2.0) circle (1.6pt);
    \draw[red!72!black] (5.25,2.0) circle (1.6pt);
    \draw[fall] plot[smooth] coordinates {(5.25,2.0) (5.65,1.5) (6.0,1.18) (6.3,1.0)};
    \node[tagg] at (6.55,2.05) {price falling};
    \node[tagc,anchor=west] at (6.45,1.2) {current price};
    \fill[blue!65!black] (6.3,1.0) circle (1.6pt);
    \draw[fb] (6.3,1.0) -- (6.3,-1.9);
        \node[tags,anchor=west] at (6.48,-0.5)
      {stale force\\ $\delta_i\epsilon\,\Delta x_i(t{-}d)$};
    \draw[mirr] (4.75,-1.0) -- (8.55,-1.0);
    \draw[bud] (8.3,1.0) -- (8.3,-1.0);
    \draw[bud] (8.22,1.0) -- (8.38,1.0);
    \draw[bud] (8.22,-1.0) -- (8.38,-1.0);
    \node[tagb,rotate=90,anchor=center] at (8.53,0) {budget $2\,\Delta x_i(t)$};
       \node[tagg] at (6.5,-2.45)
      {stability condition:\\ $\delta_i\epsilon\,\Delta x_i(t{-}d)<2\,\Delta x_i(t)$};
    \node[ttl] at (6.65,-3.1) {(b) Multi-chain};
  \end{tikzpicture}
  \caption{Delay-amplified over-correction}
  \label{fig: pathway1}
\end{figure}
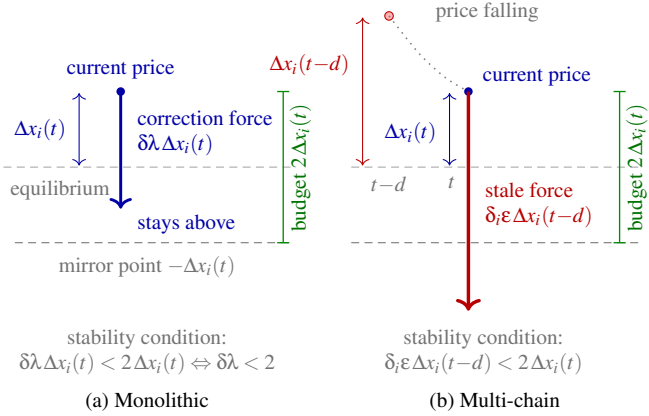

\noindent\textbf{Pathway 1: Delay-amplified over-correction (Channel (III), Figure~\ref{fig: pathway1}).}
In a multi-chain system, the stabilizing force of \eip~becomes composite. Without delay, the total effective feedback would be a convex combination of the underlying elasticities, each multiplied by its load ratio, and would stay bounded by the maximum of its constituents, so the single-chain stability logic would carry over. This is also why Channel~(I) cannot weaken instability compared with the monolithic case.
Latency, however, acts as an amplifier. $L_i(t)$ reflects not only the current price deviation $\Delta x_i(t)$ but also the delayed inbound transactions generated under past deviations $\Delta x_i(t\!-\!d)$. The effect enters the governing recurrence through Channel~(III) with strength $\Lambda_i^{\text{delay}}$, which grows with the destination-side elasticities and the inbound traffic shares.

To show why the delay creates a strictly smaller stability area, we compare the correction \emph{force} against the \emph{budget} that convergence allows, as illustrated in Figure~\ref{fig: pathway1}. 
In the monolithic case, a fee sitting above equilibrium by $\Delta x_i(t)$ shrinks demand and pulls the load below the target by $\Delta L_i(t) \approx -\lambda\,T_i\,\Delta x_i(t)$. \eip~reads this deviation and lowers the fee by $\delta_i\,|\Delta L_i(t)|/T_i = \delta_i\lambda\,|\Delta x_i(t)|$, which is the correction force. The budget is twice the current deviation, since a force of $2|\Delta x_i(t)|$ lands the fee exactly on the mirror point $-\Delta x_i(t)$, and any larger force moves the fee farther from equilibrium than before. 
The monolithic force is sized by the current deviation, so the requirement $\delta\lambda\,|\Delta x_i(t)| < 2|\Delta x_i(t)|$ recovers $\delta\lambda<2$. In the multi-chain case, consider a \emph{monotonic return} to equilibrium, where the price steadily corrects back toward $x_i^*=1$ and thus $|\Delta x_i(t\!-\!d)| > |\Delta x_i(t)|$. Assuming an extreme case where a chain's load is purely inbound flow, the force $\epsilon\Delta x_i(t{-}d)$ arriving at $t$ was fixed $d$ blocks ago by the larger past deviation, and it strikes after the system has already partially self-corrected, while the budget is still set by the smaller current deviation. Stability then requires $\delta_i\epsilon\,|\Delta x_i(t{-}d)| < 2|\Delta x_i(t)|$, so the tolerable $\delta_i\epsilon$ falls strictly below the single-chain bound of $2$.

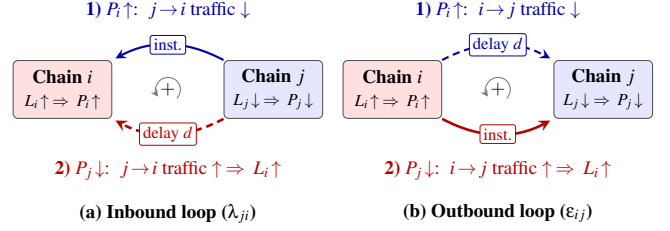
\begin{figure}[t]
  \centering
  \resizebox{\columnwidth}{!}{%
  \begin{tikzpicture}[
    font=\footnotesize,
    chain/.style={draw=black!70, rounded corners=2pt, align=center, inner sep=3pt,
                  minimum width=1.45cm, minimum height=0.85cm},
    supp/.style={-stealth, draw=blue!65!black, line width=0.9pt, shorten >=1.5pt},
    feed/.style={-stealth, draw=red!72!black, line width=1.1pt, shorten >=1.5pt},
    tsup/.style={font=\footnotesize, text=blue!65!black, inner sep=0.5pt},
    tfeed/.style={font=\footnotesize, text=red!72!black, inner sep=0.5pt},
    tag/.style={solid, draw=red!72!black, line width=0.4pt, fill=white, rounded corners=1pt,
                font=\scriptsize, text=red!72!black, inner sep=1.5pt},
    tagb/.style={solid, draw=blue!65!black, line width=0.4pt, fill=white, rounded corners=1pt,
                font=\scriptsize, text=blue!65!black, inner sep=1.5pt},
    ttl/.style={font=\footnotesize\bfseries}
  ]
    \begin{scope}
      \node[chain, fill=red!12]  (ia) at (0,0)
            {\textbf{Chain $i$}\\ {\scriptsize $L_i\!\uparrow\;\Rightarrow\;P_i\!\uparrow$}};
      \node[chain, fill=blue!10] (ja) at (3.1,0)
            {\textbf{Chain $j$}\\ {\scriptsize $L_j\!\downarrow\;\Rightarrow\;P_j\!\downarrow$}};
      \draw[supp] (ja.north west) to[bend right=28]
          node[tagb, pos=0.5] {inst.}
          node[tsup, above=10pt, pos=0.5]
            {\textbf{1)} $P_i\!\uparrow$:\; $j\!\to\!i$ traffic $\downarrow$}
          (ia.north east);
      \draw[feed, densely dashed] (ja.south west) to[bend left=28]
          node[tag, pos=0.5] {delay $d$}
          node[tfeed, below=10pt, pos=0.5]
            {\textbf{2)} $P_j\!\downarrow$:\; $j\!\to\!i$ traffic $\uparrow\;\Rightarrow\;L_i\!\uparrow$}
          (ia.south east);
      \node[font=\footnotesize] at (1.55,0) {$+$};
      \draw[-stealth, black!55, line width=0.4pt]
          (1.68,-0.13) arc[start angle=-45, end angle=215, radius=0.18];
      \node[ttl, anchor=north] at (1.55,-1.55) {(a) Inbound loop ($\lambda_{ji}$)};
    \end{scope}
    \begin{scope}[xshift=4.85cm]
      \node[chain, fill=red!12]  (ib) at (0,0)
            {\textbf{Chain $i$}\\ {\scriptsize $L_i\!\uparrow\;\Rightarrow\;P_i\!\uparrow$}};
      \node[chain, fill=blue!10] (jb) at (3.1,0)
            {\textbf{Chain $j$}\\ {\scriptsize $L_j\!\downarrow\;\Rightarrow\;P_j\!\downarrow$}};
      \draw[supp, densely dashed] (ib.north east) to[bend left=28]
          node[tagb, pos=0.5] {delay $d$}
          node[tsup, above=10pt, pos=0.5]
            {\textbf{1)} $P_i\!\uparrow$:\; $i\!\to\!j$ traffic $\downarrow$}
          (jb.north west);
      \draw[feed] (ib.south east) to[bend right=28]
          node[tag, pos=0.5] {inst.}
          node[tfeed, below=10pt, pos=0.5]
            {\textbf{2)} $P_j\!\downarrow$:\; $i\!\to\!j$ traffic $\uparrow\;\Rightarrow\;L_i\!\uparrow$}
          (jb.south west);
      \node[font=\footnotesize] at (1.55,0) {$+$};
      \draw[-stealth, black!55, line width=0.4pt]
          (1.68,-0.13) arc[start angle=-45, end angle=215, radius=0.18];
      \node[ttl, anchor=north] at (1.55,-1.55) {(b) Outbound loop ($\epsilon_{ij}$)};
    \end{scope}
  \end{tikzpicture}%
  }
  \captionsetup{justification=justified}
  \caption{Positive feedback pathway. Each loop starts from a congested
  chain $i$ ($L_i\uparrow$). Edge~1 suppresses cross-chain traffic and
  lowers the peer's load and fee, and edge~2 converts the cheaper peer
  fee back into extra load on chain $i$. }
  \label{fig: pathway2}
\end{figure}

\noindent\textbf{Pathway 2: Cross-chain positive feedback (Channels (II) and (IV), Figure~\ref{fig: pathway2}).} We illustrate how Channels (II) and (IV) potentially impose a positive feedback loop, though Section~\ref{sec: validation} validates them as unreachable.

\textit{Pathway 2-1 (via $\Lambda_{ji}^{\text{delay}}$, Channel~(IV), inbound loop).} $P_{i}\uparrow$ raises the destination-side cost of workflows from $j$ to $i$. This increase suppresses the outbound load of $j$, lowering $L_j$ and hence $P_j$ through \eip. The cheaper $P_j$ then stimulates outbound traffic from $j$ toward $i$ via $\lambda_{ji}$. A reinforcement loop forms when the induced traffic arrives $d$ blocks later with $P_i(t)$ and $P_j(t-d)$ on opposite sides of their equilibria. The delay strengthens the effect because the arriving traffic reflects an older and potentially more extreme price state on $j$.

\textit{Pathway 2-2 (via $\Lambda_{ij}^{\text{inst}}$, Channel~(II), outbound loop).} $P_i\uparrow$ also reduces outbound traffic to destination chain $j$, lowering $L_j$ and $P_j$ after $d$ steps. If the destination-side elasticity $\epsilon_{ij}$ is large, attraction of the now-cheaper destination $j$ can overwhelm the suppression from the expensive source $i$, increasing outbound volume and further inflating $L_i$.

\subsection{Validation of Potential Pathways}\label{sec: validation}

The pathways above are structural candidates derived from the recurrence, and whether either drives instability must be verified empirically. We defer the full validation to Appendix~\ref{sec: appendix-pathway-validation} and summarize the conclusion here. Only Pathway~1 is a practical instability driver. 
We track the delay amplification ratio $|\Delta x(t\!-\!d)/\Delta x(t)|$ that the delayed feedback delivers to the current step. For Pathway~1, during a monotonic return to equilibrium, the past deviation always exceeds the current one, so the delayed self-feedback of Channel~(III) can over-correct. 
Pathway~2 fails in both sub-pathways. Pathway~2-1 requires the source's past deviation and the destination's current deviation to keep opposite signs, but stronger coupling synchronizes the two prices. Pathway~2-2 is largely self-defeating, as raising the outbound share to strengthen Channel~(II) simultaneously strengthens the negative feedback of Channel~(I), and raising $\epsilon_{ij}$ alone destabilizes only an extreme corner. 

\noindent\textbf{Pathway 1 in closed form.} Suppose chain $i$'s load is purely inbound flow, and let all delay weights sit at a single lag $L$. The governing recurrence reduces to $\Delta x_i(t\!+\!1)=\Delta x_i(t)-\delta_i\Lambda_i^{\text{delay}}\,\Delta x_i(t\!-\!L)$. A classical root analysis of its characteristic polynomial~\cite{kuruklis1994asymptotic} gives stability if and only if $\delta_i\,\Lambda_i^{\text{delay}} < 2\sin\bigl(\pi/(2(2L+1))\bigr)$. The right-hand side is the budget under latency.
It recovers the monolithic $2$ at $L=0$, falls to $1$ at $L=1$, and decays as $\pi/(2L)$. Intuitively, the controller keeps correcting for $L$ blocks before any correction returns through the inbound flow, so the tolerable per-block gain shrinks with $L$. The bound also explains why the spike shape is the most dangerous. A spike places all weight at the largest lag, where the budget is smallest. The above budget predicts the inbound-dominated boundary well. At an inbound ratio of $0.99$ and $\epsilon_{j0}=16$ under spike latency ($L=5$), it gives $\delta<0.018$, one grid step from the simulated boundary of $0.019$ (Section~\ref{sec: boundary}).

\section{Quantifying Instability Consequences}\label{sec: oscillation-analysis}

\noindent\textbf{Evaluation overview.} Sections~\ref{sec: boundary} and~\ref{sec: evaluation-implication} assume a uniform $\delta$. 
Section~\ref{sec: boundary} quantifies how delayed coupling erodes the stability boundary and 
measures which oscillation stage the resulting configurations land in, and 
Section~\ref{sec: evaluation-implication} maps the load pattern to consensus security.
Section~\ref{sec: non-uniform-delta} then turns to non-uniform update rates, and shows that each chain's boundary is governed by its own update rate and load structure, and an unstable chain destabilizes peers that stay inside their own boundaries. The results are not very sensitive to $d_{\text{max}}$ or chain numbers, and we defer their experiments to Appendix~\ref{sec: additional-experiment-varying-dmax-and-chain-number}.

\subsection{Experimental Setup}\label{sec: experimental-setup}
We use a discrete-time nonlinear simulator, producing a full trajectory of blocks. Simulating such a trajectory requires evaluating demand at every price the system may enter, so we extend the local elasticity into a global one, $\mathcal{G} \propto x^{-\lambda}$ for all prices $x$. Appendix~\ref{sec: varying-elasticity} evaluates a variant whose elasticity varies with the price and shows that the extension does not affect our results. 
Because the \eip~controller observes only the load value, the simulator models demand directly without generating exact transactions, and base fees start at $x_i^*=1$.
Moreover, when the price falls well below equilibrium, the demand arriving in a single step can be many times a full block, so we assume a mempool to cache the loads. A real mempool is finite, so we cap ours at $Q$ block gas limits and drop arrivals beyond it. The default pools of go-Ethereum, Cosmos CometBFT, and Polkadot Substrate hold a few thousand transactions, tens of blocks at current occupancies, so we set $Q = 20$. Our results are insensitive to $Q$ (Appendix~\ref{sec: mempool-size}).

We use simulation rather than historical traces for two reasons. First, on-chain traces are system \textit{outputs} realized under a specific trajectory. Replaying them under different control parameters (e.g., a higher $\delta$) violates feedback integrity, because user decisions frozen under past prices would be forced to execute regardless of new costs. Second, quantifying risk requires iterating each critical parameter, while realistic configurations only sparsely sample the parameter space. Intrinsic variables such as elasticity are further unobservable, and only synthetic simulation can iterate them across their full theoretical ranges.

We initialize a 3-chain system with each chain's target load set to one half of its gas limit. All chains start at equilibrium, and we apply only a one-block \emph{$1.01\times$} multiplier to Chain~0's source-side demand at the first active step of each run, after which demand follows the calibrated function.
With the source-side demand multiplied, the inbound traffic of Chain~0 is not directly amplified and evolves with delay, so instability arises from the delayed feedback loop rather than from an injected inbound surge.
Compared with the $15\times$ stress in Ethereum's tests~\cite{eip-1559-experiment}, our stress is tiny, yet sufficient to move the dynamics away from equilibrium.
We run each configuration for $50{,}000$ steps and classify a trajectory as converged if its load over the final 50 steps stays within $0.1\%$ of the target. Since any finite-step simulation cannot exactly trace asymptotic stability, we cross-check this classification against the exact criterion $\rho(\mathbf{M}_{\mathrm{ext}})$ in Section~\ref{sec: boundary}, which covers the slow-convergence cases that a finite run may misclassify.

\begin{figure*}[!tp]
 \begin{center}
    \setlength{\abovecaptionskip}{0.mm}
    \centering
        \begin{subfigure}{0.32\linewidth}
            \centering
            \includegraphics[width=\linewidth]{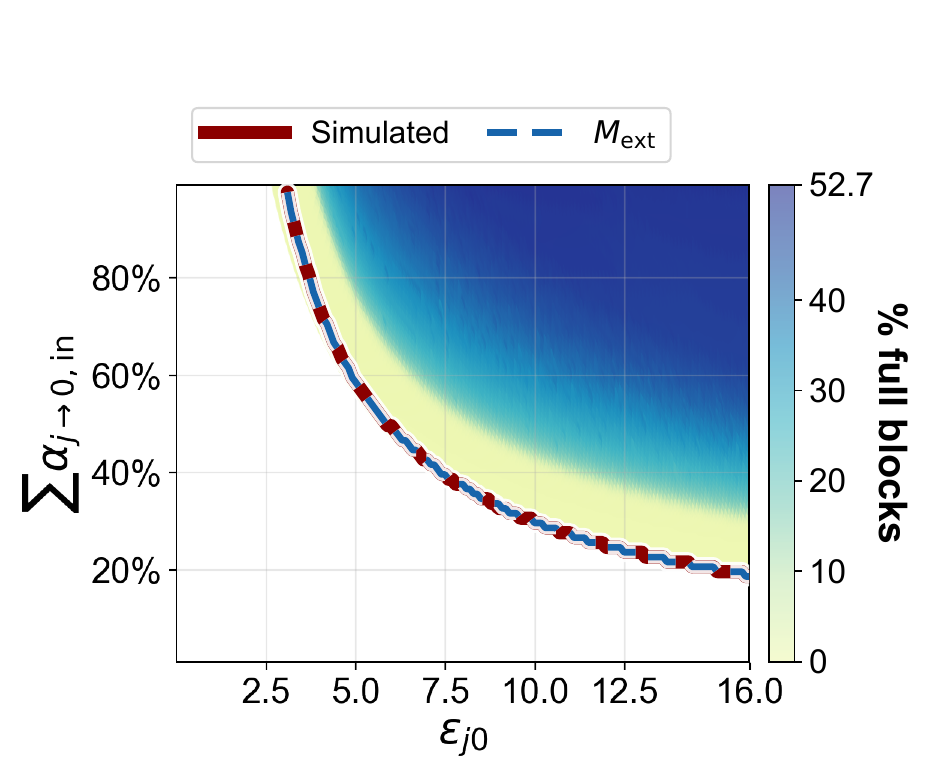}
            \caption{Boundary ($\delta=0.125$)}
            \label{fig: over-correction-spike}
        \end{subfigure}
        \hfill
        \begin{subfigure}{0.32\linewidth}
            \centering
            \includegraphics[width=\linewidth]{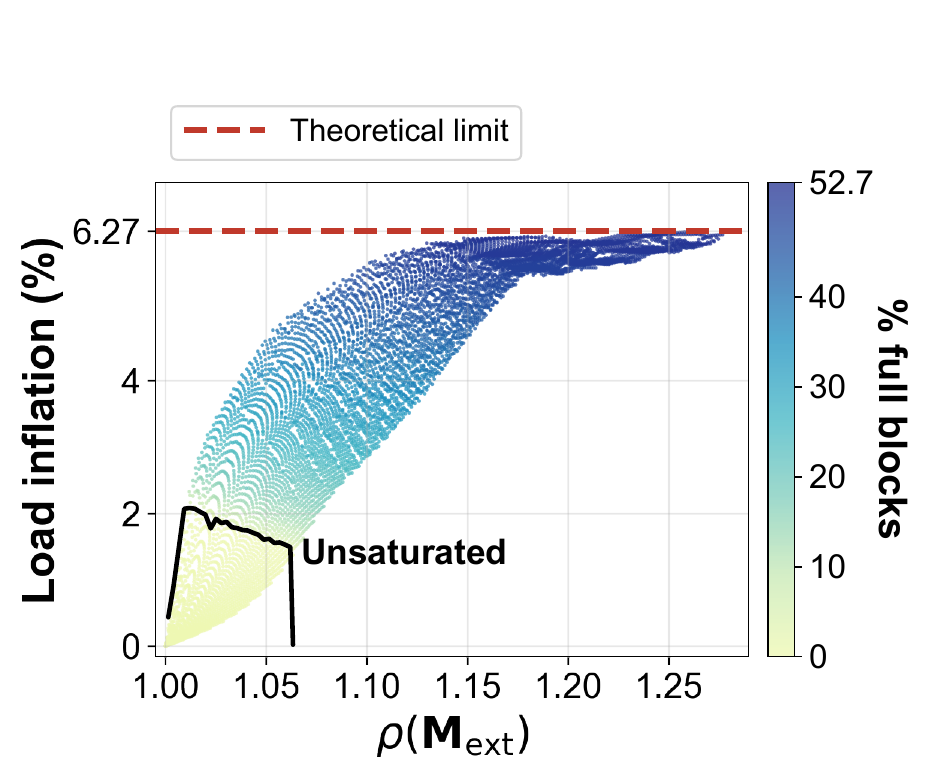}
            \caption{Load pattern of unstable points ($\delta=0.125$)}
            \label{fig: load-pattern}
        \end{subfigure}
        \hfill
        \begin{subfigure}{0.32\linewidth}
            \centering
            \includegraphics[width=\linewidth]{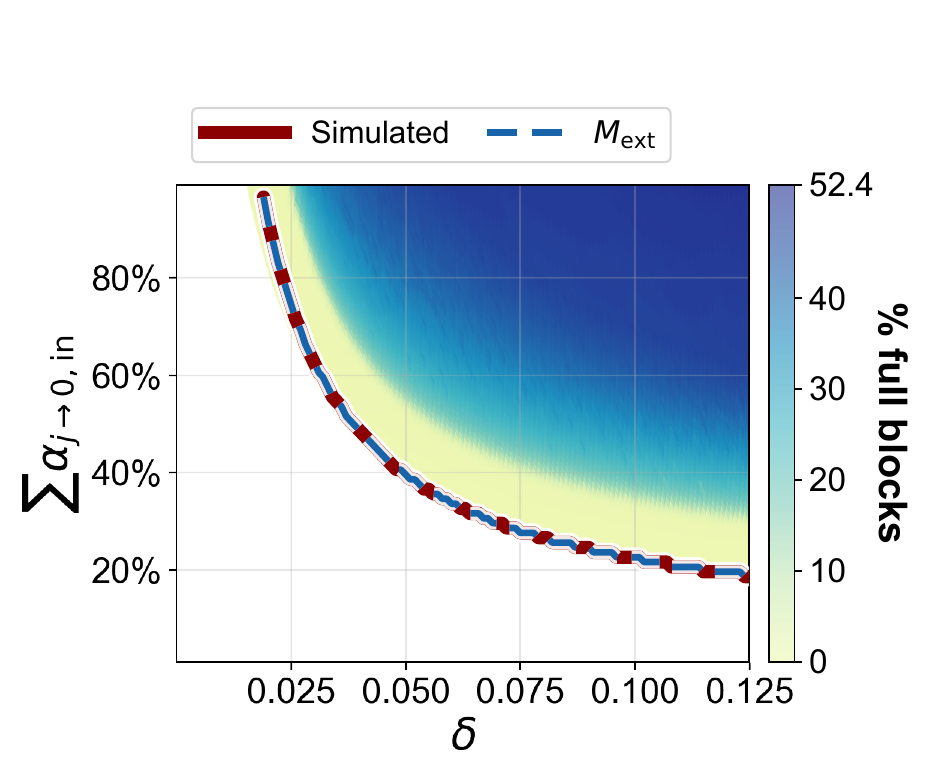}
            \caption{$\Delta_{\text{safe}}$ (elasticity $16$)}
            \label{fig: over-correction-delta-spike}
        \end{subfigure}
    \caption{Pathway~1 under spike latency}
    \label{fig: over-correction-boundary}
 \end{center}
\end{figure*}

\subsection{Stability Boundary and Load Pattern}\label{sec: boundary}
\noindent\textbf{Experimental design and parameter setting.} Since the multi-chain recurrence has no closed-form boundary, we locate it by simulation. We organize the parameter iteration around Pathway~1 and its two structural parameters, the inbound ratio and the destination-side elasticity, and we observe Chain~0 of the three chains.
The first question is whether instability emerges as Pathway~1 strengthens, and what the resulting load pattern is. Panel~\ref{fig: over-correction-spike} fixes $\delta=0.125$ and iterates $\sum\alpha_{j\to 0,\text{in}}\in[1\%,99\%]$ at step $1\%$ and $\epsilon_{j0}\in[0,16]$ at step $0.1$, with Chains~1 and~2 contributing inbound traffic symmetrically. All other elasticities stay at the base level $1.5$, and Chain~0 carries no outbound load, so the iteration strengthens Channel~(III) of Chain $0$ alone. The panel maps the stable region and colors each unstable point by the share of Chain~0 blocks that reach the gas limit. Panel~\ref{fig: load-pattern} takes those unstable points and plots the full block share and the load inflation against $\rho(\mathbf{M}_{\mathrm{ext}})$, which we will cross-validate against the simulated boundary and use to measure the instability level.

The second question is how much the delayed inbound flow shrinks $\Delta_{\text{safe}}$. At elasticity $16$, the monolithic bound $\delta<2/\lambda$ is exactly $0.125$, so Panel~\ref{fig: over-correction-delta-spike} fixes $\epsilon_{j0}=16$, iterates $\delta$ against the inbound ratio under the same coloring, and reads the erosion as the distance from $0.125$.
Both iterations use $d_{\max}=5$ and three delay distributions, \textit{spike} $[0,0,0,0,1]$, \textit{uniform} $[0.2,\ldots,0.2]$, and \textit{bimodal} $[0.4,0,0.2,0,0.4]$.
The full parameters are listed in the open science section. We simulate under spike latency. The other two shapes give slightly milder instability, and we defer them to Appendix~\ref{sec: additional-experiment-varying-dmax-and-chain-number}.

\noindent\textbf{Fixed $\delta=0.125$.} Figure~\ref{fig: over-correction-spike} shows
that instability emerges once Pathway~1 is strengthened. The simulated boundary agrees
with $\rho(\mathbf{M}_{\mathrm{ext}})$ on $15{,}938$ of the panel's $15{,}939$
configurations. The single disagreement sits at
$\rho(\mathbf{M}_{\mathrm{ext}})=0.999993$, which is too close to the boundary to call within a
finite simulation step. The criterion $\rho(\mathbf{M}_{\mathrm{ext}})>1$ therefore
identifies the experimentally unstable configurations, and we use $\rho$ to order them
below. Note that $\rho$ does not determine the instability severity. Two configurations with the
same $\rho$ can land in different stages.
Figure~\ref{fig: load-pattern} splits the unstable configurations between the two
stages. The configurations just past the boundary sit in the unsaturated stage. The
black line encloses them, approximately $16.4\%$ of the unstable points. Outside the
black line, the oscillation widens until it spans almost the whole block, and the share
of full blocks rises before flattening at $52.7\%$. The deepest configurations sit at
the far end of the saturated stage. Their load alternates between full blocks and
blocks that carry $0.82\%$ of the gas limit, and the load inflation reaches $6.26\%$,
just under the $6.27\%$ bound. In contrast, a monolithic chain at the same setting
stays stable.

\noindent\textbf{Fixed elasticity $16$.} Figure~\ref{fig: over-correction-delta-spike} shows that $\Delta_{\text{safe}}$ shrinks as coupling intensifies.
At $\alpha=0.99$, $\Delta_{\text{safe}}$ collapses to $\delta<0.019$, an $84.8\%$ erosion relative to the monolithic $\delta<0.125$. 

\subsection{Security Implication}\label{sec: evaluation-implication}

\noindent\textbf{Consensus degradation.} Our simulations show that instability can deliver a severe load pattern. However, whether it arises depends on the elasticity, for which these realistic chains provide no reliable estimate as their base fee is inactive. We therefore ask a conditional question, namely how much consensus risk the instability introduces once the chain becomes active.
As Cosmos chains enforce an aggressive $\delta$, we evaluate the security implications for them. Cosmos chains typically run CometBFT, whose core is Tendermint~\cite{buchman2018latest}, and execution lies on the critical path of its consensus. A validator can vote on the current block only after finishing execution of the preceding block. Larger blocks therefore prolong the execution, so more validators miss the voting deadline, and each block collects a lower proportion of votes.

Among Cosmos chains, we select Cosmos Hub and Osmosis, whose on-chain data allows us to relate load to the block failure rate. Concretely, each block carries a failure tag indicating whether it gathered sufficient votes. The block is not committed without enough votes.
This failure is similar to an Ethereum block reorg, whose rate is low, measured at $0.112\%$\footnote{\url{https://ethresear.ch/t/post-pectra-effects-on-ethereum-reorg-rate-propagation-and-block-size/23018}}.
Moreover, cross-chain transactions carry a timeout deadline, so even if a transaction in a failed block can be reproposed, the added delay can make it past that deadline.

We build a per-chain mapping from a block to the probability of failure.
We use approximately 9.5 million Cosmos Hub blocks produced between October 30, 2024 and August 5, 2026, reporting a failure rate of 0.166\%, and 10.2 million Osmosis blocks produced between March 24 and August 6, 2026, which report 5.25\%.
Osmosis exhibits a high failure rate because it sets an aggressive block interval.
Ideally, the mapping should key on a block's exact load, the gas its transactions actually consume (\emph{GasUsed}). However, Cosmos chains pack blocks by the user-declared \emph{GasWanted}, which typically exceeds the actual gas consumption. Empirically, GasUsed never exceeds $83\%$ of the limit in our investigated Cosmos Hub traces, so a GasUsed mapping is unreliable. We instead build the mapping from the declared gas and feed our load trajectories into it. The derived failure rate is therefore a conservative lower bound. We defer the full mapping to Appendix~\ref{sec: mapping}.
To construct different levels of instability, we iterate the Pathway~1 parameters of Chain $0$ and $\delta$.
We keep $100$ configurations at each $0.1\%$ inflation level from $0$ to $6.27\%$.
Each configuration is one deterministic $50{,}000$-step simulation. We look up each step's failure probability on the mapping and report the fraction of the $50{,}000$ steps that fail.

\begin{figure}[t]
  \centering
  \includegraphics[width=0.92\linewidth]{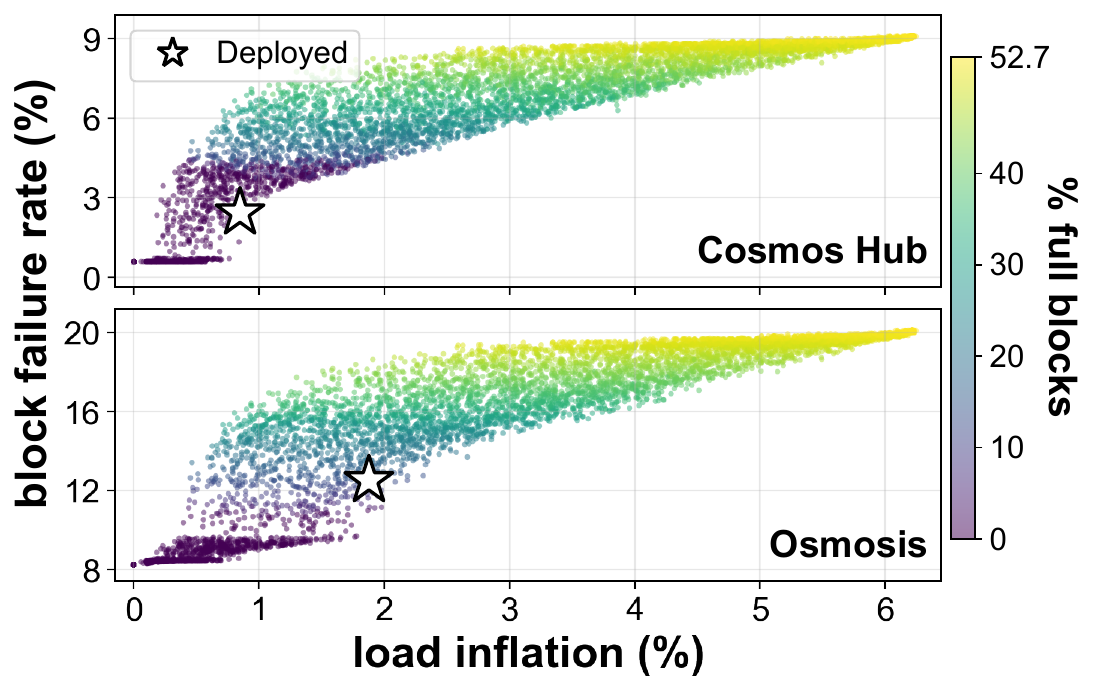}
  \caption{Block failure rate at each level}
  \label{fig: missing-rate-inflation}
\end{figure}

Figure~\ref{fig: missing-rate-inflation} shows the failure rate. Even at zero inflation, both 
curves sit above the on-chain rates, with Cosmos Hub being $0.589\%$ and Osmosis being $8.249\%$, because the two chains have low load and are short in demand, whereas the simulation assumes active demand.
Then, the failure rate climbs quickly at low inflation, 
reaching $4.8\%$ on Cosmos Hub at $1\%$ inflation. As the 
instability deepens, the large blocks first grow toward the gas limit, and each of them fails with 
a rising probability. Once they reach the limit, their number keeps rising with inflation. At the $6.27\%$ bound, more than half of all blocks sit at the limit, and the failure rate there reaches $9.1\%$ on Cosmos Hub, $55\times$ its historical rate, and $20.1\%$ on Osmosis, $3.8\times$ its historical rate.

\noindent\textbf{Consequence under current settings.} To evaluate whether realistic settings will 
induce instability, we select six representative chains (Appendix~\ref{sec: statistics}), and rerun each configuration assuming the same elasticity $16$ as 
Ethereum. The four Polkadot chains remain stable.
Both Cosmos chains are unstable at their deployed settings. The maximum stable update rate is 
$0.056$ for Cosmos Hub and $0.019$ for Osmosis, whereas the deployed rates are $0.125$ and $0.1$, 
sitting $2.2\times$ and $5.3\times$ above the bounds.
The two chains are in different stages. Cosmos Hub is in the unsaturated stage. Its load swings 
between $0.24$ and $0.84$, no block reaches the limit yet, and the load inflation stays at 
$0.85\%$. Osmosis is in the saturated stage. Its load swings between $0.05$ and the limit, $16\%$ 
of its blocks sit at the limit, and the load inflation reaches $1.88\%$.
The cost is already visible in both stages. The $50{,}000$-block traces place Cosmos Hub at a 
failure rate of $2.4\%$, $15\times$ the rate observed in its on-chain history, and Osmosis at 
$12.5\%$, $2.4\times$ its history.

\begin{figure*}[!tbp]
 \begin{center}
    \setlength{\abovecaptionskip}{0.mm}
    \centering
        \begin{subfigure}{0.28\linewidth}
            \centering
            \includegraphics[width=\linewidth]{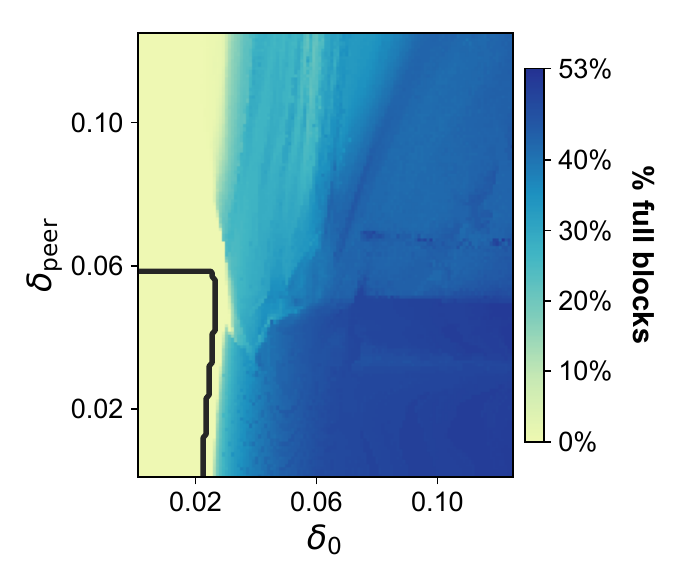}
            \caption{Baseline, Chain~0's load}
            \label{fig: nonuniform-map-a}
        \end{subfigure}
        \begin{subfigure}{0.28\linewidth}
            \centering
            \includegraphics[width=\linewidth]{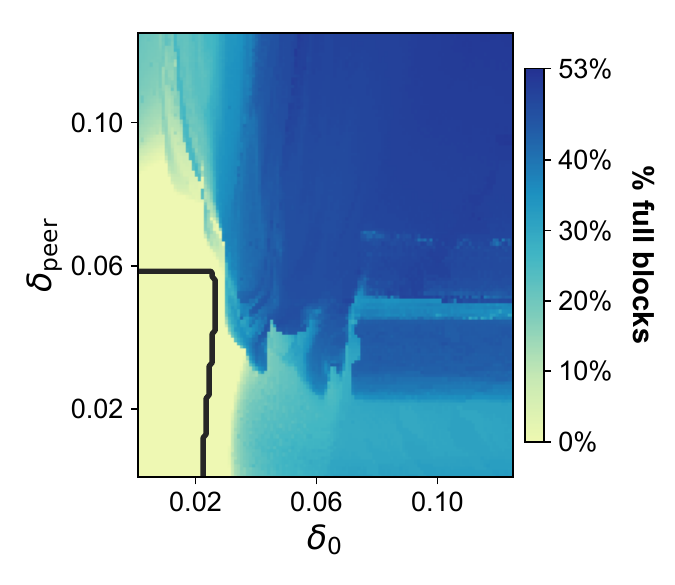}
            \caption{Baseline, peers' load}
            \label{fig: nonuniform-map-b}
        \end{subfigure}
        \begin{subfigure}{0.28\linewidth}
            \centering
            \includegraphics[width=\linewidth]{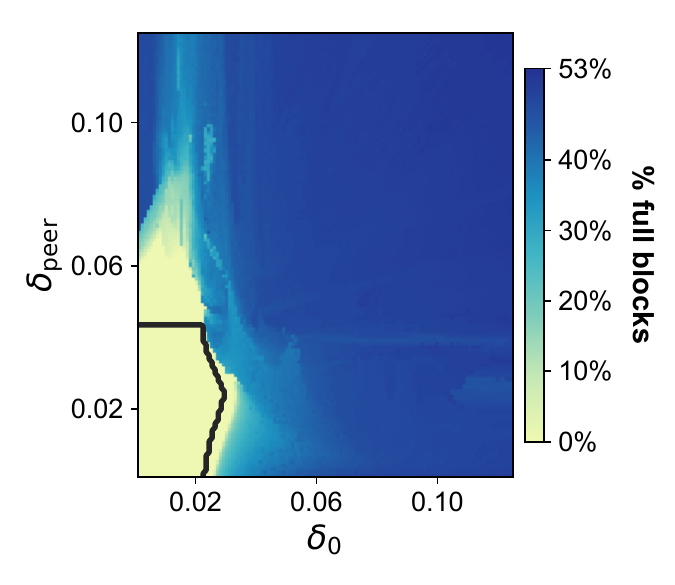}
            \caption{$20\%$ outbound, Chain~0's load}
            \label{fig: nonuniform-map-c}
        \end{subfigure}

        \begin{subfigure}{0.28\linewidth}
            \centering
            \includegraphics[width=\linewidth]{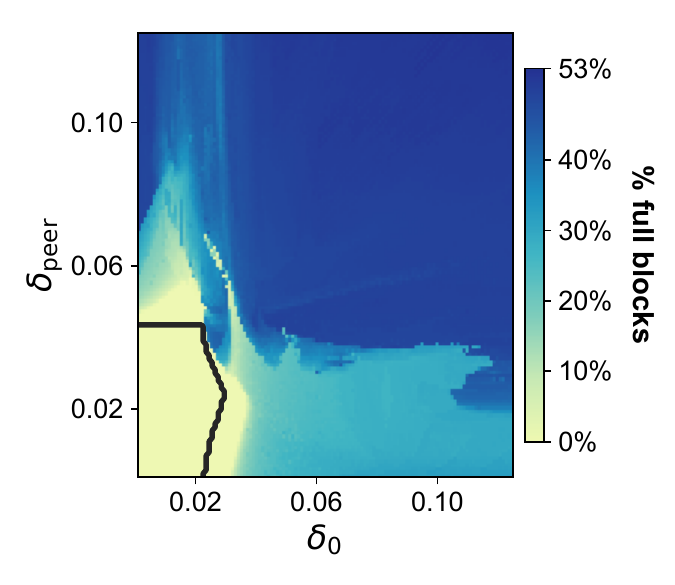}
            \caption{$20\%$ outbound, peers' load}
            \label{fig: nonuniform-map-d}
        \end{subfigure}
        \begin{subfigure}{0.28\linewidth}
            \centering
            \includegraphics[width=\linewidth]{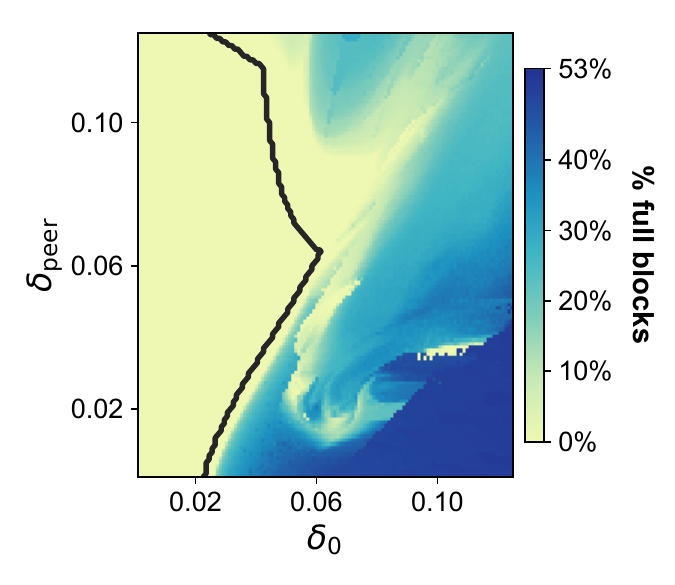}
            \caption{$\lambda_{j0}=16$, Chain~0's load}
            \label{fig: nonuniform-map-e}
        \end{subfigure}
        \begin{subfigure}{0.28\linewidth}
            \centering
            \includegraphics[width=\linewidth]{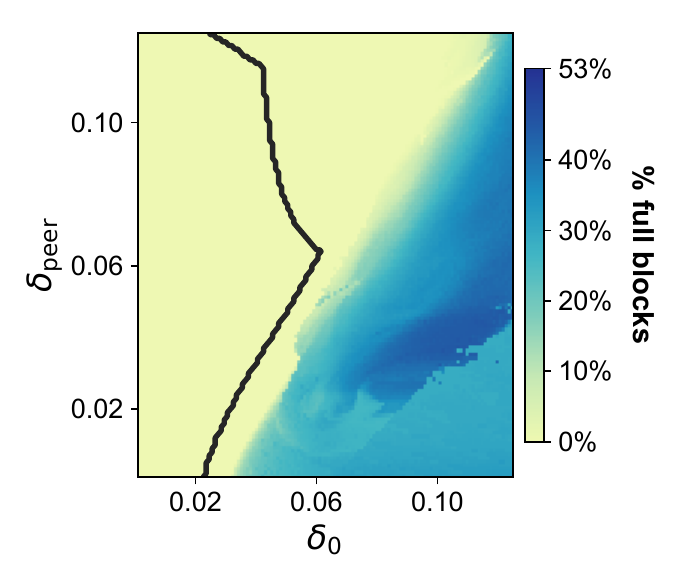}
            \caption{$\lambda_{j0}=16$, peers' load}
            \label{fig: nonuniform-map-f}
        \end{subfigure}

    \caption{Load pattern over the $(\delta_0,\delta_{\text{peer}})$ plane under the three configurations. The black line plots the boundary.}
    \label{fig: nonuniform-inflation-map}
 \end{center}
\end{figure*}

\subsection{Evaluating Non-Uniform Update Rate}\label{sec: non-uniform-delta}
All experiments above scan the uniform $\delta_i\equiv\delta$. This section examines the non-uniform case, and two questions matter.
First, whose update rate binds each chain's stability, the chain's $\delta_i$ or a peer's $\delta_j$? Second, is there a mechanism that \emph{propagates} the instability, or does the instability stay within the unstable chain?
The answers are as follows. First, each chain's boundary is governed by its own $\delta_i$ and load structure. Second, an unstable chain saturates its peers, and the propagation runs \emph{not} through the unstable chain's outbound traffic, but through the stable chains' outbound traffic toward it, which is their Channel~(II), the same route as Pathway~2-2. Channel~(IV) neither creates nor forwards instability, and strengthening Channel~(I) even makes a chain more stable.

\noindent\textbf{Setup.} We use a three-chain system with $d_{\max}=5$ spike latency.
For baseline evaluations in Figure~\ref{fig: nonuniform-map-a} and~\ref{fig: nonuniform-map-b}, Chain~$0$ has an inbound ratio of $80\%$, a local ratio of $20\%$. Each peer contributes $40\%$ of the inbound traffic, with the corresponding elasticity $\epsilon_{j0}$ at $16$.
Each peer has $10\%$ local load, and its outbound load contains $40\%$ targeting Chain~$0$ and $25\%$ targeting the other peer, so each peer's inbound ratio is also $25\%$.
We also extend $\epsilon_{0j}$ to $16$ for the peers' inbound flows, so that every chain also has a boundary.
We keep other elasticities at $1.5$.
Chain~0 runs $\delta_0$, and both peers run $\delta_{\text{peer}}$. We iterate $(\delta_0,\delta_{\text{peer}})$ over $[0.001,0.125]^2$ at step $0.001$, which yields $15{,}625$ configurations.
Each panel of Figure~\ref{fig: nonuniform-inflation-map} colors the share of blocks that reach the gas limit. 

\noindent\textbf{Boundary.} The pale region inside the black line is stable. Also, the simulated boundary and the $M_{\text{ext}}$ boundary agree on all $15{,}625$ configurations.
The boundary shape is approximately a rectangle. The horizontal edge sits at $\delta_{\text{peer}}^{\max}=0.058$ and does not move as $\delta_0$ varies. The vertical edge sits at $\delta_0^{\max}=0.022$ when the peers are nearly static, and it widens to $0.026$ as $\delta_{\text{peer}}$ increases.
Each chain's boundary is therefore governed by its own $\delta_i$ and load structure.

\noindent\textbf{Peer chains cannot saturate Chain $0$.} The boundary raises a natural question. If 
each chain's bound depends only on its own rate, can a chain protect itself by choosing a 
conservative $\delta_i$, no matter how its peers behave?
Consider first the area where the peers are unstable while Chain~0 stays inside its own bound 
($\delta_{\text{peer}}>0.058$, $\delta_0<0.022$). Panel~\ref{fig: nonuniform-map-a} stays pale 
throughout this band.
Chain~0's full-block share never exceeds $0.16\%$.
However, only $0.4\%$ of the band converges, and Chain~0's load oscillates around the target. The 
oscillation simply never reaches the gas limit. Chain~0 therefore sits in the unsaturated stage of 
Figure~\ref{fig: load-pattern}, where a chain is unstable but produces no full blocks.

\noindent\textbf{Chain $0$ can saturate peer chains.} Conversely, considering Chain~0 is unstable while the peers stay conservative ($\delta_{\text{peer}}\leq0.058$), the contamination succeeds (peer chains' load in Figure~\ref{fig: nonuniform-map-b}).
Panel~(b) is dark there although the peers' \eip~is inside the bound. Up to $49.5\%$ of the peers' blocks reach the gas limit, and the median over the band is $32.6\%$. Even at $\delta_{\text{peer}}=0.001$, where the peers are extremely conservative, $30.9\%$ of their blocks are full at the median. 

\noindent\textbf{Reason for the asymmetry.} Chain~0 sends no cross-chain traffic to peer chains. The peers' instability therefore reaches it only through its inbound flow, and these flows respond to the peers' prices with $\lambda_{j0}=1.5$.
Such a weak response can swing the arriving load, but it cannot push the load against the gas limit, so no saturation is observed. The peers face the opposite situation. They direct $40\%$ of their load toward Chain~0, and this demand responds to Chain~0's price with the destination-side elasticity $\epsilon=16$. When Chain~0 is unstable, the demand term explodes, and the source-side execution of these workflows saturates the peers' own blocks.

\noindent\textbf{Channel (II) of Chain $0$ propagates peer chains' instability.} To verify this attribution, we convert Chain~0's $20\%$ local load into outbound load, $10\%$ toward each peer, and keep the peers symmetric. Chain~0's inbound stays at $80\%$, and each peer now receives the $10\%$ inbound from Chain~0. The stable region remains a rectangle, and the peers' bound tightens from $0.058$ to $0.043$ (this is because each peer chain now has more inbound load). Over the same band, Chain~0's full-block share jumps from a maximum of $0.16\%$ to a median of $36.8\%$ and a maximum of $48.0\%$ (Figure~\ref{fig: nonuniform-map-c}), and it also raises peer chains' load (Figure~\ref{fig: nonuniform-map-d}).
This confirms that activating Channel~(II) of Chain~$0$ propagates the instability of peer chains to Chain~$0$.

\noindent\textbf{Channel (I) strengthens stability.} The baseline of Figure~\ref{fig: nonuniform-map-a} and~\ref{fig: nonuniform-map-b} leaves one question. Can Channel~(IV) of Chain~$0$ propagate the peers' instability to Chain~$0$? In Figure~\ref{fig: nonuniform-map-a}, the forwarding fails because $\lambda_{j0}=1.5$ is too weak, so we raise $\lambda_{j0}$ to $16$ (Figures~\ref{fig: nonuniform-map-e} and~\ref{fig: nonuniform-map-f}). However, it demonstrates that the peer chains become more stable, with their bound moving from $0.058$ to beyond the top of the scanned grid at $0.125$. Chain~0 now converges everywhere in the band and produces no full blocks at all.
To locate the cause, note that the coefficient of Chain~0's Channel~(IV) is $\lambda_{j0}\,\alpha_{j\to 0,\text{in}}$, in which the only parameter tied to the peers' response is $\lambda_{j0}$. Raising $\lambda_{j0}$ simultaneously raises the peers' Channel~(I).

We discover that strengthened Channel~(I) of the peer suppresses the peers' own Channel~(III). Consider a price regressing to equilibrium. While the price is above equilibrium, both channels push downward, and the stronger Channel~(I) merely speeds up the return. The difference appears when the price crosses the equilibrium. Channel~(III), still reading the larger pre-crossing deviations, keeps pushing downward for $d$ more blocks and drives the price into an overshoot. Channel~(I) instead reverses immediately, because the low price stimulates load and turns its negative feedback upward. During this window, the two channels oppose each other, and the stronger Channel~(I), the faster it erodes the overshoot of Channel~(III). The echo returning $d$ blocks later is therefore weaker, the oscillation dies out, and the peers tolerate a larger $\delta\Lambda_j^{\text{delay}}$.

\section{Protocol-level mitigation}\label{sec: mitigation}
Recall that the cause of the instability is that the load mixes in \emph{inbound flow that responds to stale prices}, causing overshoot. Inspired by Figures~\ref{fig: nonuniform-map-e} and~\ref{fig: nonuniform-map-f}, we aim to damp and even remove this overshoot by strengthening Channel~(I) and weakening Channel~(III).  We find that one added term in the update rule~\eqref{eq:multi-chain-eip-1559} achieves exactly this rebalancing: $x_i(t\!+\!1)=x_i(t)\bigl(1+\delta_i\bigl[(L_i(t)-T_i)/T_i-H_i\sum_{d=1}^{d_{\max}}w_d\,(P_i(t)-P_i(t\!-\!d))/P_i(t)\bigr]\bigr)$, where $H_i\ge 0$ is a compensation gain. 
The reason is that the added counter-force \emph{re-prices} the stale inbound load. In the governing recurrence, the term adds $H_i$ to Channel~(I) and subtracts $H_i$ from Channel~(III). 
Every input it needs is locally observable so that each chain can deploy the term without cross-chain coordination. 

The crux is deriving $H_i$. We find that $H_i$ can be calculated with the load topology and elasticities, and restoring to stability does \emph{not} require an accurate estimation of elasticities.
We validate experimentally that the augmented rule mitigates the instability. We defer the full analysis to Appendix~\ref{sec: protocol-level-mitigation}.

\section{Discussion}\label{sec: discussion}

\noindent\textbf{Methodology.} Our analysis is \emph{ex ante} because the weak demand keeps the base fees at the floor, so the controller is not active and the traces leave the elasticity unidentified. Ethereum's blob market spent its first year in the same regime\footnote{\url{https://eips.ethereum.org/EIPS/eip-7918}}, described as an inelasticity horizon.
Note that elasticity governs the slope of the demand curve, not the demand level, so low demand on Cosmos does not indicate a low elasticity. The demand level is a separate degree of freedom, which is extremely low on Cosmos and rarely produces a load above the target even at the price floor. Once instability arises after demand activates, it can shift the load between empty and the gas limit, affecting about two million monthly active users and several billion dollars of value across the two chains.

\noindent\textbf{Price-dependent elasticity.} Our experiments above hold the elasticity constant in the price. Prior analysis has a similar assumption. Roughgarden fixes the demand curve as a static primitive~\cite{roughgarden2020transaction}, Ndiaye assumes a Pareto valuation distribution whose elasticity is constant~\cite{ndiaye2024blockchain}, and Crapis et al. evaluate the elasticity only at the target point~\cite{crapis2024optimal}. 
In realistic chains, elasticity may vary with prices. 
We redo Figure~\ref{fig: over-correction-delta-spike} in Appendix~\ref{sec: varying-elasticity} with different price-elasticity shapes. The experiment demonstrates that the shape affects neither the boundary nor the severity of the instability past it.

\noindent\textbf{Extension to the current model.}
We keep the smallest model in which we analyze how the instability arises and propagates, namely
two-hop workflows under fixed delays with non-strategic demand. 
Each extension below would push the system toward more instability.
First, our measured delay is based on the long-term statistics, while finality and relaying slow
down under congestion. Since the correction budget shrinks with lag, load-dependent delay can only tighten the bound.
Second, though two-hop workflows dominate, longer workflows carry more delayed loops and should amplify the mechanism. Third, we assume demand is memoryless and non-strategic, matching the monolithic analyses~\cite{leonardos2021dynamical}. Holding the demand model fixed attributes the instability to coupling and delay, rather than to a new behavioral assumption. Real participants, in contrast, respond with lag, for example through wallet fee estimators. Such lag is additional delayed feedback, and we expect it to widen the unstable region.

\section{Related Work}\label{sec: related-work}

\noindent\textbf{Transaction fee mechanism.} Research on transaction fee mechanisms (TFMs) pivoted with the introduction of \eip. 
Prior work documented the high fee volatility and poor user experience of Ethereum's first-price auction~\cite{roughgarden2020transaction, de2021analysis, karaivanov2024economic, chung2023foundations}, and subsequent analysis confirmed that \eip~stabilizes gas prices and reduces fee variance~\cite{karaivanov2024economic, liu2022empirical, roughgarden2020transaction}. Within the monolithic setting, several works refine \eip, for example, through adaptive update rates~\cite{reijsbergen2021transaction}. A separate line extends TFMs to multi-dimensional resource pricing through a control-theoretic approach~\cite{crapis2024optimal}. This paradigm now informs upgrades such as EIP-4844~\cite{eip-4844-proposal} and EIP-8037~\cite{eip-8037-proposal}.

\noindent\textbf{Dynamical analysis of \eip.} Most relevant to our work is the dynamical 
analysis, all confined to monolithic settings. Leonardos et al.~\cite{leonardos2021dynamical} model 
monolithic \eip~and characterize the stability condition, with a progression of the oscillation patterns. 
Follow-ups~\cite{leonardos2023optimality, leonardos2026dynamics} bound the resulting load inflation at $6.27\%$. An empirical study~\cite{liu2022empirical} confirms that 
monolithic \eip~are stable. 
We take the oscillation stages and the load cap from this line and do not reprove them. 
What prior analysis leaves unexplored is the multi-chain setting. 
We identify the mechanism that drives and propagates instability, and show that the load oscillation reaches the far end of the saturated stage at an update rate at which a monolithic chain is stable. We further quantify the consensus consequences, which prior monolithic analysis leaves unmeasured.

\noindent\textbf{Game-theoretic fee mechanism design.} A separate line targets a fee mechanism design problem.
Lavi et al.~\cite{lavi2022redesigning} first analyze the Bitcoin fee market with strategic participants, covering miner manipulation and side payments.
Roughgarden~\cite{roughgarden2020transaction} develops a general formalism and introduces specific incentive compatibility notions. Chung and Shi~\cite{chung2023foundations, chung2024collusion} refine collusion-resilience and establish impossibility results. Other works extend the analysis to base-fee manipulation~\cite{azouvi2023base}, MEV-aware block producers~\cite{bahrani2024transaction}, and multi-proposer architectures~\cite{garimidi2025transaction}. This line is orthogonal to our analysis.

\section{Conclusion}\label{sec: Conclusion}
In this work, we examine the stability of \eip~in multi-chain systems. We model the chains as a delayed dynamical system, and show that delayed cross-chain feedback strictly shrinks the safe regime relative to the monolithic baseline. We identify and validate the mechanisms that drive instability and propagate it across chains. On realistic configurations, the deployed update rates of Cosmos Hub and Osmosis already exceed their safe bounds. We finally propose a protocol-level mitigation that cancels the destabilizing delayed feedback.

\newpage

\section*{Open Science}\label{sec: open-science}

We release the evaluation artifact behind this paper\footnote{\url{https://github.com/CGCL-codes/FeeMechanism-inst}}. The artifact holds the code and the data to reproduce every experimental figure and table in the paper.

\noindent\textbf{Layout.}
The artifact has three parts. The first part is the simulator. It is a discrete-time nonlinear model of $N$ coupled \eip~chains, written in Go, and it runs every simulation in the paper. The second part is the exact criterion. It is a script that builds the matrix $\mathbf{M}_{\mathrm{ext}}$ for a given configuration and checks whether $\rho(\mathbf{M}_{\mathrm{ext}})<1$. It computes the theoretical boundary, and it also writes $\rho$, the verdict, and whether the verdict agrees with the simulation. The third part is one directory per figure. Each directory holds the raw results, the scripts that turn them into panels, and a README stating what the element shows, in which we specify the command that reproduces it.

\noindent\textbf{Requirements.}
The simulator needs Go~1.24 or newer and depends only on a YAML parser. The scripts need Python~3.9 or newer with \texttt{numpy}, \texttt{pandas}, \texttt{matplotlib}, \texttt{pyyaml}, and \texttt{requests}.

\noindent\textbf{Reproduction.}
Replotting every figure from the released raw data in each folder takes a few minutes and does not run any simulations. Rerunning the simulations is the expensive part. Each boundary figure evaluates $15{,}939$ trajectories, so one figure costs tens of minutes to a few hours. We therefore do not list the rerun commands here in this section. Each figure's README includes the command to rerun its own simulation, next to the numbers it should produce. Every simulation is deterministic.

\noindent\textbf{Experiment settings.}
Recall the setup of Section~\ref{sec: experimental-setup}. A run uses three chains unless stated otherwise. Each chain has a gas limit of $2{,}000{,}000$ and a target load of $0.5$, so each chain carries $G=1{,}000{,}000$ gas of demand at equilibrium. 
As proved in Section~\ref{sec: stability-analysis}, the results hold regardless of whether the gas limits and the target loads are uniform across chains.
Let $\mathbf{D}$ be the demand matrix at equilibrium, where $D_{ij}$ is the demand that starts on chain $i$ and executes on chain $j$. Recall that a two-hop flow consumes gas twice, so the load of a chain is the demand that starts on it plus the demand that arrives at it. The cross-chain share sets how much of a chain's demand leaves it. The artifact uses three topologies. Under the inbound-coupled topology, Chain~0 only receives cross-chain load and sends none, in which $\alpha$ denotes the inbound load share of Chain $0$:
\begin{equation*}
\mathbf{D}=
\begin{pmatrix}
1-\alpha & 0 & 0\\[2pt]
\alpha/2 & 1-\alpha & \alpha/4\\[2pt]
\alpha/2 & \alpha/4 & 1-\alpha
\end{pmatrix}.
\end{equation*}
The demand function is calibrated to meet the above matrix at equilibrium. Figures~\ref{fig: over-correction-boundary}, \ref{fig: kappa-defense}, \ref{fig: kappa-attack}, \ref{fig: over-correction-appendix-latency}, \ref{fig: over-correction-dominance-additional},~\ref{fig: over-correction-boundary-under-varying-chain-number}, ~\ref{fig: varying-elasticity-boundary}, \ref{fig: mempool-sensitivity}, and~\ref{fig: mitigation} use this topology, and so does the top row of Figure~\ref{fig: coupling-dominance-merged}.

Figure~\ref{fig: missing-rate-inflation} uses the same topology. Its inflation levels come from random samples of the three Pathway-1 parameters, the inbound ratio $\alpha\in[0.01,0.99]$, the inbound elasticity $\epsilon\in[0,16]$, and the update rate $\delta\in[0.001,0.125]$. Each sample is one deterministic $50{,}000$-step run.

Under the outbound-coupled topology, the direction is reversed, and Chain~0 only sends, where $\alpha$ denotes the outbound ratio of Chain $0$:
\begin{equation*}
\mathbf{D}=
\begin{pmatrix}
1-\alpha & \alpha/2 & \alpha/2\\[2pt]
0 & 1-\alpha & \alpha/4\\[2pt]
0 & \alpha/4 & 1-\alpha
\end{pmatrix}.
\end{equation*}
The demand function is calibrated to meet the above matrix at equilibrium. The bottom row of Figure~\ref{fig: coupling-dominance-merged} and Figure~\ref{fig: boundary-inst-attack-without-latency} use it. 

The third topology is the one of Figure~\ref{fig: nonuniform-inflation-map}. Chain~0's inbound load is $80\%$, $40\%$ from each peer. The peers do not follow the inbound-coupled matrix, because each of them is given a finite boundary of its own within the iterated range. In Figures~\ref{fig: nonuniform-map-a}, \ref{fig: nonuniform-map-b}, \ref{fig: nonuniform-map-e} and~\ref{fig: nonuniform-map-f}, Chain~0 keeps $20\%$ local and sends nothing, and each peer keeps $10\%$ local and sends $25\%$ to the other peer:
\begin{equation*}
\mathbf{D}=
\begin{pmatrix}
0.20 & 0 & 0\\[2pt]
0.40 & 0.10 & 0.25\\[2pt]
0.40 & 0.25 & 0.10
\end{pmatrix}.
\end{equation*}
In Figures~\ref{fig: nonuniform-map-c} and~\ref{fig: nonuniform-map-d}, Chain~0's $20\%$ local load becomes $10\%$ outbound to each peer. Each peer's $10\%$ local load is then absorbed by the $10\%$ it receives from Chain~0, which holds every chain's load at $G$:
\begin{equation*}
\mathbf{D}=
\begin{pmatrix}
0 & 0.10 & 0.10\\[2pt]
0.40 & 0 & 0.25\\[2pt]
0.40 & 0.25 & 0
\end{pmatrix}.
\end{equation*}

Elasticity is carried by two matrices of the same shape. $\epsilon_{ij}$ is the
destination-side elasticity of the flow from $i$ to $j$, and $\lambda_{ij}$ is its
source-side elasticity. On three chains, they read
\begin{equation*}
\boldsymbol{\epsilon}=
\begin{pmatrix}
0 & \epsilon_{01} & \epsilon_{02}\\[2pt]
\epsilon_{10} & 0 & \epsilon_{12}\\[2pt]
\epsilon_{20} & \epsilon_{21} & 0
\end{pmatrix},
\quad
\boldsymbol{\lambda}=
\begin{pmatrix}
\lambda_{00} & \lambda_{01} & \lambda_{02}\\[2pt]
\lambda_{10} & \lambda_{11} & \lambda_{12}\\[2pt]
\lambda_{20} & \lambda_{21} & \lambda_{22}
\end{pmatrix}.
\end{equation*}
The diagonal of $\boldsymbol{\epsilon}$ is $0$, because a local flow has no separate
destination-side term. Every other entry of $\boldsymbol{\epsilon}$, and every entry of
$\boldsymbol{\lambda}$, is $1.5$ by default. Most experiments raise exactly one family of
entries. Figures~\ref{fig: over-correction-boundary}, \ref{fig: kappa-defense}, \ref{fig: over-correction-appendix-latency}, \ref{fig: over-correction-dominance-additional}, ~\ref{fig: over-correction-boundary-under-varying-chain-number}, \ref{fig: varying-elasticity-boundary}, \ref{fig: mempool-sensitivity}, and~\ref{fig: mitigation} raise $\epsilon_{j0}$, the destination-side elasticity of the flows into Chain~0. Figure~\ref{fig: kappa-attack}, and the top row of Figure~\ref{fig: coupling-dominance-merged} raises $\lambda_{j0}$. The bottom row of Figure~\ref{fig: coupling-dominance-merged} and Figure~\ref{fig: boundary-inst-attack-without-latency} raise $\epsilon_{0j}$. Figure~\ref{fig: nonuniform-inflation-map} is the exception. It raises every off-diagonal entry of $\boldsymbol{\epsilon}$ to $16$, so that each of the three chains has a boundary of its own, and Figures~\ref{fig: nonuniform-map-e} and~\ref{fig: nonuniform-map-f} additionally raise $\lambda_{j0}$ to $16$. For each figure, the artifact README lists which topology it uses, which family it raises, what is fixed, what is iterated, and how many grid points the iteration contains.

\noindent\textbf{Setting example.}
We write out two points of the Figure~\ref{fig: over-correction-spike} grid in full, and any other point follows the same rule. Both points fix $\delta=0.125$, $d_{\max}=5$, the spike delay weights $[0,0,0,0,1]$ and the one-block $1.01\times$ shock.

\noindent\emph{Point A:} $\epsilon_{j0}=16.0$ and $\alpha=0.99$. Chain~0 keeps $10{,}000$ gas of local demand and receives $495{,}000$ gas from each peer. Each peer keeps $10{,}000$ gas local and sends $247{,}500$ gas to the other peer:
\begin{equation*}
\mathbf{D}=
\begin{pmatrix}
10{,}000 & 0 & 0\\[2pt]
495{,}000 & 10{,}000 & 247{,}500\\[2pt]
495{,}000 & 247{,}500 & 10{,}000
\end{pmatrix},
\end{equation*}
\begin{equation*}
\boldsymbol{\epsilon}=
\begin{pmatrix}
0 & 1.5 & 1.5\\[2pt]
16 & 0 & 1.5\\[2pt]
16 & 1.5 & 0
\end{pmatrix},
\quad
\boldsymbol{\lambda}=
\begin{pmatrix}
1.5 & 1.5 & 1.5\\[2pt]
1.5 & 1.5 & 1.5\\[2pt]
1.5 & 1.5 & 1.5
\end{pmatrix}.
\end{equation*}

\noindent\emph{Point B:} $\epsilon_{j0}=4.0$ and $\alpha=0.50$. Chain~0 keeps $500{,}000$ gas local and receives $250{,}000$ gas from each peer. Each peer keeps $500{,}000$ gas local and sends $125{,}000$ gas to the other peer:
\begin{equation*}
\mathbf{D}=
\begin{pmatrix}
500{,}000 & 0 & 0\\[2pt]
250{,}000 & 500{,}000 & 125{,}000\\[2pt]
250{,}000 & 125{,}000 & 500{,}000
\end{pmatrix},
\end{equation*}
\begin{equation*}
\boldsymbol{\epsilon}=
\begin{pmatrix}
0 & 1.5 & 1.5\\[2pt]
4 & 0 & 1.5\\[2pt]
4 & 1.5 & 0
\end{pmatrix},
\quad
\boldsymbol{\lambda}=
\begin{pmatrix}
1.5 & 1.5 & 1.5\\[2pt]
1.5 & 1.5 & 1.5\\[2pt]
1.5 & 1.5 & 1.5
\end{pmatrix}.
\end{equation*}

\noindent\textbf{Data availability.}
We include the raw data to draw the figures in the artifact. However, the on-chain block raw data of Cosmos Hub and Osmosis are too large, so the two raw block data are not included.
For Figure~\ref{fig: missing-rate-inflation}, we include each chain's mapping from load to block failure rate. The dumps are re-collectable from public archive RPC endpoints with the scripts in the Figure~\ref{fig: missing-rate-inflation} directory. The on-chain statistics illustrated in Figure~\ref{fig: fee-comparison} and Table~\ref{table: data} are also re-collectable.

\bibliographystyle{plain}
\bibliography{refs}

\appendix

\section{Derivation of the Channel Coefficients}\label{sec: derivation-of-forces}

Recall the load equation for chain $i$:
\begin{equation*}
\begin{aligned}
L_i(t)=\frac{1}{g_{\mathrm{lim}}^i}\Big(
&\mathcal{G}_{i\to i}(x_i(t))
+\sum_{j\neq i}\mathcal{G}_{i\to j}(x_i(t),x_j(t)) \\
&+\sum_{j\neq i}\sum_{d=1}^{d_{\max}}
w_d\,\mathcal{G}_{j\to i}(x_j(t-d),x_i(t-d))
\Big).
\end{aligned}
\end{equation*}
The demand components are approximated around the equilibrium as:
\begin{align*}
\mathcal{G}_{i\to i}(x_i)
&\approx
\alpha_{i,\text{local}}\,T_i\,g_{\mathrm{lim}}^i\,x_i^{-\lambda_{ii}}, \\
\mathcal{G}_{i\to j}(x_i,x_j)
&\approx
\alpha_{i\to j,\text{out}}\,T_i\,g_{\mathrm{lim}}^i\,
x_i^{-\lambda_{ij}}x_j^{-\epsilon_{ij}}, \\
\mathcal{G}_{j\to i}(x_j,x_i)
&\approx
\alpha_{j\to i,\text{in}}\,T_i\,g_{\mathrm{lim}}^i\,
x_j^{-\lambda_{ji}}x_i^{-\epsilon_{ji}}.
\end{align*}
We now derive the four channels one by one.

\noindent\textbf{Channel (I): Instantaneous self-feedback.}
The current price $x_i(t)$ affects $L_i(t)$ through the local component $\mathcal{G}_{i\to i}$ and the source-side price of outbound traffic $\mathcal{G}_{i\to j}$. Therefore,
\begin{equation*}
\frac{\partial L_i}{\partial x_i(t)}
=
\frac{1}{g_{\mathrm{lim}}^i}
\left(
\frac{\partial \mathcal{G}_{i\to i}(x_i(t))}{\partial x_i(t)}
+
\sum_{j\neq i}
\frac{\partial \mathcal{G}_{i\to j}(x_i(t),x_j(t))}{\partial x_i(t)}
\right).
\end{equation*}
Evaluating the derivatives at $\mathbf{x}^*=\mathbf{1}$ gives
\begin{align*}
\left.\frac{\partial \mathcal{G}_{i\to i}}{\partial x_i}\right|_*
&=
-\lambda_{ii}\,\alpha_{i,\text{local}}\,T_i\,g_{\mathrm{lim}}^i, \\
\left.\frac{\partial \mathcal{G}_{i\to j}}{\partial x_i}\right|_*
&=
-\lambda_{ij}\,\alpha_{i\to j,\text{out}}\,T_i\,g_{\mathrm{lim}}^i.
\end{align*}
Substituting them back yields
\begin{equation*}
\left.\frac{\partial L_i}{\partial x_i(t)}\right|_*
=
-T_i
\left(
\lambda_{ii}\alpha_{i,\text{local}}
+
\sum_{j\neq i}\lambda_{ij}\alpha_{i\to j,\text{out}}
\right).
\end{equation*}
We therefore define
\begin{equation*}
\Lambda_i^{\text{inst}}
=
\lambda_{ii}\alpha_{i,\text{local}}
+
\sum_{j\neq i}\lambda_{ij}\alpha_{i\to j,\text{out}}.
\end{equation*}
This is the coefficient of Channel~(I).

\noindent\textbf{Channel (II): Instantaneous feedback from other chains.}
For a fixed peer chain $j\neq i$, the current price $x_j(t)$ affects $L_i(t)$ through the destination-side elasticity of outbound traffic from $i$ to $j$. Hence,
\begin{equation*}
\frac{\partial L_i}{\partial x_j(t)}
=
\frac{1}{g_{\mathrm{lim}}^i}
\frac{\partial \mathcal{G}_{i\to j}(x_i(t),x_j(t))}{\partial x_j(t)}.
\end{equation*}
Using the product-form demand and evaluating at equilibrium,
\begin{equation*}
\left.\frac{\partial \mathcal{G}_{i\to j}}{\partial x_j}\right|_*
=
-\epsilon_{ij}\,\alpha_{i\to j,\text{out}}\,T_i\,g_{\mathrm{lim}}^i,
\end{equation*}
which gives
\begin{equation*}
\left.\frac{\partial L_i}{\partial x_j(t)}\right|_*
=
-T_i\,\epsilon_{ij}\alpha_{i\to j,\text{out}}.
\end{equation*}
We therefore define
\begin{equation*}
\Lambda_{ij}^{\text{inst}}
=
\epsilon_{ij}\alpha_{i\to j,\text{out}}.
\end{equation*}
This is the coefficient of Channel~(II).

\noindent\textbf{Channel (III): Delayed self-feedback.}
The past price $x_i(t-d)$ affects the current load through the delayed inbound terms, which respond to the destination price via elasticity $\epsilon_{ji}$ and arrive under weight $w_d$. The relevant demand component is
$
w_d\,\alpha_{j\to i,\text{in}}\,T_i\,g_{\text{lim}}^i\,
x_j(t\!-\!d)^{-\lambda_{ji}}\,x_i(t\!-\!d)^{-\epsilon_{ji}}.
$
Differentiating with respect to $x_i(t\!-\!d)$ at $\mathbf{x}^*=\mathbf{1}$, dividing by $g_{\text{lim}}^i$, and summing over all $j\neq i$:
$$
\left.\frac{\partial L_i}{\partial x_i(t\!-\!d)}\right|_* 
= -w_d\,T_i\!\sum_{j\neq i}\epsilon_{ji}\,\alpha_{j\to i,\text{in}}.
$$
We define $\Lambda_{i}^{\text{delay}} = \sum_{j \neq i} \epsilon_{ji} \alpha_{j \to i, \text{in}}$, which is the coefficient of Channel~(III).

\noindent\textbf{Channel (IV): Delayed feedback from other chains.}
For a fixed peer chain $j\neq i$, the past price $x_j(t-d)$ affects the current load through delayed inbound traffic from $j$ to $i$. Hence,
\begin{equation*}
\frac{\partial L_i}{\partial x_j(t-d)}
=
\frac{1}{g_{\mathrm{lim}}^i}
\,w_d\,
\frac{\partial \mathcal{G}_{j\to i}(x_j(t-d),x_i(t-d))}{\partial x_j(t-d)}.
\end{equation*}
Evaluating the derivative at equilibrium gives
\begin{equation*}
\left.\frac{\partial \mathcal{G}_{j\to i}}{\partial x_j}\right|_*
=
-\lambda_{ji}\,\alpha_{j\to i,\text{in}}\,T_i\,g_{\mathrm{lim}}^i,
\end{equation*}
and therefore
\begin{equation*}
\left.\frac{\partial L_i}{\partial x_j(t-d)}\right|_*
=
-w_d\,T_i\,\lambda_{ji}\alpha_{j\to i,\text{in}}.
\end{equation*}
We define
\begin{equation*}
\Lambda_{ji}^{\text{delay}}
=
\lambda_{ji}\alpha_{j\to i,\text{in}},
\end{equation*}
which is the coefficient of Channel~(IV).

\section{Additional Experiments}\label{sec: additional-experiments}
In this section, we validate both potential pathways and rerun the experiment under several varying parameters. For the reruns, we mainly aim to report that the results are insensitive to these varying parameters, such as delay shapes, value of $d_{max}$, number of chains, the price-dependent elasticity, and mempool sizes.

\subsection{Validation of Potential Pathways}\label{sec: appendix-pathway-validation}\label{sec: additional-evaluation-boundary-coupling-dominance}
This section presents the full validation deferred from Section~\ref{sec: validation}. We first define a diagnostic criterion, then validate Pathway~1, and finally evaluate both sub-channels of Pathway~2. All experiments follow the setup of Section~\ref{sec: experimental-setup}.

\noindent\textbf{Diagnostic criterion.}
The delayed terms in Eq.~\ref{eq:governing-recurrence} involve past deviations $\Delta x(t\!-\!d)$, whose relationship to the current deviation $\Delta x(t)$ is not fixed by the model or the system parameters but emerges from the realized trajectory. If a delayed pathway (Pathway~1 or Pathway~2-1) is an effective instability mechanism, the delayed feedback it delivers must systematically amplify the current-step dynamics. We capture this through the ratio $|\Delta x(t\!-\!d)/\Delta x(t)|$, where a consistently large ratio indicates an effective pathway.

We define two diagnostic indicators, one for each delayed pathway. For Pathway~1, the delayed self-feedback amplifies the current action when past and present deviations share the same sign, under constructive interference during monotonic return. We define the amplified defense factor $\kappa_{\text{def}}$ as
\begin{equation}\label{eq:kappa-def}
\kappa_{\text{def},i}(t)=\sum_{d=1}^{d_{\max}} w_d \cdot \left|\frac{\Delta x_i(t\!-\!d)}{\Delta x_i(t)}\right| \cdot \mathbf{1}\!\left(\Delta x_i(t\!-\!d)\,\Delta x_i(t)>0\right).
\end{equation}
The same-sign indicator isolates the constructive episodes where delayed inbound feedback reinforces the current correction. When the system is monotonically returning to equilibrium, $|\Delta x_i(t\!-\!d)| > |\Delta x_i(t)|$ by construction, so $\kappa_{\text{def}} > 1$ and the delayed feedback overshoots.

For Pathway~2-1, the delayed cross-chain feedback acts as a destabilizing attack when a past low price on chain $j$ sends a large demand wave to chain $i$ at a time when $i$ is already congested and needs to cool down. This amplification requires opposite signs between the source's past deviation and the destination's current deviation. We define the corresponding factor $\kappa_{\text{att}}$ as
\begin{equation}\label{eq:kappa-att}
\kappa_{\text{att},j\to i}(t)\!=\!\sum_{d=1}^{d_{\max}} w_d \!\left|\frac{\Delta x_j(t\!-\!d)}{\Delta x_j(t)}\right| \!\cdot\mathbf{1}\!\big(\Delta x_i(t)\!\cdot\!(-\Delta x_j(t\!-\!d))\!>\!0\big).
\end{equation}
Pathway~2-2 is instantaneous and involves no historical state, so $\kappa$ cannot capture it. We evaluate Pathway~2-2 directly through stability scans at the end of this section. To summarize, if a pathway is a viable instability mechanism, its indicator should grow systematically with the corresponding channel strength. If it does not, the pathway is unlikely to be a practical driver of instability.

\begin{figure}[!tp]
    \centering
    \begin{subfigure}{0.95\linewidth}
        \centering
        \includegraphics[width=\linewidth]{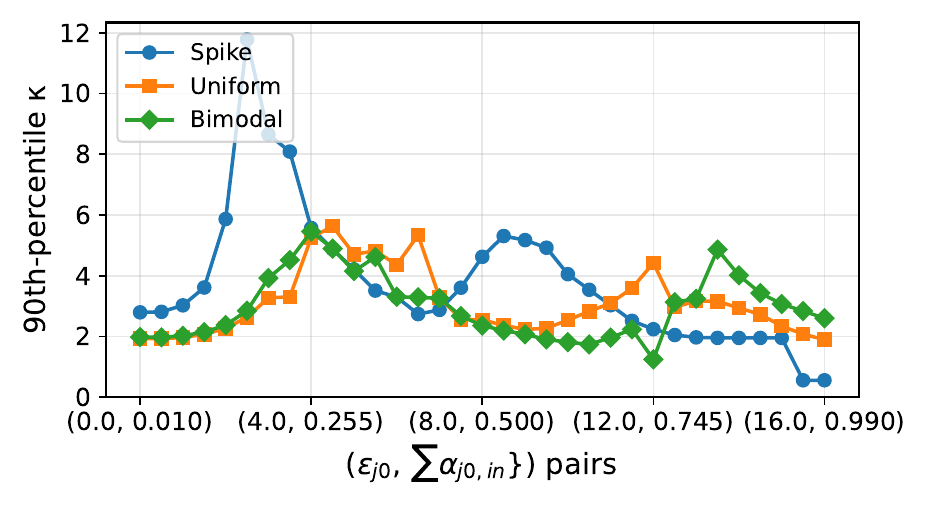}
        \caption{$90$th-percentile $\kappa_{\text{def}}$.}
        \label{fig: p90-kappa-defense}
        \vspace{0.5cm}
    \end{subfigure}
    \begin{subfigure}{0.95\linewidth}
        \centering
        \includegraphics[width=\linewidth]{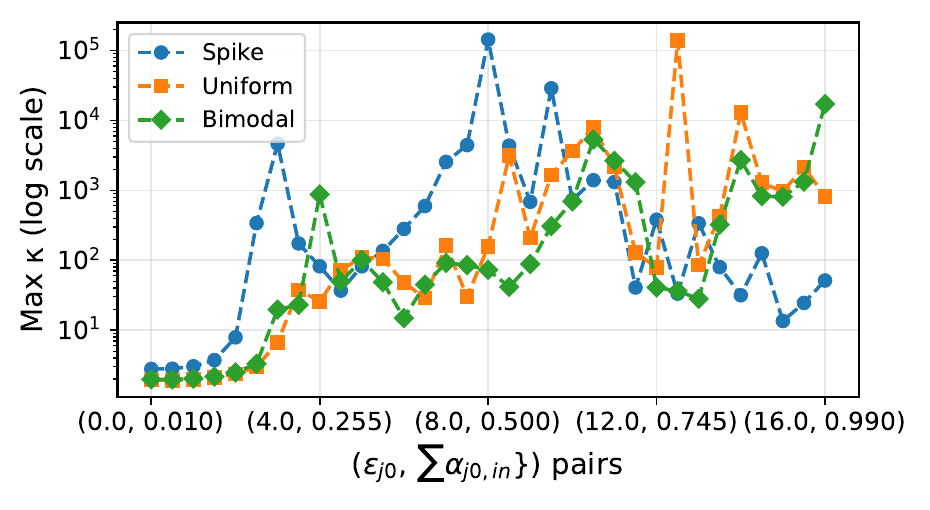}
        \caption{Maximum $\kappa_{\text{def}}$.}
        \label{fig: max-kappa-defense}
    \end{subfigure}
    \caption{Correlation between $\kappa_{\text{def}}$ and the delayed defense channel $\Lambda_i^{\text{delay}}$.}
    \label{fig: kappa-defense}
\end{figure}

\noindent\textbf{Validation of Pathway~1.}
Recall that $\kappa_{\text{def}}$ divides a past deviation by the current one. The current deviation crosses zero whenever the trajectory passes through equilibrium, so the ratio diverges on those steps. The mean and the maximum are both dominated by a handful of such steps. We therefore report the $90$th percentile of the per-step $\kappa_{\text{def}}$ series, which is insensitive to the crossings.

Figure~\ref{fig: p90-kappa-defense} shows the result. Under spike latency, $\kappa_{\text{def}}$ reaches $11.8$. The uniform and bimodal shapes reach only $5.6$ and $5.5$. Spike latency therefore amplifies about twice as strongly, and it reaches its peak at $\epsilon_{j0}=2.5$, against $4.5$ and $4.0$ for the two dispersed shapes. The reason is mechanical. A spike distribution concentrates the force of a past deviation into a single step, while a dispersed distribution spreads the same force over the delay window.

\begin{figure}[!tp]
    \centering
    \begin{subfigure}{0.95\linewidth}
        \centering
        \includegraphics[width=\linewidth]{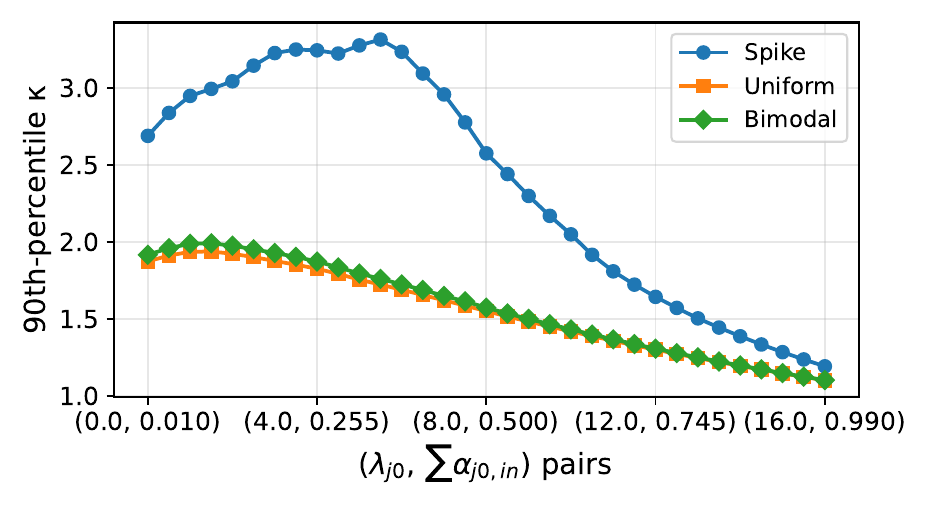}
        \caption{$90$th-percentile $\kappa_{\text{att}}$.}
        \label{fig: p90-kappa-attack}
        \vspace{0.5cm}
    \end{subfigure}
    \begin{subfigure}{0.95\linewidth}
        \centering
        \includegraphics[width=\linewidth]{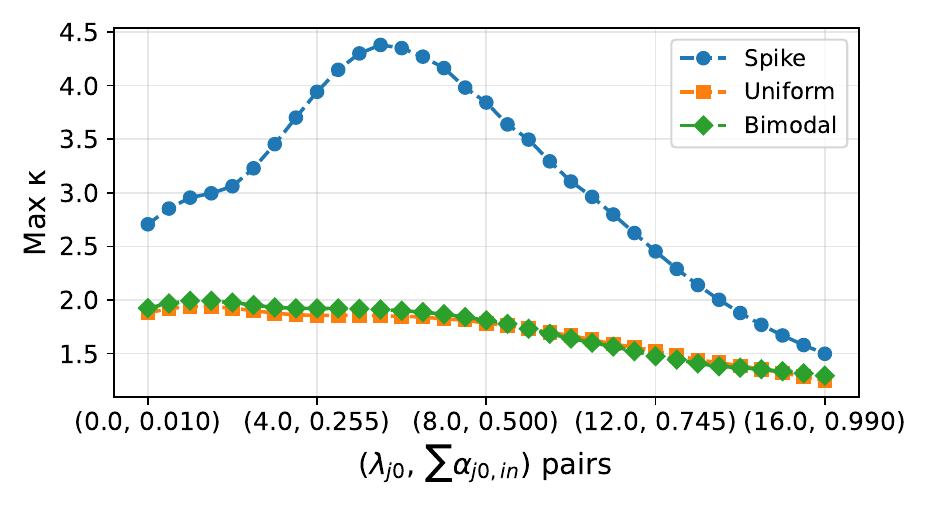}
        \caption{Maximum $\kappa_{\text{att}}$.}
        \label{fig: max-kappa-attack}
    \end{subfigure}
    \caption{Correlation between $\kappa_{\text{att}}$ and the delayed attack channel $\Lambda_{ji}^{\text{delay}}$.}
    \label{fig: kappa-attack}
\end{figure}

Figure~\ref{fig: max-kappa-defense} reports the maximum of $\kappa_{\text{def}}$ along the same trajectories, plotted on a log scale. The maximum exceeds the $90$th percentile by four orders of magnitude and reaches $1.4\times10^{5}$ under spike latency. Note that the gap has the same mechanical origin. The ratio $|\Delta x_i(t\!-\!d)/\Delta x_i(t)|$ peaks when the current deviation approaches zero while the delayed deviations remain large. These steps are exactly the moments when the system has almost recovered and the stale inbound force strikes again, so the delayed feedback most severely exceeds the correction budget of Figure~\ref{fig: pathway1}. The maximum therefore describes the worst single step rather than a trend, and we read the trend from the $90$th percentile alone.

Together, the two panels confirm that delay-amplified over-correction is an effective
mechanism, and that its strength depends on the shape of the delay distribution.

\noindent\textbf{Validation of Pathway~2.}
We next evaluate Pathway~2 empirically. Pathway~2-1 is fundamentally self-limiting.
Figure~\ref{fig: kappa-attack} reports $\kappa_{\text{att}}$ under the same protocol and
the same $90$th-percentile statistic. Under spike latency $\kappa_{\text{att}}$ peaks at
$3.3$, and the uniform and bimodal shapes peak at $1.9$ and $2.0$. All three peaks sit far
below the $11.8$ that $\kappa_{\text{def}}$ reaches. As for the shape of $\kappa_{\text{att}}$, strengthening Channel~(IV) does not drive $\kappa_{\text{att}}$ upward. Under spike
latency the indicator rises only from $2.7$ at $\lambda_{j0}=0$ to its shallow peak at
$\lambda_{j0}=5.5$, and it then falls monotonically to $1.2$ at $\lambda_{j0}=16$, below
where it started. The reason is mechanical.
Amplification through this channel requires the price perturbations of the source and
destination chains to have opposite signs. However, stronger coupling causes their prices
to synchronize and reduces the opposite-sign intervals required for amplification.

\begin{figure*}[!tbp]
    \begin{center}
    \setlength{\abovecaptionskip}{0.mm}
    \centering
    \begin{subfigure}{0.32\linewidth}
        \centering
        \includegraphics[width=\linewidth]{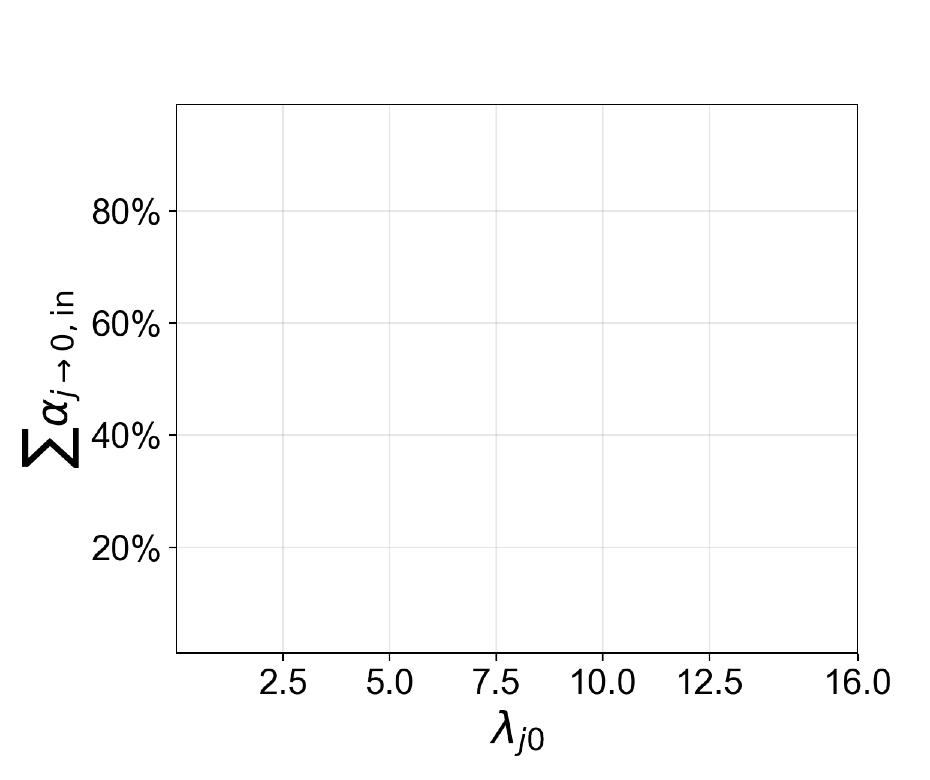}
        \caption{$\Lambda_{ji}^{\text{delay}}$, bimodal, all stable}
        \label{fig: converged-coupling-dominance-bimodal}
    \end{subfigure}
    \begin{subfigure}{0.32\linewidth}
        \centering
        \includegraphics[width=\linewidth]{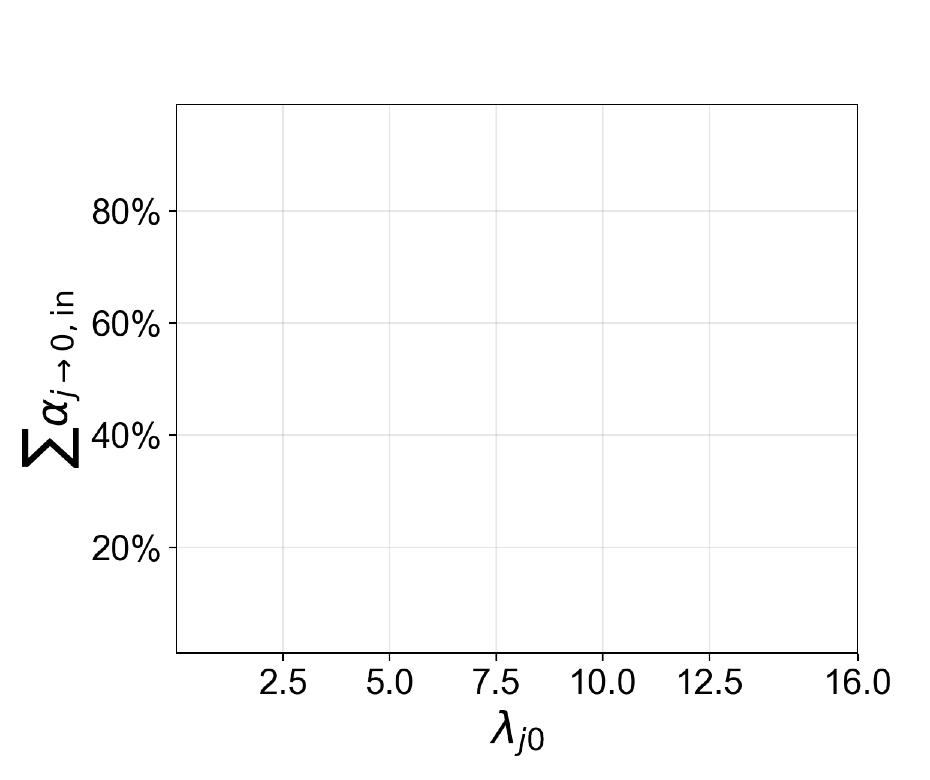}
        \caption{$\Lambda_{ji}^{\text{delay}}$, uniform, all stable}
        \label{fig: converged-coupling-dominance-uniform}
    \end{subfigure}
    \begin{subfigure}{0.32\linewidth}
        \centering
        \includegraphics[width=\linewidth]{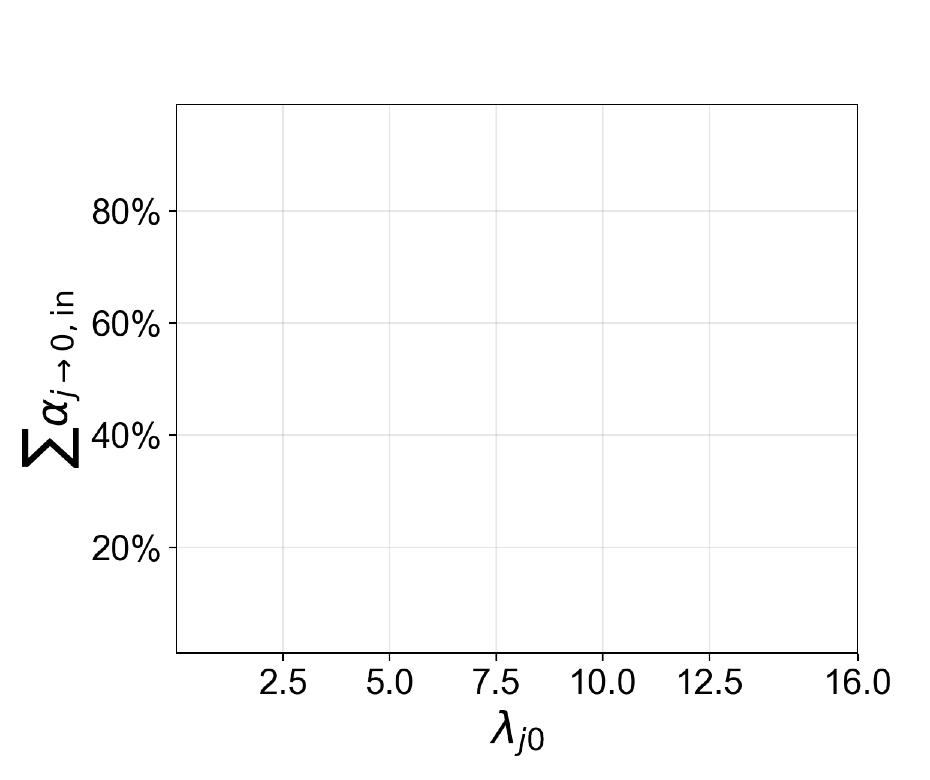}
        \caption{$\Lambda_{ji}^{\text{delay}}$, spike, all stable}
        \label{fig: converged-coupling-dominance-spike}
    \end{subfigure}

    \begin{subfigure}{0.32\linewidth}
        \centering
        \includegraphics[width=\linewidth]{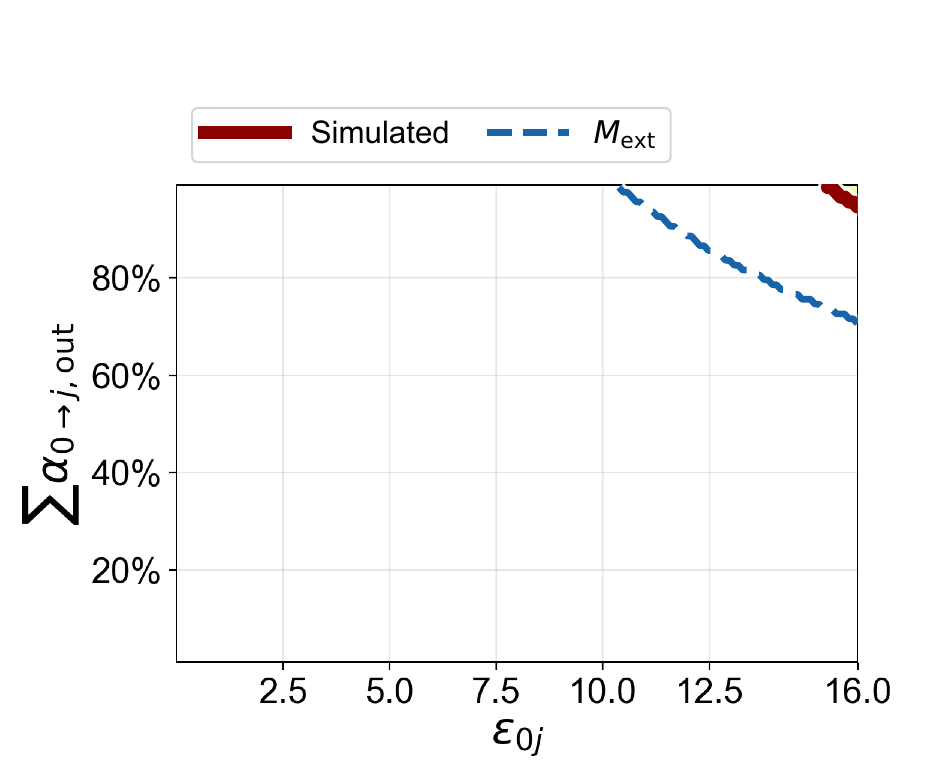}
        \caption{$\Lambda_{j}^{\text{inst}}$, bimodal latency}
        \label{fig: coupling-dominance-bimodal}
    \end{subfigure}
    \begin{subfigure}{0.32\linewidth}
        \centering
        \includegraphics[width=\linewidth]{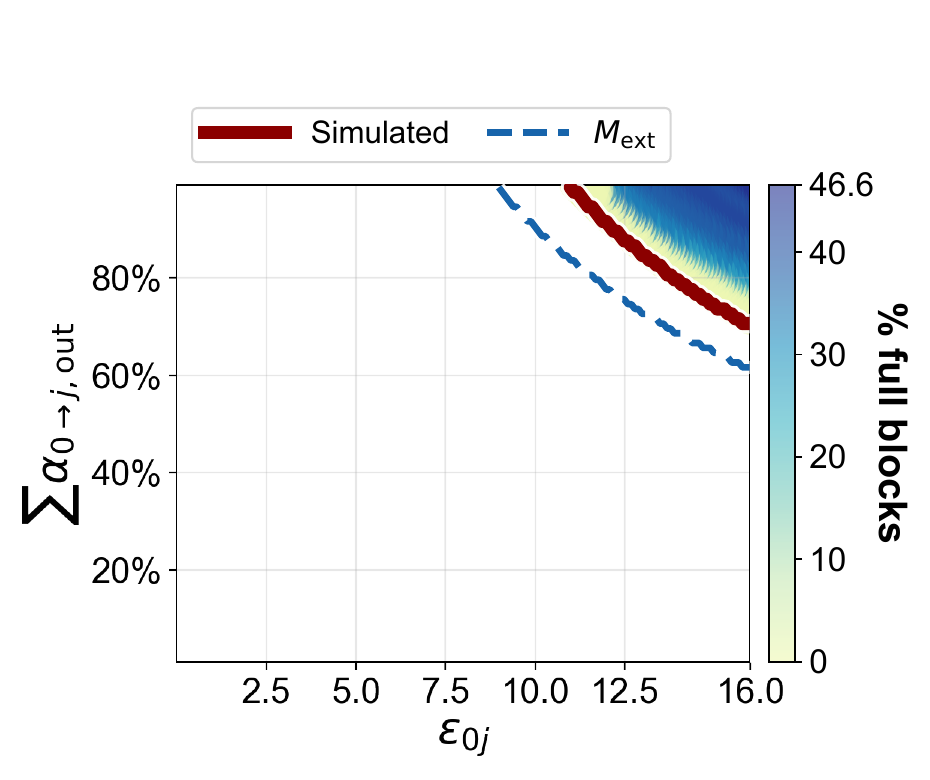}
        \caption{$\Lambda_{j}^{\text{inst}}$, uniform latency}
        \label{fig: coupling-dominance-uniform}
    \end{subfigure}
    \begin{subfigure}{0.32\linewidth}
        \centering
        \includegraphics[width=\linewidth]{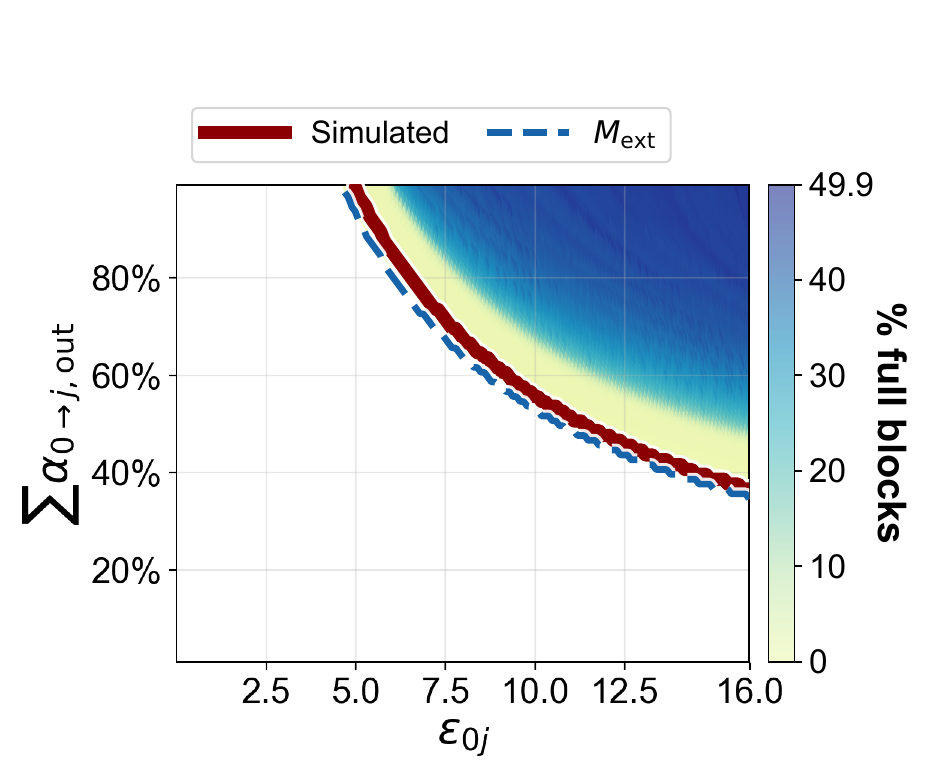}
        \caption{$\Lambda_{j}^{\text{inst}}$, spike latency}
        \label{fig: coupling-dominance-spike}
    \end{subfigure}

\caption{Stability boundary while enhancing the delayed attack channel $\Lambda_{ji}^{\text{delay}}$ (top row) and the instantaneous attack channel $\Lambda_{j}^{\text{inst}}$ (bottom row) under three latency distributions ($d_{\max}=5$)}
\label{fig: coupling-dominance-merged}
    \end{center}
\end{figure*}

\begin{figure}
    \centering
    \includegraphics[width=\linewidth]{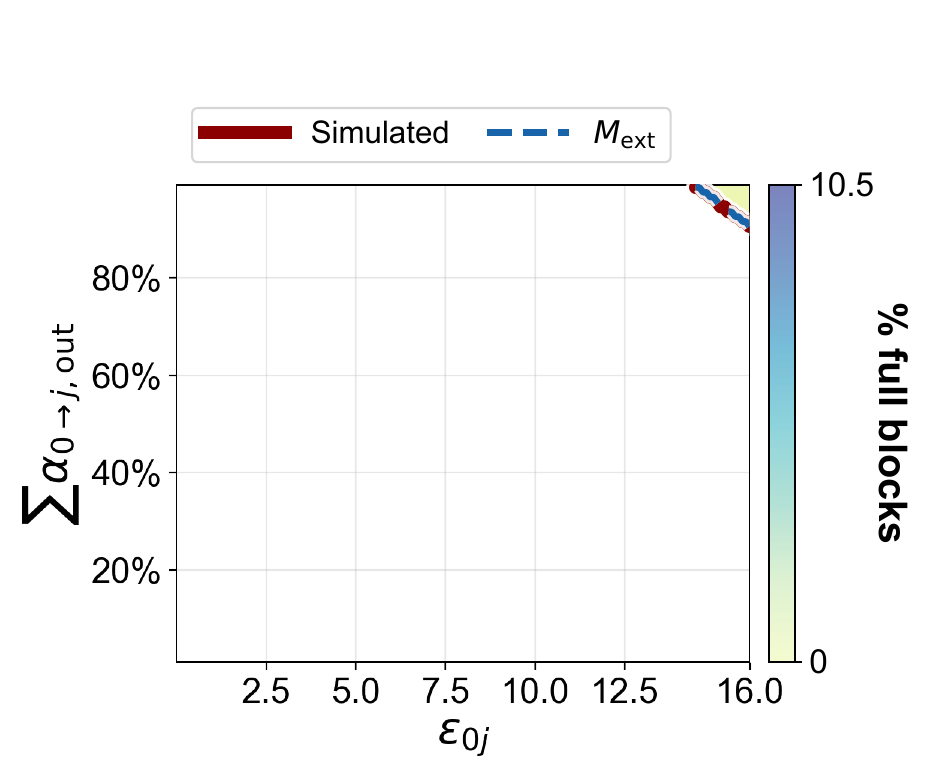}
    \captionsetup{justification=justified}
    \caption{Stability boundary and load pattern while enhancing $\Lambda_{ij}^{\text{inst}}$ under $d_{\max}=1$}
    \label{fig: boundary-inst-attack-without-latency}
\end{figure}

Next, we iterate the parameters associated with $\Lambda_{ji}^{\text{delay}}$, including $\lambda_{j0}$ and $\alpha_{j\to 0,\text{in}}$, and observe whether the system becomes unstable. The top row of Figure~\ref{fig: coupling-dominance-merged} shows that all $47{,}817$ tested configurations converge. Channel~(IV) therefore produces no simulated stability boundary within the scanned range. Note that the three panels are empty in a stronger sense than convergence alone. No chain fills a single block in any of the $47{,}817$ runs, so the full-block share is zero everywhere and does not grow as $\Lambda_{ji}^{\text{delay}}$ grows.

Pathway~2-2 also fails to produce comparable amplification. We iterate the critical parameters of $\Lambda_{ij}^{\text{inst}}$ to evaluate whether its positive feedback can overwhelm the system's negative feedback. Counter-intuitively, some configurations do fail to converge. The bottom row of Figure~\ref{fig: coupling-dominance-merged} reports the outcome. Under spike latency, $4{,}625$ of the $15{,}939$ configurations fail. The first failure appears at $\epsilon_{0j}=5.0$ and a $99\%$ outbound ratio, and the worst configuration fills $49.9\%$ of its blocks. The uniform shape fails at $852$ configurations and reaches $46.6\%$. The bimodal shape is the informative case. Only $23$ configurations fail, and not one of them ever fills a block, so that panel carries no colour bar at all.

\begin{figure*}[!tp]
\begin{center}
\setlength{\abovecaptionskip}{0.mm}
\centering

    \begin{subfigure}{0.24\linewidth}
        \centering
        \includegraphics[width=\linewidth]{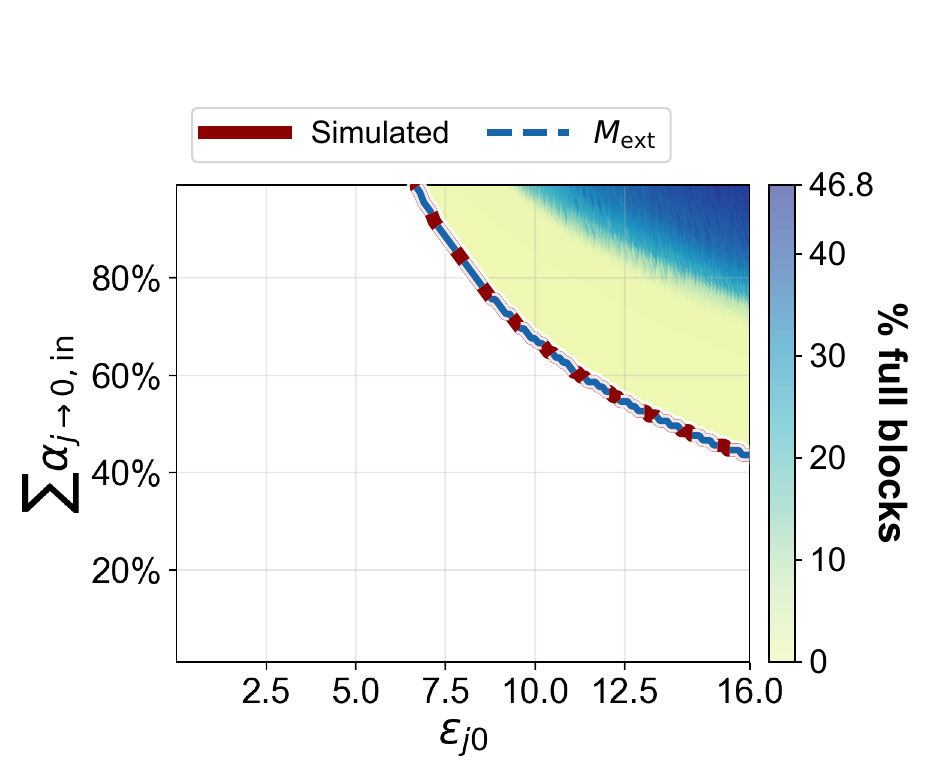}
        \caption{Bimodal, fixed $\delta$}
        \label{fig: over-correction-bimodal}
    \end{subfigure}
    \hfill
    \begin{subfigure}{0.24\linewidth}
        \centering
        \includegraphics[width=\linewidth]{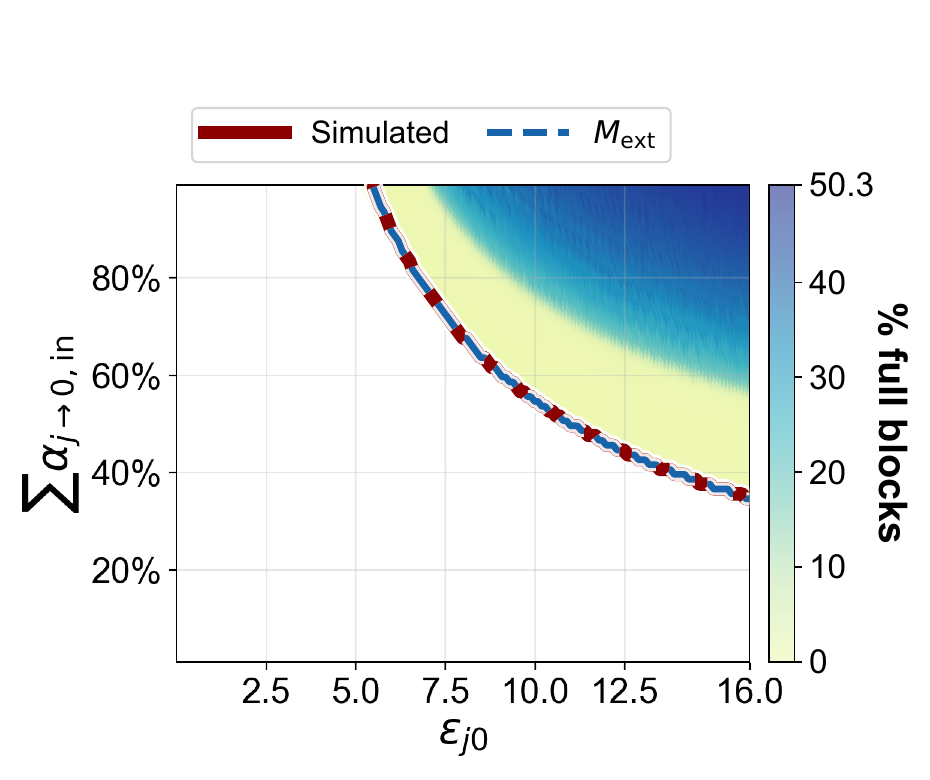}
        \caption{Uniform, fixed $\delta$}
        \label{fig: over-correction-uniform}
    \end{subfigure}
    \hfill
    \begin{subfigure}{0.24\linewidth}
        \centering
        \includegraphics[width=\linewidth]{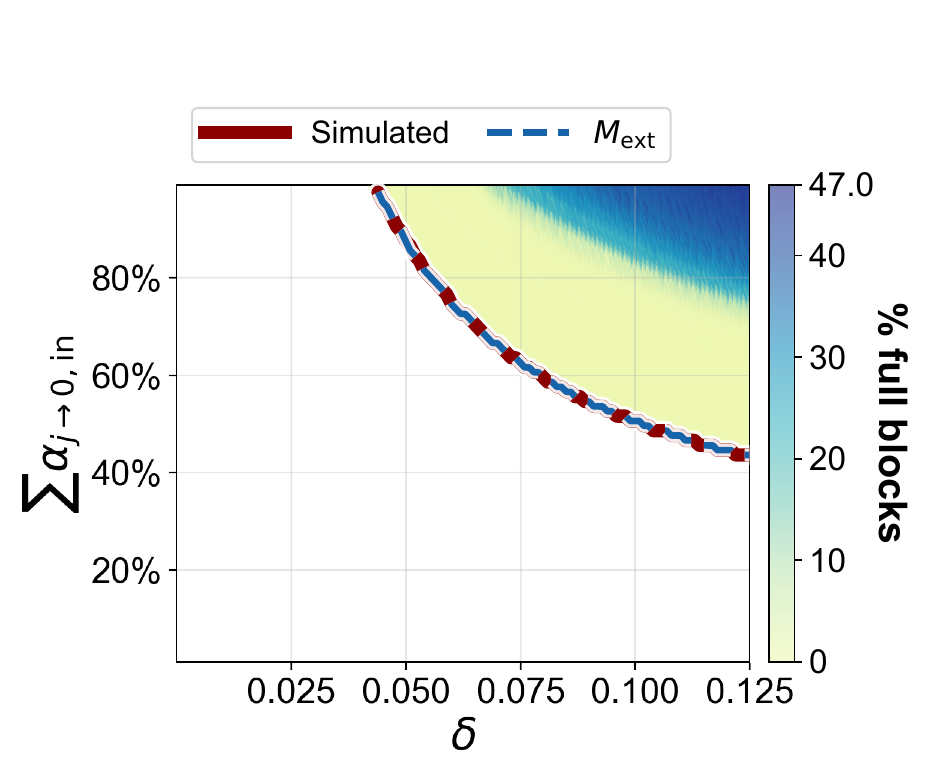}
        \caption{Bimodal, fixed elasticity}
        \label{fig: over-correction-delta-bimodal}
    \end{subfigure}
    \hfill
    \begin{subfigure}{0.24\linewidth}
        \centering
        \includegraphics[width=\linewidth]{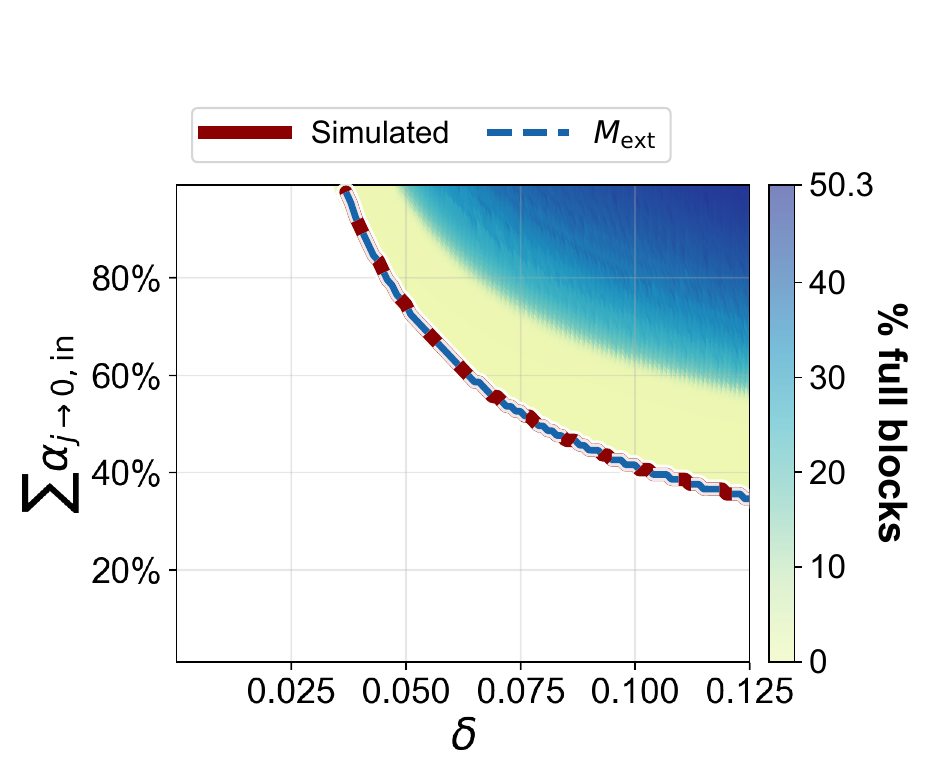}
        \caption{Uniform, fixed elasticity}
        \label{fig: over-correction-delta-uniform}
    \end{subfigure}

\caption{Over-correction under bimodal and uniform latency. Stable region under standard parameters with fixed $\delta=0.125$~(a)-(b), and under fixed elasticity ($16$)~(c)-(d).}
\label{fig: over-correction-appendix-latency}
\end{center}
\end{figure*}

However, strengthening $\Lambda_{ij}^{\text{inst}}$ on one chain also strengthens the negative-feedback channels of the connected chains. The delayed pathways of other chains therefore remain active, and the instability in the bottom row cannot be attributed to Channel~(II) alone. Unlike in the other figures, the $M_{\mathrm{ext}}$ boundary in the bottom row does not closely track the simulated boundary. The disagreement is one-sided. Under spike latency, $354$ configurations are $M_{\mathrm{ext}}$-unstable and still converge, while no configuration that $M_{\mathrm{ext}}$ calls stable fails in simulation. The same one-sided gap appears under the uniform and bimodal shapes, at $760$ and $906$ configurations. The reason is that the symmetric perturbation used in this experiment does not excite all system modes captured by $M_{\mathrm{ext}}$.

To isolate Channel~(II), we set the maximum latency to $d_{\max}=1$ so that other chains' delayed feedback channel~(III) is not activated, and we repeat the experiment. Figure~\ref{fig: boundary-inst-attack-without-latency} shows that only $77$ of $15{,}939$ configurations ($0.48\%$) fail the convergence criterion. Simulated nonconvergence first appears at $\epsilon_{0j}=14.5$ and a $99\%$ outbound ratio. The failures are also mild. $71$ of the $77$ unstable points never fill a block, and the worst configuration fills $10.5\%$ of its blocks.
This confirms that Channel~(II) alone produces only a narrow instability region under extreme conditions, and it does not reproduce the broad amplification observed with Pathway~1.

\subsection{Experimental Analysis Varying Other System Configurations}\label{sec: additional-experiment-varying-dmax-and-chain-number}
We evaluate Pathway~1 under varying latency shapes, $d_{\max}$, and chain numbers, using Figure~\ref{fig: over-correction-boundary} as the baseline.

\noindent\textbf{Varying the latency shape.} We repeat the two experiments of Section~\ref{sec: boundary} under bimodal and uniform latency, and Figure~\ref{fig: over-correction-appendix-latency} reports the results. The top row fixes $\delta=0.125$ and iterates the structural parameters. Both shapes preserve the boundary contraction, but the contraction is milder than under spike latency. At $\alpha=0.97$, the maximum stable elasticity drops from the monolithic $16$ to $6.8$ under bimodal latency and to $5.5$ under uniform latency, compared with $3.1$ under spike latency. The share of the grid that fails follows the same ordering. Spike latency breaks $8{,}008$ of the $15{,}939$ configurations, uniform latency breaks $4{,}644$, and bimodal latency breaks $3{,}405$. Note that the three shapes nearly agree once a configuration sits past its own boundary. The worst configuration fills $52.7\%$ of its blocks under spike latency, $50.3\%$ under uniform latency, and $46.8\%$ under bimodal latency. The latency shape therefore decides more how much of the parameter space is unstable.
The bottom row fixes the elasticity at $16$ and iterates $\delta$. $\Delta_{\text{safe}}$ again shrinks monotonically as the inbound ratio grows, while the erosion is weaker than the $84.8\%$ collapse under spike latency. It reaches $70.4\%$ under uniform latency and $64.8\%$ under bimodal latency. 

\begin{figure*}[!tp]
 \begin{center}
    \setlength{\abovecaptionskip}{0.mm}
    \centering
        \begin{subfigure}{0.24\linewidth}
            \centering
            \includegraphics[width=\linewidth]{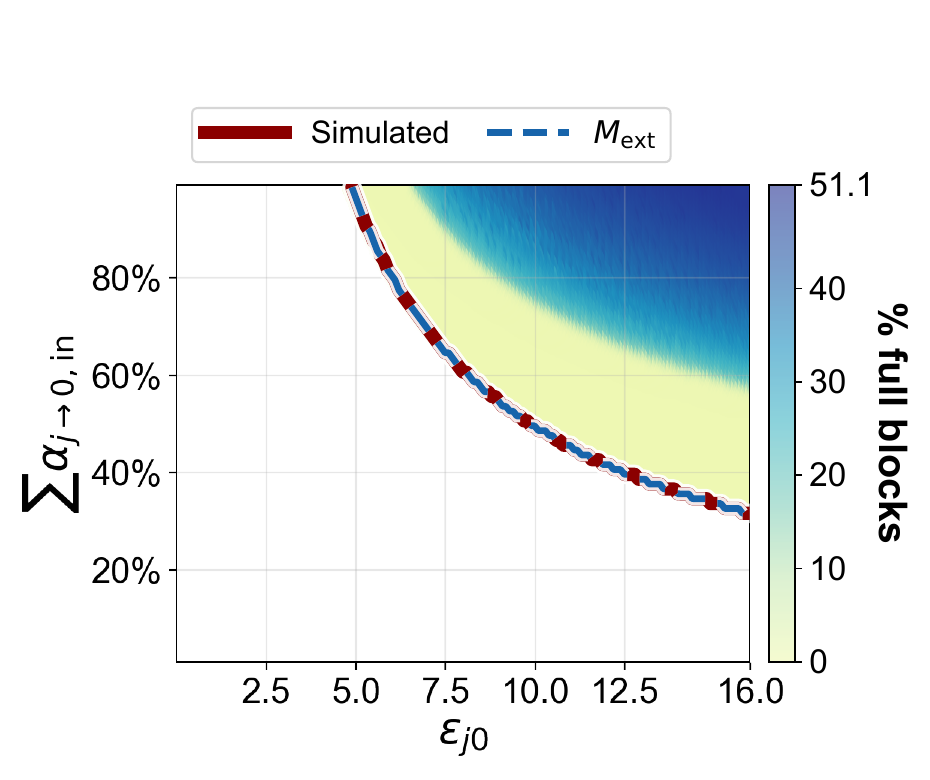}
            \caption{Boundary (OC-uniform-$7$)}
            \label{fig: latency-7-over-correction-uniform}
        \end{subfigure}
        \begin{subfigure}{0.24\linewidth}
            \centering
            \includegraphics[width=\linewidth]{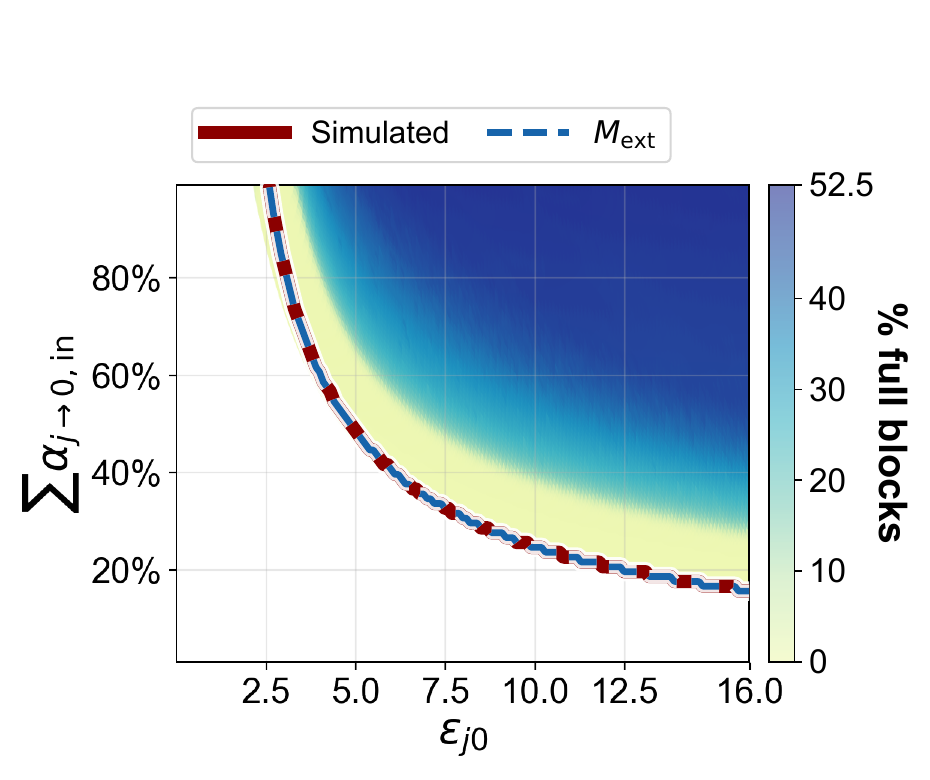}
            \caption{Boundary (OC-spike-$7$)}
            \label{fig: latency-7-over-correction-spike}
        \end{subfigure}
        \begin{subfigure}{0.24\linewidth}
            \centering
            \includegraphics[width=\linewidth]{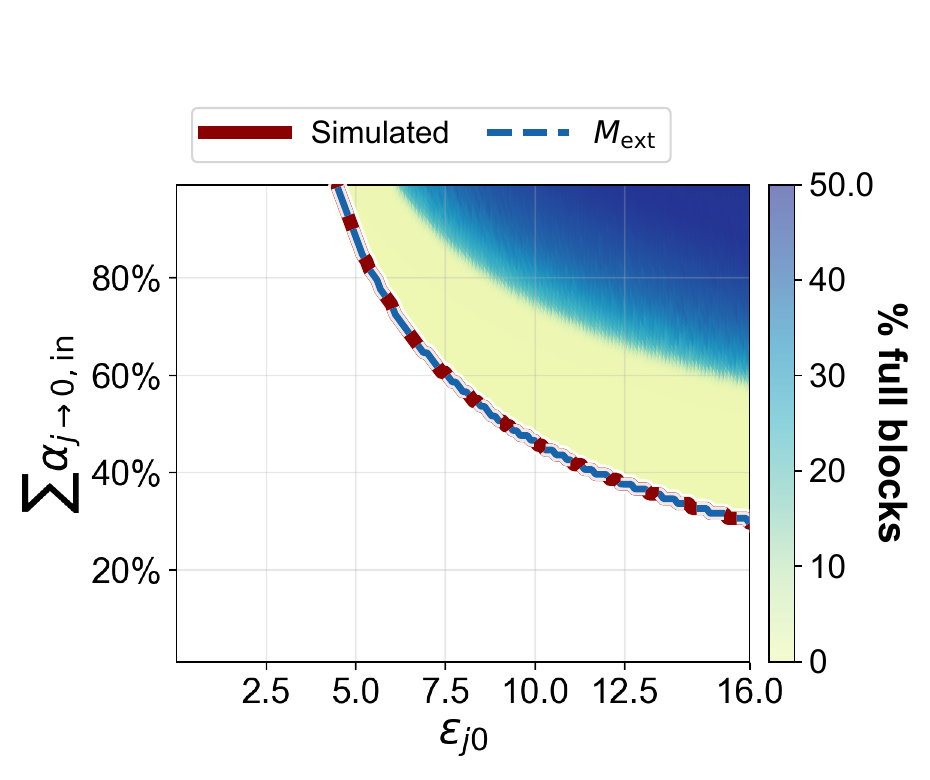}
            \caption{Boundary (OC-uniform-$10$)}
            \label{fig: latency-10-over-correction-uniform}
        \end{subfigure}
        \begin{subfigure}{0.24\linewidth}
            \centering
            \includegraphics[width=\linewidth]{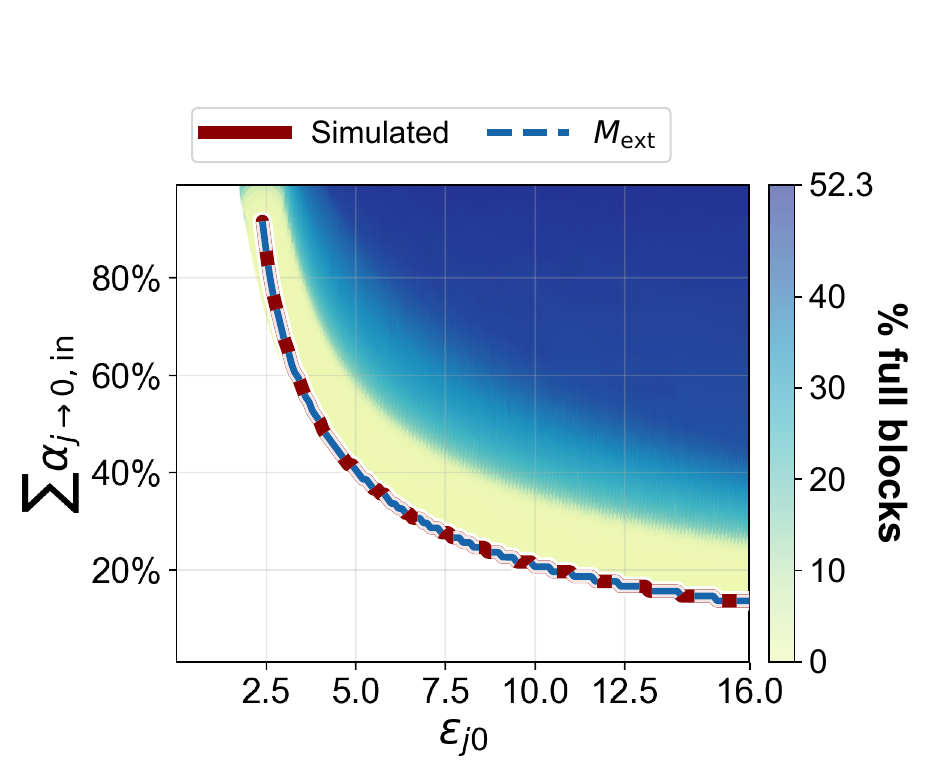}
            \caption{Boundary (OC-spike-$10$)}
            \label{fig: latency-10-over-correction-spike}
        \end{subfigure}

    \caption{Stability boundary for over-correction (OC) under two latency distributions with $d_{\max}=7$ and $10$}
    \label{fig: over-correction-dominance-additional}
 \end{center}
\end{figure*}

\begin{figure*}[!tp]
 \begin{center}
    \setlength{\abovecaptionskip}{0.mm}
    \centering
        
        \begin{subfigure}{0.32\linewidth}
            \centering
            \includegraphics[width=\linewidth]{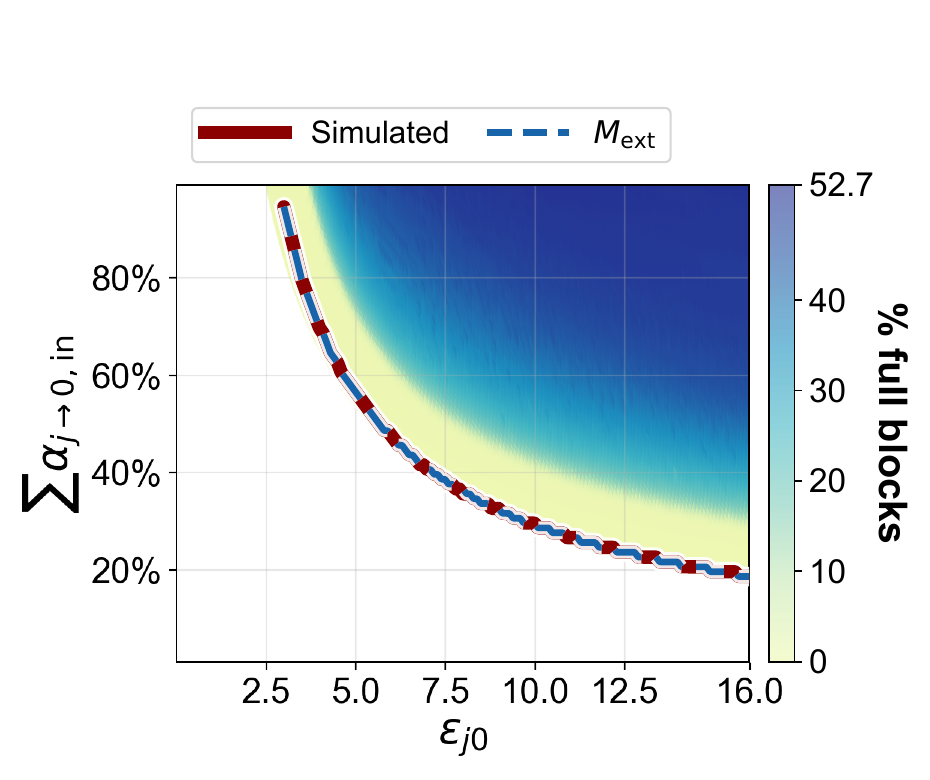}
            \caption{Boundary (\textit{OC-spike}) with $5$ chains}
            \label{fig: over-correction-spike-5-chains}
        \end{subfigure}
        \begin{subfigure}{0.32\linewidth}
            \centering
            \includegraphics[width=\linewidth]{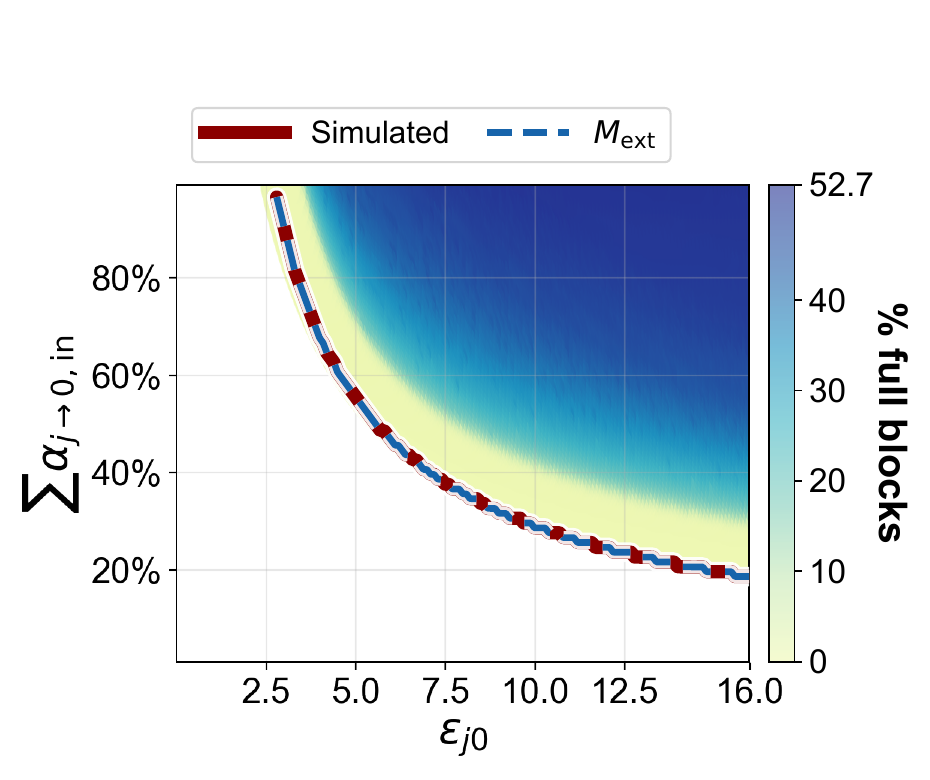}
            \caption{Boundary (\textit{OC-spike}) with $7$ chains}
            \label{fig: over-correction-spike-7-chains}
        \end{subfigure}
        \begin{subfigure}{0.32\linewidth}
            \centering
            \includegraphics[width=\linewidth]{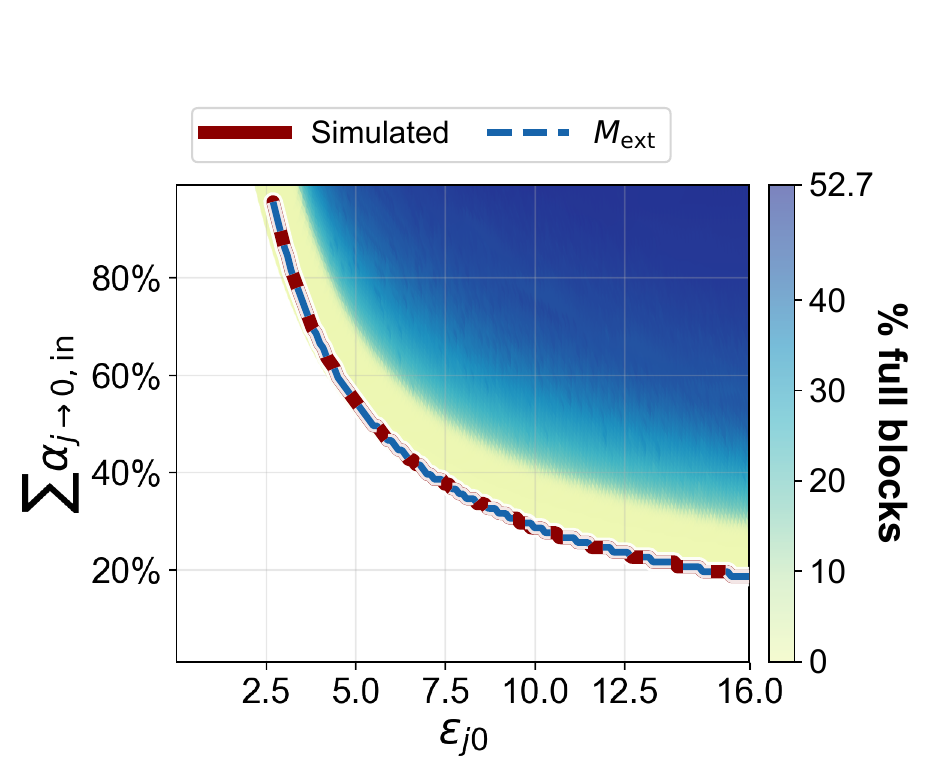}
            \caption{Boundary (\textit{OC-spike}) with $10$ chains}
            \label{fig: over-correction-spike-10-chains}
        \end{subfigure}

    \caption{Stability boundaries for over-correction (\textit{OC}) under spike latency with $5$, $7$, and $10$ chains ($d_{\max}=5$) }
    \label{fig: over-correction-boundary-under-varying-chain-number}
 \end{center}
\end{figure*}

\noindent\textbf{Varying $d_{\max}$.} To further probe the latency amplification mechanism, we repeat the experiment with $d_{\max}\in\{7,10\}$. For each $d_{\max}$, we evaluate the uniform and spike distributions. Figure~\ref{fig: over-correction-dominance-additional} shows the resulting stability boundaries.

The results reveal an inverse relation between $d_{\max}$ and system stability. This degradation is particularly evident under spike latency. The critical elasticity decreases from $3.1$ at $d_{\max}=5$ to $2.6$ at $d_{\max}=7$ and $2.4$ at $d_{\max}=10$. At $d_{\max}=10$, non-convergence at $\epsilon_{j0}=2.4$ occurs for inbound ratios of $92\%$--$99\%$, representing an $85.0\%$ contraction from the monolithic boundary of $16$. The same boundary contraction appears under uniform latency. Its critical elasticities are $5.5$, $4.9$, and $4.5$, corresponding to contractions of $65.6\%$, $69.4\%$, and $71.9\%$. As for the full block share, the maximum full-block share under spike latency is $52.7\%$, $52.5\%$, and $52.3\%$, and under uniform latency it is $50.3\%$, $51.1\%$, and $50.0\%$. 

\noindent\textbf{Varying the chain number.} To investigate the effect of network scale on stability, we expand our analysis to systems with $5$, $7$, and $10$ chains. Based on our prior findings, we conduct these experiments using spike latency with $d_{\max}=5$, which represents the most severe tested distribution. Figure~\ref{fig: over-correction-boundary-under-varying-chain-number} shows how the stability boundary changes as the number of chains increases. The boundary remains nearly invariant to the chain count. The critical elasticities are $3.0$, $2.8$, and $2.7$ for $5$, $7$, and $10$ chains, corresponding to contractions of $81.3\%$, $82.5\%$, and $83.1\%$ from the monolithic boundary. The maximum full-block share is $52.7\%$ at all three chain counts. 

\subsection{Evaluation Under Price-Dependent Elasticity}\label{sec: varying-elasticity}

\begin{figure*}[!tp]
\begin{center}
\setlength{\abovecaptionskip}{0.mm}
\centering
    \begin{subfigure}{0.32\textwidth}
        \centering
        \includegraphics[width=0.85\linewidth]{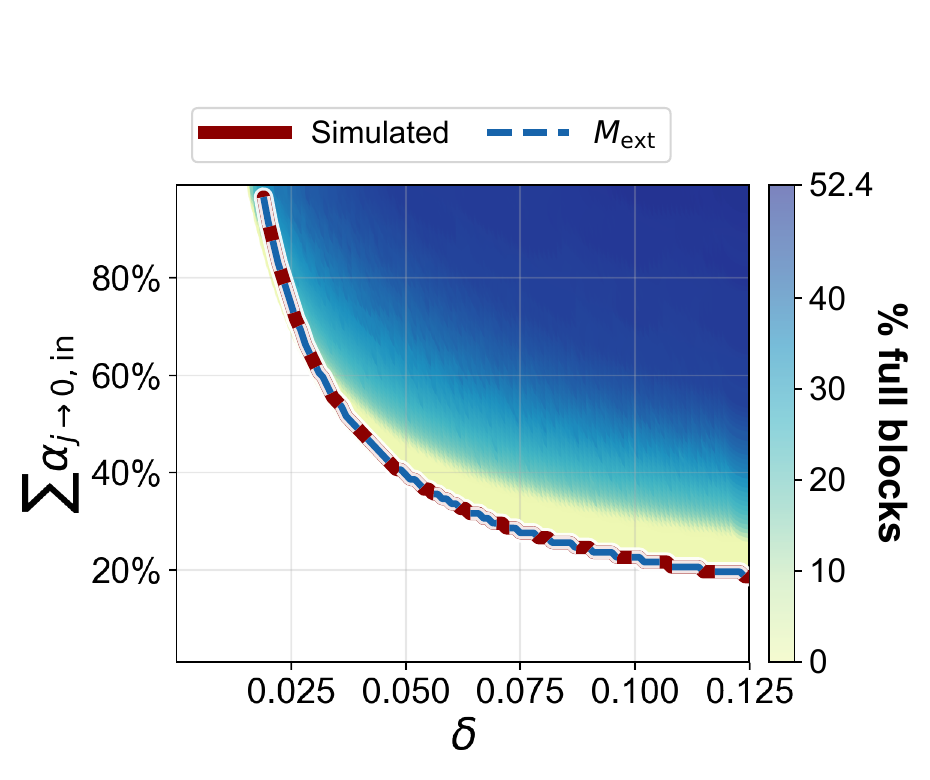}
        \caption{$\epsilon^{\text{log}}(x)=\max(0,\,16+\ln x)$}
        \label{fig: varying-elasticity-boundary-log}
    \end{subfigure}
    \begin{subfigure}{0.32\textwidth}
        \centering
        \includegraphics[width=0.85\linewidth]{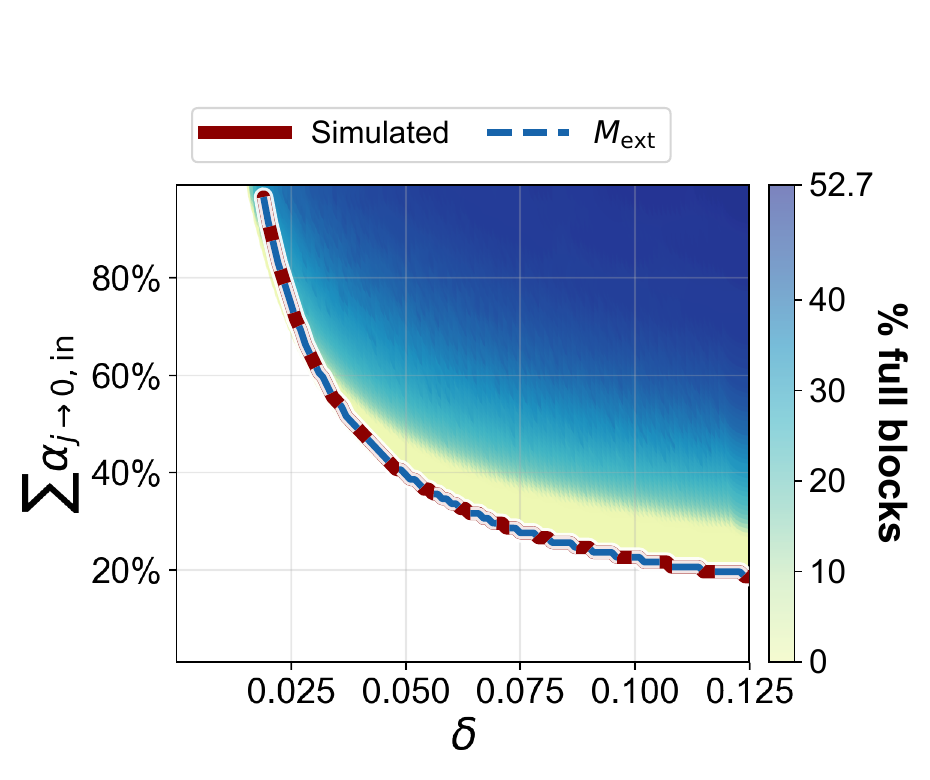}
        \caption{$\epsilon^{\text{lin}}(x)=\min(16x,\,32)$}
        \label{fig: varying-elasticity-boundary-linear}
    \end{subfigure}
    \begin{subfigure}{0.32\textwidth}
        \centering
        \includegraphics[width=0.85\linewidth]{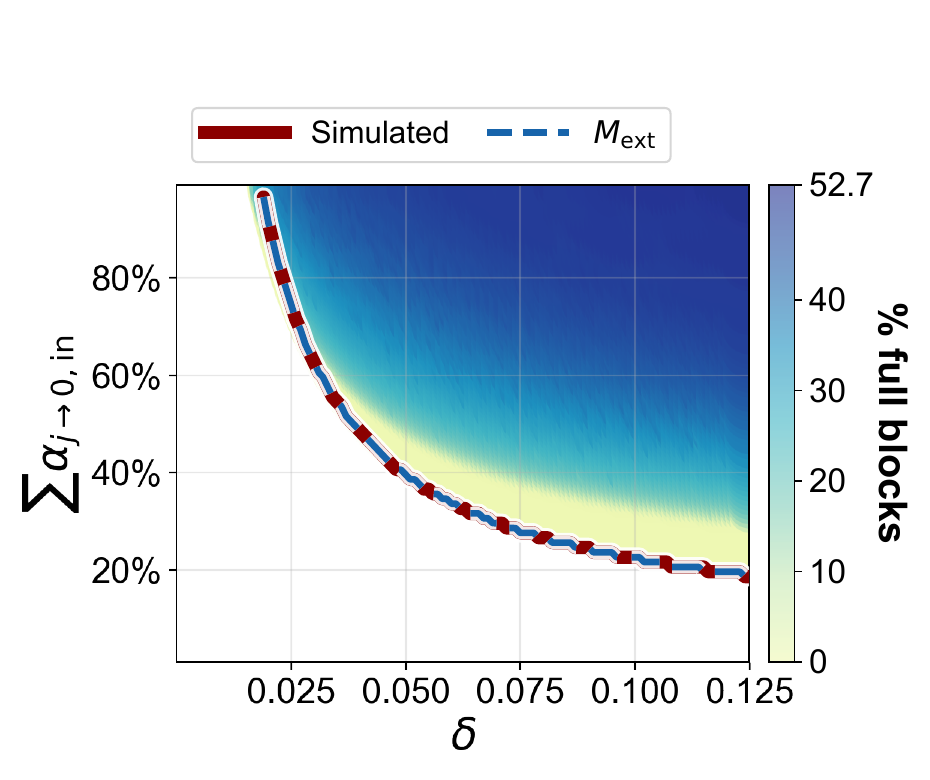}
        \caption{$\epsilon^{\text{logi}}(x)=32x^{2}/(x^{2}+1)$}
        \label{fig: varying-elasticity-boundary-logistic}
    \end{subfigure}
\caption{Stability under three price-dependent elasticity shapes (spike latency, $d_{\max}=5$)}
\label{fig: varying-elasticity-boundary}
\end{center}
\end{figure*}

Previous experiments hold the inbound elasticity $\epsilon_{j0}$ constant at every price. This section repeats the two main experiments with a price-dependent elasticity. Let $x$ denote Chain~0's base fee normalized by its equilibrium value. We replace the constant $\epsilon_{j0}$ with a function $\epsilon_{j0}(x)$ of the price of Chain $0$ and keep every other parameter of Section~\ref{sec: boundary} unchanged.

We evaluate three shapes:
\begin{align*}
\epsilon^{\text{log}}(x)&=\max\!\left(0,\,16+\ln x\right)\!,\\
\epsilon^{\text{lin}}(x)&=\min\!\left(16x,\,32\right)\!,\\
\epsilon^{\text{logi}}(x)&=32\,x^{2}/(x^{2}+1).
\end{align*}
The three shapes share two properties. First, each shape is truncated below at zero. The truncation binds only at prices many orders of magnitude below the equilibrium. Second, each shape passes through $16$ at the equilibrium price $x=1$, the constant used in Section~\ref{sec: boundary}, so all three systems present the same elasticity at the target point. 

\noindent\textbf{Stability boundary.} Figure~\ref{fig: varying-elasticity-boundary} repeats the $\Delta_{\text{safe}}$ scan of Figure~\ref{fig: over-correction-delta-spike} under each shape. The boundary is identical across the three panels and matches the constant-elasticity boundary, including the collapse to $\delta<0.019$ at an inbound ratio of $0.99$. 
As for the consistency between simulated boundary and $M_{\text{ext}}$, the logarithmic and the linear shape return the same verdict on all $12{,}375$ configurations, and the logistic shape differs on a single one. The match follows from the linearization of Section~\ref{sec: decomposition}, which states that stability is only governed by the dynamics around the target. All three shapes present the same elasticity at the target point, and the shape therefore does not affect whether the instability occurs.

\begin{figure*}[!tp]
 \begin{center}
    \setlength{\abovecaptionskip}{0.mm}
    \centering
        \begin{subfigure}{0.245\textwidth}
            \centering
            \includegraphics[width=\linewidth]{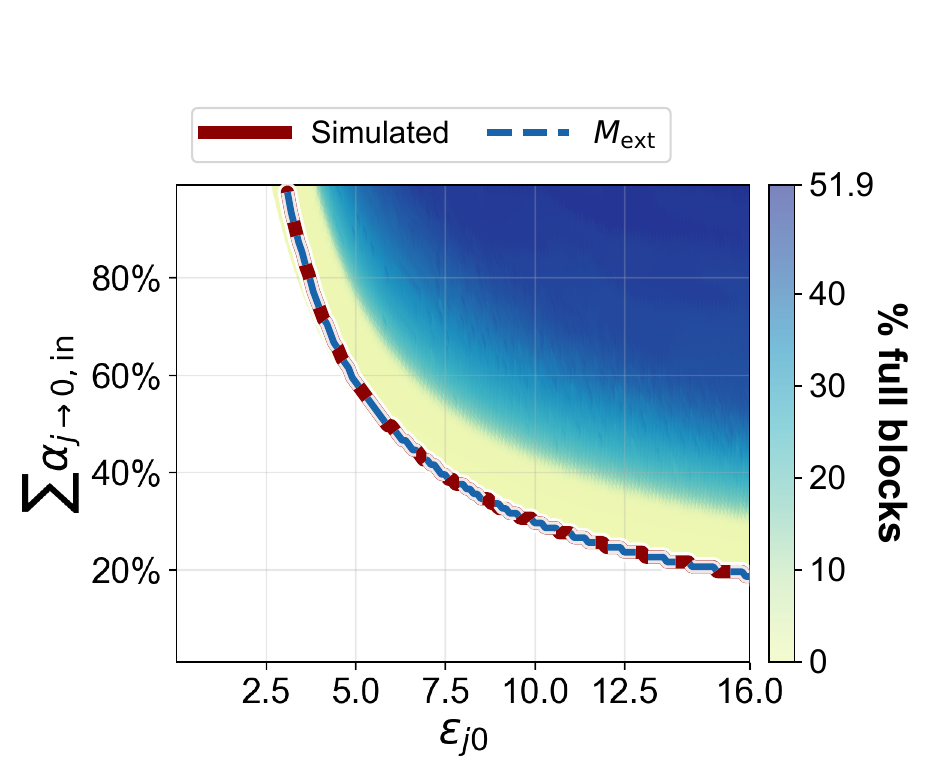}
            \caption{$Q=10$}
            \label{fig: mempool-q10}
        \end{subfigure}
        \hfill
        \begin{subfigure}{0.245\textwidth}
            \centering
            \includegraphics[width=\linewidth]{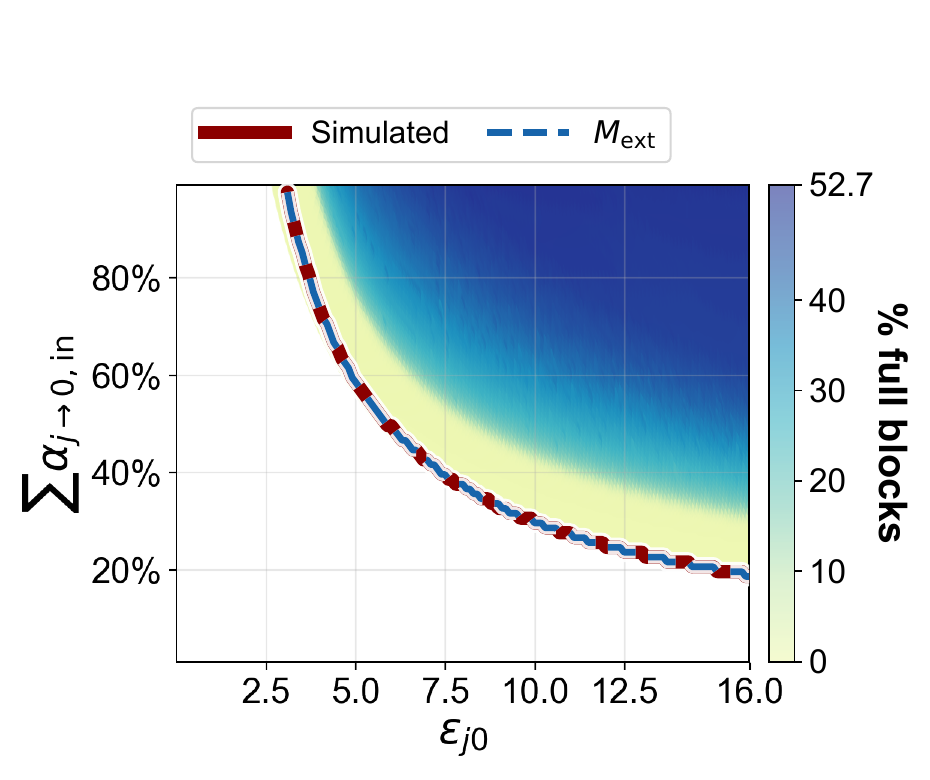}
            \caption{$Q=20$ (main body)}
            \label{fig: mempool-q20}
        \end{subfigure}
        \hfill
        \begin{subfigure}{0.245\textwidth}
            \centering
            \includegraphics[width=\linewidth]{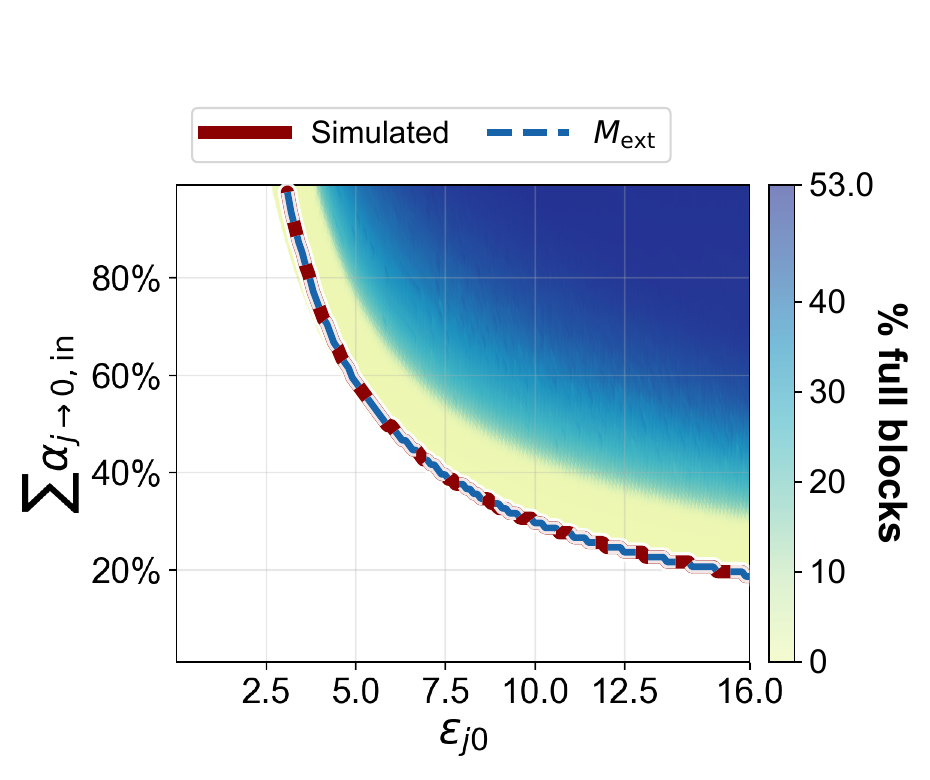}
            \caption{$Q=50$}
            \label{fig: mempool-q50}
        \end{subfigure}
        \hfill
        \begin{subfigure}{0.245\textwidth}
            \centering
            \includegraphics[width=\linewidth]{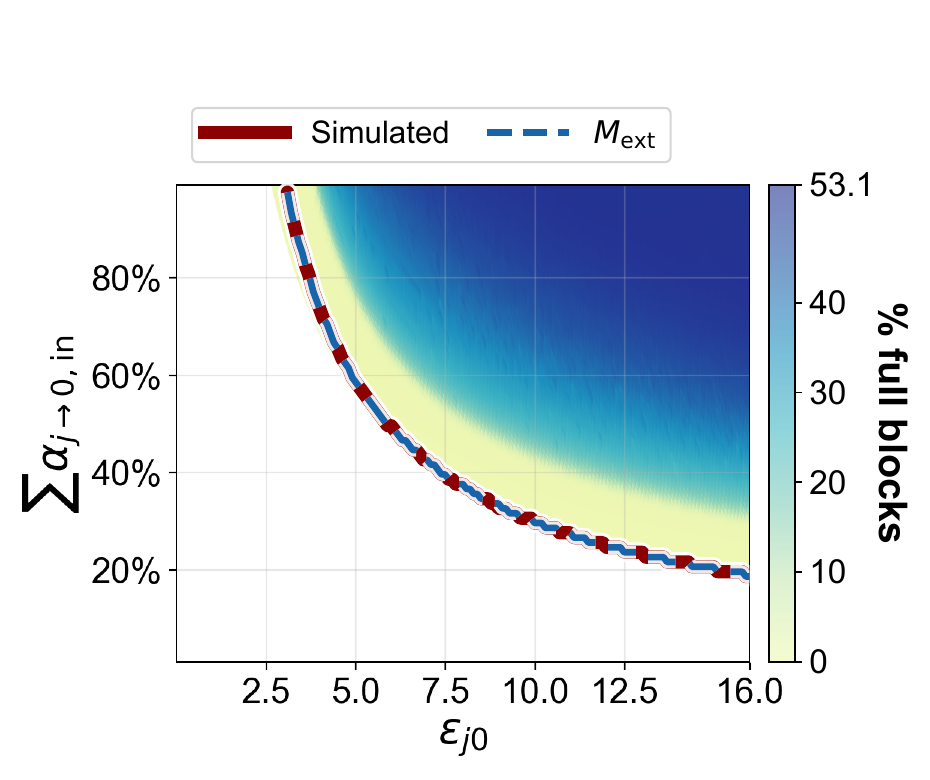}
            \caption{$Q=100$}
            \label{fig: mempool-q100}
        \end{subfigure}
    \caption{The stable region of Figure~\ref{fig: over-correction-spike} at four mempool capacities, under spike latency at $\delta=0.125$. Each panel carries its own color scale, topped at the largest share of full blocks that the panel itself reaches.}
    \label{fig: mempool-sensitivity}
 \end{center}
\end{figure*}
\noindent\textbf{Severity.} The severity of instability is almost the same on the three shapes, even though the three trajectories visit very different elasticities. Over the scanned grid, the logarithmic shape stays within $[15.4, 21.1]$, while the linear shape spans $[8.3, 32.0]$ and the logistic shape spans $[6.8, 32.0]$. However, the realized severity barely moves. The maximum full-block share is $52.5\%$ under the logarithmic shape, $52.7\%$ under the linear shape, and $52.7\%$ under the logistic shape. 
Moreover, all three shapes leave about $6{,}806$ configurations unstable. Of these, $1{,}114$ never fill a block under the logarithmic shape, against $1{,}198$ under the linear shape and $1{,}208$ under the logistic shape. Intuitively, the logarithmic shape still presents a high elasticity at a reduced fee, so the cheap fee attracts a surge that reaches the gas limit more often. The linear and the logistic shape cut the elasticity quickly as the fee falls, so a slightly larger part of their unstable region oscillates without ever saturating. The gap is about one percentage point, and it moves neither the boundary nor the peak.

\subsection{Experiment on Varying Mempool Sizes}\label{sec: mempool-size}
Section~\ref{sec: experimental-setup} gives each chain a mempool of capacity $Q=20$ block gas limits. The capacity belongs to a client implementation rather than to the fee mechanism, so a natural question is how much of Section~\ref{sec: boundary} depends on the value we pick. We therefore rerun the $\delta=0.125$ scan of Figure~\ref{fig: over-correction-spike} at $Q\in\{10,20,50,100\}$, leaving the grid, the shock, the latency shape, and the convergence test of Section~\ref{sec: boundary} untouched. 

\noindent\textbf{The boundary does not move.} Figure~\ref{fig: mempool-sensitivity} shows the stable regions of four settings. All four runs mark the same $8{,}008$ of the $15{,}939$ configurations unstable, and no cell changes verdict. The reason is that the capacity never enters the linearized recurrence of Eq.~\ref{eq:governing-recurrence}. Near the target load, no backlog accumulates, and the mempool stays empty, so the pool begins to bind only after the trajectory has already left the target.

\noindent\textbf{Only the reported severity slightly moves.} A $10\times$ increase in $Q$ raises the largest share of full blocks from $51.9\%$ to $53.1\%$. 
Most of the map does not move at all. Between $Q=10$ and $Q=100$, $59.8\%$ of the unstable configurations report exactly the same share. No configuration moves by more than $5.2$ percentage points. 

\noindent\textbf{Inflation level at different $Q$.} The $6.27\%$ overshoot bound~\cite{leonardos2023optimality} is an asymptotic statement about the average load, while the number we report is the average over a finite window. A deeper pool takes longer to fill, and the load runs above its long-run level for as long as the pool is filling, so the gap between the two grows with $Q$. Take the most saturated configuration of the grid as an example. At $Q=100$, its first $25{,}000$ blocks report an inflation of $6.76\%$. Every $25{,}000$-block window after block $50{,}000$ reports at most $6.44\%$. We extend the simulation steps, and the average over $5{,}000{,}000$ blocks settles at $6.23\%$. At $Q=20$ the same configuration shows no such transient, and its first $25{,}000$ blocks already report $6.20\%$ against $6.14\%$ over $500{,}000$ blocks. A reading above the bound therefore reflects the position of the measurement window rather than the behavior of the mechanism.

\section{Real-World Statistics}\label{sec: appendix-real-world-data}

\subsection{Load-Failure Mapping}\label{sec: mapping}

Section~\ref{sec: evaluation-implication} collects block statistics from Cosmos Hub and Osmosis, and maps a simulated load onto a block failure probability. We describe the mapping here.
A block fails when it does not gather two-thirds of the votes, which CometBFT records as a non-zero round index in the commit message. We count these events over each chain's history, which yields $15{,}831$ events on Cosmos Hub over $9{,}515{,}098$ blocks and $534{,}882$ events on Osmosis over $10{,}186{,}290$ blocks. 

Execution lies on the critical path, so the quantity that delays a vote on block $h$ is the load of block $h-1$. We key the load on GasWanted, for the reason given in Section~\ref{sec: evaluation-implication}, and normalize it by the block gas limit. We group the block history into $100$ load bins of width $0.01$. For each bin, we divide the failures whose predecessor falls in that bin by the blocks in that bin, thereby estimating one failure probability per bin.
The numerator is exact, because we hold every failure event. For the denominator, we count blocks at two rates. At low load (less than $0.5$) blocks are plentiful, so we count every $50$th block of the history and multiply the count by $50$. At high load, there are fewer blocks, so we count all of them. 
The per-bin rate rises with load over most of the range. Where it does not, we pool the affected bins by isotonic regression, because a heavier block is expected to receive fewer votes.

\begin{table*}[ht]
\centering
\begin{threeparttable}
\caption{Empirical configuration and data sources of representative parachains and zones. \textit{Inbound Ratio} is explicitly isolated because it governs the Channel (III) ($\Lambda_i^{\text{delay}}$), the primary driver of \textit{delay-amplified over-correction} and the critical metric for stability analysis.}
\label{table: data}
\small
\setlength{\tabcolsep}{0pt}
\renewcommand{\arraystretch}{1.15}
\begin{tabular*}{\textwidth}{@{\extracolsep{\fill}} l cccc @{}}
\toprule
\textbf{Chain / Zone} &
\textbf{Total Ratio (\%)} &
\textbf{Inbound Ratio (\%)} &
\textbf{Update Rate ($\delta$)} &
\textbf{Delay ($d_{\max}$)} \\
\midrule
\multicolumn{5}{@{}l}{\textit{Polkadot}\tnote{1}} \\
\quad Asset Hub & 27.4 & \textbf{13.4} & $7.5 \times 10^{-5}$ & 5 \\
\quad Bifrost   & 18.6 & \textbf{6.0}  & $7.5 \times 10^{-5}$ & 5 \\
\quad Hydration & 21.6 & \textbf{13.3} & $7.5 \times 10^{-5}$ & 5 \\
\quad Moonbeam  & 43.3 & \textbf{25.5} & $0.004$              & 5 \\
\addlinespace[4pt]
\multicolumn{5}{@{}l}{\textit{Cosmos}\tnote{2}} \\
\quad Cosmos Hub & 23.9 & \textbf{21.6} & $0.125$ & 10 \\
\quad Osmosis    & 26.1 & \textbf{24.8} & $0.1$   & 25 \\
\bottomrule
\end{tabular*}
\begin{tablenotes}
    \footnotesize
    \item[1] \textbf{Polkadot:} traffic ratios from Subscan
    (\shorturl{assethub-polkadot.subscan.io}, \shorturl{bifrost.subscan.io},
    \shorturl{hydration.subscan.io}, \shorturl{moonbeam.subscan.io}). Update
    rates from runtime code:
    \shorturl{docs.rs/polkadot-runtime-common/latest/src/polkadot_runtime_common/lib.rs.html}
    for Asset Hub, Bifrost, and Hydration, and
    \shorturl{github.com/moonbeam-foundation/moonbeam/blob/runtime-4401/runtime/moonbeam/src/lib.rs}
    for Moonbeam. Cross-chain latency from~\cite{fujihara2025understanding},
    with block intervals measured from chain headers.
    \item[2] \textbf{Cosmos:} traffic ratios measured on-chain by sampling
    $60{,}000$ Cosmos Hub blocks and $80{,}000$ Osmosis blocks and classifying
    each transaction's gas by its IBC message content.
    Update rates from Proposal~970
    (\shorturl{forum.cosmos.network/t/proposal-970-accepted-update-feemarket-max-block-utilization-to-75m-gas/14227})
    for Cosmos Hub and \shorturl{docs.osmosis.zone/overview/features/eip-1559}
    for Osmosis. Cross-chain latency measured on-chain from every inbound IBC
    packet, reported as the mean delay in destination-chain blocks. Cosmos Hub
    is measured over the 2025 calendar year, and Osmosis from April 15 to
    August 6, 2026, the earliest date a public archive indexes its transactions.
\end{tablenotes}
\end{threeparttable}
\end{table*}

\subsection{Cosmos and Polkadot Statistics}\label{sec: statistics}
The collection of real-world operational data serves a critical objective in our study. 
It validates the parameter ranges used in our general stability analysis, ensuring that our iterations of cross-chain ratios, latencies, and elasticities cover the realistic operating conditions of modern multi-chain blockchains. 

However, acquiring high-quality, explicit data for individual chains presents significant challenges. While systems like Polkadot and Cosmos host hundreds of parachains and zones, a vast majority remain inactive, and centralized data aggregators often report only system-wide statistics, which obscures the specific load dynamics of individual chains. 
To address this, we conducted a detailed investigation. 
Specifically, we combine blockchain explorer data with direct on-chain measurement. For Polkadot, we extract XCM message counts and total load from \textit{Subscan}. For Cosmos, we sample blocks from archive nodes and classify every transaction by its IBC message content, which yields gas-denominated traffic shares. 
We take the update rates $\delta$ from official protocol documentation and runtime code on GitHub, and we measure block intervals and cross-chain latencies directly from on-chain data. This effort
allowed us to reconstruct the operational profiles of the most active and representative chains, with a summary presented in Table~\ref{table: data}.

\noindent\textbf{Cross-chain ratio.} A straightforward proxy for cross-chain ratio is the message count. However, it is an imperfect proxy for load. An XCM message of Polkadot does not map one-to-one onto a sub-transaction, and IBC packets of Cosmos batch transfers. We therefore measure the ratio directly in gas wherever the chain records it.
For Cosmos chains, every transaction's gas and message types are recoverable on-chain. 
We can classify every transaction by the IBC message types its body carries, and compute the inbound share of executed gas directly.
However, the total blocks consume more than $1$ TB of data for both Cosmos Hub and Osmosis.
Therefore, we sample blocks uniformly at random inside each chain's measurement window. The samples hold $60{,}000$ blocks of Cosmos Hub's 2025 calendar year and $80{,}000$ blocks of Osmosis between April 15 and August 6, 2026.
For Cosmos Hub, packet-receive transactions execute $12.9\%$ of all gas, and IBC infrastructure transactions (client updates, acknowledgements, timeouts) execute another $8.7\%$. We thereby assume an inbound ratio of $21.6\%$. Outbound transfers execute $2.4\%$, so the total cross-chain share measures $23.9\%$ in gas.
For Osmosis, the same three classes execute $15.7\%$, $9.1\%$, and $1.3\%$, giving an inbound ratio of $24.8\%$ and a total cross-chain share of $26.1\%$.

We sample this quantity because to get the gas share, we need the transaction bodies and the per-transaction gas of every block. 
Sampling is enough here. Drawing the sample again, we would land within $0.5$ points of $21.6\%$ on Cosmos Hub and within $1.1$ points of $24.8\%$ on Osmosis. We also reran the stability analysis at both ends of those two intervals. Cosmos Hub's maximum stable update rate moves between $0.061$ and $0.052$, and Osmosis's between $0.023$ and $0.017$, so both chains stay unstable at their deployed rates throughout. 

For Polkadot parachains, per-message gas attribution is not publicly recoverable, so we retain the message-count proxy from \textit{Subscan}. The proxy overcounts multi-message actions and therefore only overestimates coupling. We use it as an explicit upper bound.

\noindent\textbf{Update rates.} 
We extract the precise \eip~update rates currently employed by these parachains and zones from their official documentation and codebases. 
For the selected Cosmos zones, the update parameters are explicitly defined. 
Conversely, determining the exact rates for Polkadot parachains presents a challenge, as many projects do not enforce documentation for internal fee parameters. We attribute this opacity to the modular nature of Polkadot, which provides ready-to-deploy Substrate templates that developers often adopt without modification.
To address this, we traced the initial default configurations inherited by these chains (sources verified in Table~\ref{table: data}). Our investigation confirms the universal adoption of an extremely conservative $\delta$, a design choice that is also explicitly acknowledged within the protocol's technical specifications\footnote{\url{https://research.web3.foundation/Polkadot/overview/token-economics}}.

\noindent\textbf{$d_{max}$ and latency shape.} In our model, $d_{max}$ measures the number of block steps, in destination-chain blocks, between the commitment of a sub-transaction on the source chain and the inclusion of its counterpart on the destination chain. Generally, the process involves three sequential phases: 1) Proof generation, where the source chain generates a commitment proof after a
stabilization period to ensure finality (e.g., the \emph{delayPeriodBlocks} parameter in Cosmos); 2) Cross-chain relay, where the proof is forwarded across the network; and 3) Destination inclusion, where validators verify the proof and execute the transaction.
We measure this delay through on-chain IBC packets for Cosmos. IBC packets carry a channel identifier and a sequence number, so the transaction on the destination chain can be matched exactly to the original send. The delay in blocks is then the elapsed time between the two commitments divided by the destination chain's block interval. 
For both chains, we measure packets across each chain's active channels, and report the average number of steps it takes to receive a packet from other chains. 

For Cosmos Hub, we observe every IBC packet that Cosmos Hub received during the 2025 calendar year, and every packet that Osmosis received between April 15 and August 6, 2026.
The windows differ because no public archive indexes Osmosis transactions earlier. 
For Cosmos Hub, the statistics cover $822{,}514$ packets across $102$ channels. The mean delay is $59.3$ seconds, which is $10.07$ blocks at the $5.89$-second Cosmos Hub block interval. The traffic is concentrated and uneven. The largest channel carries $47.2\%$ of the packets and delivers in $7.9$ blocks on average, while the $97$ smallest channels carry $27.9\%$ of the packets and average $12.6$ blocks.
For Osmosis, the statistics cover $286{,}717$ packets across $119$ channels, and the mean delay is $29.2$ seconds.  Osmosis's $1.15$-second block interval places the latency at $25.47$ blocks.
We set $d_{max}$ to the measured mean, $10$ for Cosmos Hub and $25$ for Osmosis.
The weights $w_d$ are a separate question. They describe how the load generated in one source block spreads over destination blocks, which the destination-side record cannot recover. Recovering it would require pairing every arrival with its source block, and therefore an archive for every source chain, which is practically less feasible. We treat the shape as a sensitivity parameter. The main text places all weight at $d_{max}$. Appendix~\ref{sec: additional-experiment-varying-dmax-and-chain-number} repeats the experiments under uniform and bimodal weights, and both give milder instability.

For Polkadot, XCM messages lack an equally clean on-chain pairing, so we rely on a recent empirical study of cross-parachain latency~\cite{fujihara2025understanding}, which reports 25 to 60 seconds.
Parachain block intervals are 5.9 to 6.6 seconds. We therefore use the median of the reported cross-chain latency and set $d_{max}=7$ for Polkadot.

\section{Protocol-Level Migation}\label{sec: protocol-level-mitigation}

This section develops the mitigation of Section~\ref{sec: mitigation}. Inspired by the observations in Figures~\ref{fig: nonuniform-map-e} and~\ref{fig: nonuniform-map-f}, we add the compensation term and show how it affects the governing recurrence.
We then fix the exact term assuming accurate elasticities, and explain how precise the estimation of elasticities must be.

\noindent\textbf{Synthesizing the counter-force.} Recall the mechanism of Pathway~1. The load that chain $i$ reads at time $t$ mixes in the inbound flow priced at $x_i(t\!-\!d)$. 
Figures~\ref{fig: nonuniform-map-e} and~\ref{fig: nonuniform-map-f} point to the remedy. In that experiment, we raised a source-side elasticity to probe Channel~(IV), and the peer chains became more stable instead. The raised elasticity also strengthened the peers' Channel~(I), and Channel~(I) reverses at the moment the price crosses the equilibrium. From that moment on, it opposes the stale push of Channel~(III) and erodes the overshoot. In that experiment, however, the counter-force was an accident of a peer-facing elasticity, which no chain controls. The natural question is whether a chain can synthesize the same counter-force from data it owns.

We synthesize the counter-force by re-pricing. The inbound load that arrives at time $t$ with lag $d$ was generated under the price $x_i(t\!-\!d)$. The inbound flow responds to the destination price with $\Lambda_i^{\text{delay}}$, so this load and the load that the same users would generate under the current price $x_i(t)$ differ by $\Lambda_i^{\text{delay}}\,(\Delta x_i(t)-\Delta x_i(t\!-\!d))$ in the normalized load signal. We subtract this gap from the \eip~controller's signal, at every lag and weighted by $w_d$. The controller then reads the load as if the inbound flow had been priced at $x_i(t)$, and the stale component is removed from the feedback. The gap contains only price differences, which the chain computes from its own base-fee history. The compensated rule is
\begin{equation}\label{eq:compensated-update}
\begin{aligned}
x_i(t\!+\!1)=x_i(t)\Big(1+\delta_i\Big[\;&\frac{L_i(t)-T_i}{T_i}\\
&-H_i\sum_{d=1}^{d_{\max}} w_d\,\frac{P_i(t)-P_i(t\!-\!d)}{P_i(t)}\Big]\Big),
\end{aligned}
\end{equation}
$H_i \ge 0$ is chain $i$'s compensation gain. The strength is left as a parameter because $\Lambda_i^{\text{delay}}$ contains the elasticities and is not directly known. We set $H_i$ after the linearization. The fees $P_i(t)$ and $P_i(t\!-\!d)$ are the chain's own base-fee history, and the weights $w_d$ are its measured inbound latency distribution. A chain therefore deploys the rule without cross-chain coordination.

\noindent\textbf{The term in the recurrence.} We linearize Eq.~\ref{eq:compensated-update} around $\mathbf{x}^*=\mathbf{1}$. To first order, $[P_i(t)-P_i(t\!-\!d)]/P_i(t)\approx\Delta x_i(t)-\Delta x_i(t\!-\!d)$, and the equilibrium price cancels. Since $\sum_d w_d=1$, the compensation splits into an instantaneous part, $H_i\,\Delta x_i(t)$ with full weight, and a delayed part, $H_i\,\Delta x_i(t\!-\!d)$ with weight $w_d$ at each lag. Substituting the load decomposition of Eq.~\ref{eq:decomposition} and grouping terms by lag gives
\begin{align}\label{eq:compensated-recurrence}
\Delta x_i(t\!+\!1) \approx{}\;
& \bigl(1 - \delta_i[\Lambda_i^{\text{inst}}+H_i]\bigr)\,\Delta x_i(t)
 - \delta_i\!\sum_{j\neq i}\Lambda_{ij}^{\text{inst}}\,\Delta x_j(t) \nonumber\\
& - \delta_i\!\sum_{d=1}^{d_{\max}} w_d\,\bigl[\Lambda_i^{\text{delay}}-H_i\bigr]\,\Delta x_i(t\!-\!d) \nonumber\\
& - \delta_i\!\sum_{j\neq i}\sum_{d=1}^{d_{\max}} w_d\,\Lambda_{ji}^{\text{delay}}\,\Delta x_j(t\!-\!d).
\end{align}
Compared with Eq.~\ref{eq:governing-recurrence}, exactly two coefficients change. Channel~(I) grows from $\Lambda_i^{\text{inst}}$ to $\Lambda_i^{\text{inst}}+H_i$, and Channel~(III) shrinks from $\Lambda_i^{\text{delay}}$ to $\Lambda_i^{\text{delay}}-H_i$. Channels~(II) and~(IV) are unchanged, because the term contains no peer price. The term therefore performs the intended strengthening and weakening, in equal amounts of $H_i$. Setting $H_i=0$ recovers plain \eip. Substituting the two new diagonal coefficients into $\mathbf{M}_{\mathrm{ext}}$ gives the exact criterion for the compensated system.
\begin{figure*}[t]
    \centering
    \includegraphics[width=0.9\textwidth]{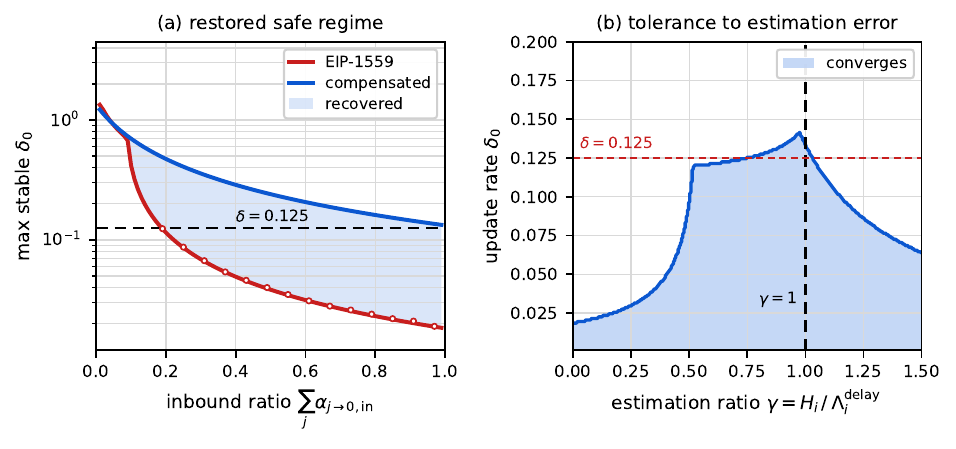}
    \captionsetup{justification=justified}
    \caption{The compensated controller applied to the setting of Figure~\ref{fig: over-correction-boundary}, under spike latency with $d_{\max}=5$. Panel~(a) plots the largest stable update rate of Chain~0 against the inbound ratio at elasticity $16$, and the shaded band is the regime that compensation recovers. Panel~(b) fixes an inbound ratio of $0.99$ and maps the stable region over the update rate $\delta_0$ and the estimation ratio $\gamma=H_i/\Lambda_i^{\text{delay}}$.}
    \label{fig: mitigation}
\end{figure*}
\noindent\textbf{The exact gain.} We now set $H_i$, assuming for this step that the elasticities are known. A nonzero delayed coefficient still leaves residual over-correction that some update rate can excite, so we set the coefficient to zero. The choice
\begin{equation}\label{eq:exact-gain}
    H_i^{*}=\Lambda_i^{\text{delay}}=\sum_{j\neq i}\epsilon_{ji}\,\alpha_{j\to i,\text{in}}
\end{equation}
cancels the delayed self-feedback at every lag, because the coefficient $w_d[\Lambda_i^{\text{delay}}-H_i]$ vanishes for each $d$ separately. The cancellation is therefore independent of the delay shape. The cancelled gain moves into Channel~(I), whose coefficient becomes $\Lambda_i^{\text{inst}}+\Lambda_i^{\text{delay}}$. Written out with Table~\ref{table: forces},
\begin{equation}\label{eq:mixed-elasticity}
\Lambda_i^{\text{inst}}+\Lambda_i^{\text{delay}}
=\lambda_{ii}\alpha_{i,\text{local}}
+\sum_{j\neq i}\lambda_{ij}\alpha_{i\to j,\text{out}}
+\sum_{j\neq i}\epsilon_{ji}\alpha_{j\to i,\text{in}},
\end{equation}
and the load shares $\alpha$ sum to one. The combined coefficient is a convex combination of the elasticities of chain $i$'s flows, so it is at most the largest elasticity $\lambda_{\max,i}$. The self-feedback condition becomes $\delta_i\,\lambda_{\max,i}<2$, which is the monolithic condition of Eq.~\ref{eq: monolithic-boundary}. The compensated chain behaves like a monolithic chain whose elasticity is the traffic-weighted mixture of its flows. 

\noindent\textbf{Why the gain is an estimate.} $H_i^{*}$ mixes two kinds of inputs. The inbound shares $\alpha_{j\to i,\text{in}}$ are measured on-chain, as in Table~\ref{table: data}. The destination-side elasticities $\epsilon_{ji}$ describe user behavior and are not observable. A deployer therefore computes the compensation term from estimated elasticities $\hat{\epsilon}_{ji}$, as $H_i=\sum_{j\neq i}\hat{\epsilon}_{ji}\,\alpha_{j\to i,\text{in}}$, while the actual delayed coefficient in the recurrence remains $\Lambda_i^{\text{delay}}$. We measure the estimation accuracy by the ratio $\gamma=H_i/\Lambda_i^{\text{delay}}$. $\gamma=1$ means the estimate is exact, and $\gamma=0$ is plain \eip. The residual Channel~(III) coefficient is $\Lambda_i^{\text{delay}}-H_i=(1-\gamma)\Lambda_i^{\text{delay}}$, and its sign determines the failure mode.
When $\gamma<1$, the residual is negative. The chain behaves like an uncompensated chain with a smaller delayed coefficient. Part of the over-correction survives, and the tolerable $\delta_i$ lies between the plain bound and the restored bound.
When $\gamma>1$, the residual is positive. The delayed feedback pushes the price in the direction of the old deviation, so successive echoes reinforce each other. The system gains a positive-feedback channel that the decomposition of Section~\ref{sec: decomposition} does not contain. Stability therefore holds on an interval of $\gamma$ around $1$. An elasticity estimate keeps the chain stable exactly when its resulting $\gamma$ falls inside the interval.

\noindent\textbf{The exact gain restores the monolithic bound.} We rerun the system of Figure~\ref{fig: over-correction-boundary} with $H_i=H_i^{*}$ on every chain. The topology, the elasticities, the spike latency, and the one-block $1.01\times$ shock stay unchanged. All $28{,}314$ configurations of the two grids converge, the maximum full-block share drops from $52.7\%$ to zero, and the classification agrees with $\rho(\mathbf{M}_{\mathrm{ext}})$ at every point. Figure~\ref{fig: mitigation}(a) plots the boundary at elasticity $16$. The plain \eip~boundary falls to $0.019$ at an inbound ratio of $0.99$. The compensated boundary follows $2/(\Lambda_0^{\text{inst}}+\Lambda_0^{\text{delay}})$, the monolithic bound at the mixed elasticity of Eq.~\ref{eq:mixed-elasticity}. The shaded band between the two curves is the recovered regime.

\noindent\textbf{Estimation tolerance.}
Next, to evaluate whether the mitigation works if we do not have an accurate estimation of elasticities, we fix the hardest-to-converge configuration, with an inbound ratio of $0.99$ at elasticity $16$, and vary $H_i$ to cover different levels of $\gamma$. We scan the plane $(\gamma,\delta_0)$, and evaluate, for each $\gamma$, the maximum tolerable $\delta_0$.
The stable region peaks near $\gamma=1$ and is asymmetric. The left edge falls gently, because under-estimation only shrinks the tolerable $\delta_0$. The right edge drops steeply, because over-estimation flips the sign of the residual.
Note that with the vanilla \eip~rule, the maximum $\delta$ is $0.019$ for this setting.
With the augmented rule, the system can tolerate a $\delta=0.125$, with $\gamma\in[0.75,1.030]$. It means if $\delta$ is set to $0.125$, an overestimate of $3\%$ destabilizes the chain, and an underestimate of up to $25\%$ is tolerated. At $\delta=0.05$, the interval widens to a lower edge of $0.405$ and an upper edge beyond the scanned limit of $1.5$.
In practice, the elasticity should be estimated low rather than high, and $\delta$ should be set with a margin.

\begin{table}
\renewcommand\arraystretch{1.15}
\caption{Compensation on the deployed configurations under elasticity $16$. The stable $\gamma$ interval is evaluated at each chain's deployed $\delta$.}
\label{table: mitigation-deployed}
\centering
\small
\setlength{\tabcolsep}{3.5pt}
\begin{tabular}{@{}lcccc@{}}
    \toprule
    \textbf{Chain} & \textbf{Deployed $\delta$} & \makecell{\textbf{Max $\delta$}\\(\eip)} & $H_i^{*}$ & \textbf{Stable $\gamma$} \\
    \midrule
    Cosmos Hub & $0.125$ & $0.056$ & $3.456$ & $[0.23,\,2.66]$ \\
    Osmosis    & $0.1$   & $0.019$ & $3.968$ & $[0.30,\,2.88]$ \\
    \bottomrule
\end{tabular}
\end{table}

\noindent\textbf{The deployed chains.} Finally, we rerun Cosmos Hub and Osmosis. As in Section~\ref{sec: evaluation-implication}, we assume Ethereum's elasticity of $16$ and use each chain's measured inbound ratio and latency from Table~\ref{table: data}. Recall that both chains are unstable at their deployed rates under plain \eip. We therefore ask how accurate the elasticity estimate must be for the compensation to make the deployed rates stable. Under the assumed elasticity, the exact terms are $H_i^{*}=3.456$ for Cosmos Hub and $3.968$ for Osmosis, and $\gamma$ measures the deployer's estimate relative to them.
Table~\ref{table: mitigation-deployed} reports the stable $\gamma$ intervals. Cosmos Hub stays stable at its deployed $\delta=0.125$ for $\gamma\in[0.23,2.66]$, and Osmosis at its deployed $\delta=0.1$ for $\gamma\in[0.30,2.88]$. The intervals are wider than in the worst case of Figure~\ref{fig: mitigation}(b), because the two chains' inbound ratios are near $25\%$ rather than $99\%$. In other words, a deployer estimates the elasticity, computes $H_i$ from Eq.~\ref{eq:exact-gain}, and the deployed rates become stable as long as the estimate is within a factor of about three of the true value.

\end{document}